\documentclass[prx,aps,twocolumn,superscriptaddress,showpacs,floatfix,longbibliography]{revtex4-2}

\usepackage{amsmath,amssymb,amsfonts,bbm,float,graphics,epsfig,epstopdf,color,verbatim,tabularx,bm,multirow,hyperref,ulem,times,subfigure}
\hypersetup{
  colorlinks, linkcolor={blue},
  citecolor={blue}, urlcolor={blue}
}

\usepackage{textcomp}
\usepackage{times}
\usepackage{tabularx}
\usepackage{graphicx}
\usepackage{float}
\usepackage{latexsym,amsmath,amssymb,bm,euscript}
\usepackage{color}
\usepackage{subfigure}
\usepackage{epstopdf}

\graphicspath{{./Figs/}}

\begin{document}

\title{From Fractional Quantum Anomalous Hall Smectics to Polar Smectic Metals: Nontrivial Interplay Between Electronic Liquid Crystal Order and Topological Order in Correlated Topological Flat Bands}

\author{Hongyu Lu}
\altaffiliation{The two authors contributed equally to this work.}
\affiliation{Department of Physics and HK Institute of Quantum Science \& Technology, The University of Hong Kong, Pokfulam Road, Hong Kong SAR, China}

\author{Han-Qing Wu}
\altaffiliation{The two authors contributed equally to this work.}
\affiliation{Guangdong Provincial Key Laboratory of Magnetoelectric Physics and Devices, School of Physics, Sun Yat-sen University, Guangzhou 510275, China }

\author{Bin-Bin Chen}
\email{bchenhku@hku.hk}
\affiliation{Department of Physics and HK Institute of Quantum Science \& Technology, The University of Hong Kong, Pokfulam Road, Hong Kong SAR, China}

\author{Kai Sun}
\email{sunkai@umich.edu}
\affiliation{Department of Physics, University of Michigan, Ann Arbor, Michigan 48109, USA}

\author{Zi Yang Meng}
\email{zymeng@hku.hk}
\affiliation{Department of Physics and HK Institute of Quantum Science \& Technology, The University of Hong Kong, Pokfulam Road, Hong Kong SAR, China}

\begin{abstract}
Symmetry-breaking orders can not only compete with each other, but also be intertwined, and the intertwined topological and symmetry-breaking orders make the situation more intriguing.
This work examines the archetypal correlated flat band model on a checkerboard lattice at filling $\nu=2/3$ and we find that the unique interplay between smectic charge order and topological order gives rise to two novel quantum states. 
As the interaction strength increases, the system first transitions from a Fermi liquid into FQAH smectic (FQAHS) state, where FQAH topological order coexists cooperatively with smectic charge order with enlarged ground-state degeneracy and interestingly, the Hall conductivity is $\sigma_{xy}=\nu=2/3$, different from the band-folding or doping scenarios. 
Further increasing the interaction strength, the system undergoes another quantum phase transition and evolves into a polar smectic metal (PSM) state. This emergent PSM is an anisotropic non-Fermi liquid, whose interstripe tunneling is irrelevant while it is metallic inside each stripe.
Different from the FQAHS and conventional smectic orders, this PSM spontaneously breaks the two-fold rotational symmetry, resulting in a nonzero electric dipole moment and ferroelectric order. 
In addition to the exotic ground states, large-scale numerical simulations are also used to study low-energy excitations and thermodynamic characteristics.  
We find that the onset temperature of the incompressible FQAHS state, which also coincides with the onset of non-polar smectic order, is dictated by the magneto-roton modes.
Above this onset temperature, the PSM state exists at an intermediate-temperature regime.
Although the $T=0$ quantum phase transition between PSM and FQAHS is first order, the thermal FQAHS-PSM transition could be continuous. 
We expect the features of the exotic states and thermal phase transitions could be accessed in future experiments.
\end{abstract}

\date{\today }
\maketitle

\section{Introduction}
Apart from competing with each other, the symmetry breaking orders can be intertwined, which broadly exists in correlated systems and is of great significance, such as in the complex finite-temperature phase diagram of high-temperature superconductors with different onset temperatures~\cite{Fradkin2015interwine, dai2015sc, Keimer2015highTc, Wang2015coexist, Vishik2018SC_CDW, Ortiz2020SC, Xu2024hubbard}. Moreover, the interplay between symmetry-breaking order and topological order has been a focal point in the study associated with quantum Hall effects.  
Traditionally, they are perceived as competing orders as well, because they're governed by different physics principles -- Landau's symmetry-breaking paradigm for the former and topologically nontrivial quantum wavefunctions for the latter. 
However, these two distinct types of orders can also coexist cooperatively, as theoretical studies have shown~\cite{Kivelson1986, Halperin1986_QHC, Halperin1989_QHC}. 
Over the past thirty years, extensive research has been devoted to investigating such coexistence, both theoretically~\cite{Balents1996_FQHC, Murthy2000_QHC, Fradkin1999_Hall_crystal, Fradkin2002_smecticHall1, Fradkin2002_smecticHall2, Murthy2000_CFC, Jain2006_CFC,You2014nematic, Jain2020_QHC,trungFractionalization2021} and experimentally~\cite{West2004_FQHC, West2010_FQHC, xiaEvidence2011, duObservation2019, xuAnomalous2020,Shingla2023_bubble,  Benjamin2016nematic, Samkharadze2016_nematicHall}, where charge order and topological order are strongly intertwined together.  
Different charge orders based on their symmetry-breaking patterns are theoretically predicted, including nematics (breaking only the rotational symmetry), smectics (breaking rotational symmetry and translational symmetry along one spatial direction), and crystals (breaking the 2D translational symmetry).
However, in Landau level systems, only the coexisting nematic order has been realized~\cite{Benjamin2016nematic, Samkharadze2016_nematicHall, yangMicroscopic2020, Pu2024nematic}.

Parallel to Landau level systems, integer and fractional quantum anomalous Hall (IQAH and FQAH) states at {\it zero magnetic field} -- known as integer and fractional Chern insulators -- have been proposed~\cite{Haldane1988_QAH,KSun2011_model,DNSheng2011_fci,Neupert2011_fci, Bernevig2011fci, XGWen2011_fci}  and realized~\cite{CZChang2013_QAH,caiSignature2023,park2023_fqah,zengThermodynamic2023,xu2023_fci,multilayer_graphene_fqah, Kang2024fqsh}. 
Different from Landau levels, can the QAH states, governed by lattice space group symmetry, coexist cooperatively with the translational symmetry breaking order?

Theoretically, such coexistence is feasible~\cite{XYSong2023fqahc}. One exotic case is in rhombohedral pentalayer graphene/hBN moir\'e superlattices, where the mean-field studies propose that the narrow $C=-1$ Chern band can exhibit interaction-driven spontaneously time-reversal and translational symmetry breaking, which might be stable even without moir\'e potential and can give rise to anomalous Hall crystals~\cite{dong2023graphene, BRZhou2023_AHC,Dong2023_AHC,kwanMoire2023, ashvin2024ahc2, dong2024ahc, tan2024ahc}.  
For fractional fillings with coexisting charge density wave (CDW), current microscopic discoveries only report the states with $\sigma_{xy}\neq \nu C_\mathrm{band}$. 
For instance, studies of twisted MoTe$_2$ bilayers at $\nu=1/2$~\cite{Sheng2024QAHC} and AB-stacked MoTe$_2$/WSe$_2$ at $\nu=2/3$~\cite{pan2022ahc} find the CDW order fold the original Chern band, leading to effective integer fillings of Chern bands and integer Hall conductance, with similar experimental signatures in twisted graphene systems under finite magnetic field~\cite{xie2021fci,polshyn2022tcdw}.
FQAH crystals with fractional Hall conductivity at fractional filling with $\sigma_{xy}\neq \nu C_\mathrm{band}$ have been reported in the flat-band model on a triangle lattice, where only a portion of the particles contribute to the formation of CDW order while the remaining particles unoccupied by the CDW order contribute to the topological order~\cite{Kourtis2014, stefanos2018fqahc}.

The FQ(A)H states are known as incompressible liquids, whose collective excitations with magneto-roton modes are closely analogous to Feynman's theory of superfluid~\cite{GMP1985, GMP1986, Zhang1989composite}. 
In the composite-boson picture, FQH states are often interpreted as superfluid of composite bosons.
Here, we can make an intuitive while less-mentioned comparison between supersolid with coexisting superfluid and CDW, and charge ordered FQ(A)H states.
Among the microscopic mechanisms to realize the supersolid state, one way is to dope from the solid phase~\cite{Sengupta2005supersolid, Raczkowski2009supersolid, Xi2011supersolid}. 
This scenario is similar to the realization of FQAH crystal from doping a CDW~\cite{Kourtis2014, stefanos2018fqahc}, but again, with Hall conductivity $\sigma_{xy}\neq \nu C_\mathrm{band}$. 
Another interesting scenario for supersolid comes from transitions triggered by the roton instability in superfluid~\cite{Chomaz2018roton, Tanzi2019supersolid, Fabien2019supersolid,Chomaz2019supersolid,  zhang2019supersolid, Bland2022supersolid, Alana2023supersolid}, which could be continuous. 
Therefore, one would expect the roton-triggered charge order in FQ(A)H states realized in a continuous way such that the Hall conductivity would not change. However, before this work, such realization is only reported in the FQH nematics where the magnetoroton mode in isotropic FQH states goes soft in the long-wavelength limit, with broken rotational symmetry but preserved translational symmetry~\cite{Pu2024nematic, You2014nematic,yangMicroscopic2020}.
The FQ(A)H state with broken translational symmetry and  $\sigma_{xy} = \nu$ is still lacking either numerically or experimentally.

Furthermore, a more critical challenge for such intertwined states lies in understanding their thermodynamic properties at finite temperatures --- a largely uncharted territory in both theoretical and numerical studies. The primary obstacle stems from the absence of unbiased theoretical and numerical tools capable of providing reliable predictions at finite temperatures. 
Since experimental studies are exclusively conducted at finite temperatures, and considering that such translational symmetry breaking has not been probed in experiments despite the recent breakthrough in realizing FQAH states~\cite{caiSignature2023,park2023_fqah,zengThermodynamic2023,xu2023_fci,multilayer_graphene_fqah, Kang2024fqsh}, such theoretical knowledge is of paramount importance. 
In the previous work, the thermodynamics of IQAH and FQAH without symmetry breaking are studied, where the proliferation of charge-neutral exciton (for IQAH) and magnetoroton (for FQAH) modes together with thermal fluctuations lead to charged excitations at temperatures much lower than the charge gap~\cite{HYL2023_thermoFQAH,linExciton2022, panThermodynamic2023}. 
The thermodynamic phase diagram and properties of the intertwined states would certainly be more intriguing.

\begin{figure}[htp!]
	\centering		
	\includegraphics[width=\columnwidth]{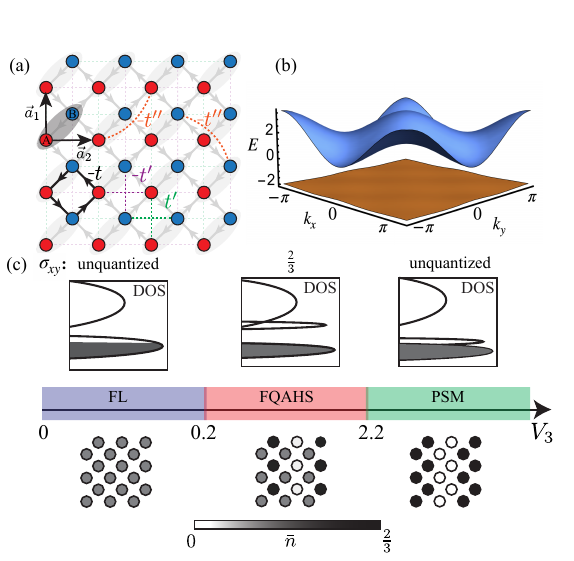}
	\caption{ \textbf{Model and phase diagram.} (a) Checkerboard lattice with the primitive vectors $\mathbf{a_1}= (0,1)$, $\mathbf{a_2}=(1,0)$. Different hoppings are denoted by different colors and the arrows represent the directions of the loop current.
(b) The band dispersion of the tight-binding Hamiltonian, with the lower band nearly flat. (c) Phase diagram at $\nu=2/3$ as NNNN interaction strength $V_3$ is varied. FL represents a $C_4$-symmetric Fermi liquid state with uniform charge distribution, and FQAHS represents a fractional quantum anomalous Hall smectic state with gapped bulk, a unidirectional stripe order and quantized $\sigma_{xy}=2/3$, and PSM represents a non-Fermi-liquid polar smectic metal phase with a ferroelectric stripe order. The schematic density of states (DOS) for the three phases are shown in the boxes above the phase diagram.}
	\label{fig_phasediagram}
\end{figure}

\begin{figure}[htp!]
	\centering		
	\includegraphics[width=\columnwidth]{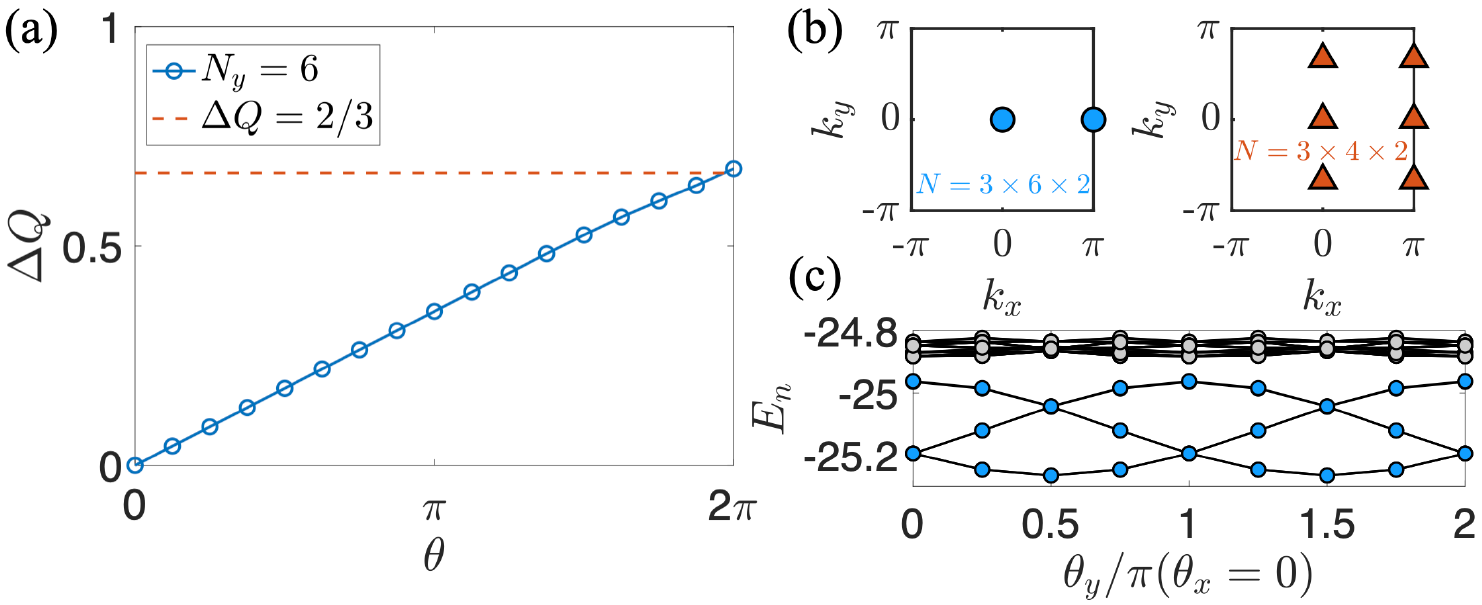}
	\includegraphics[width=\columnwidth]{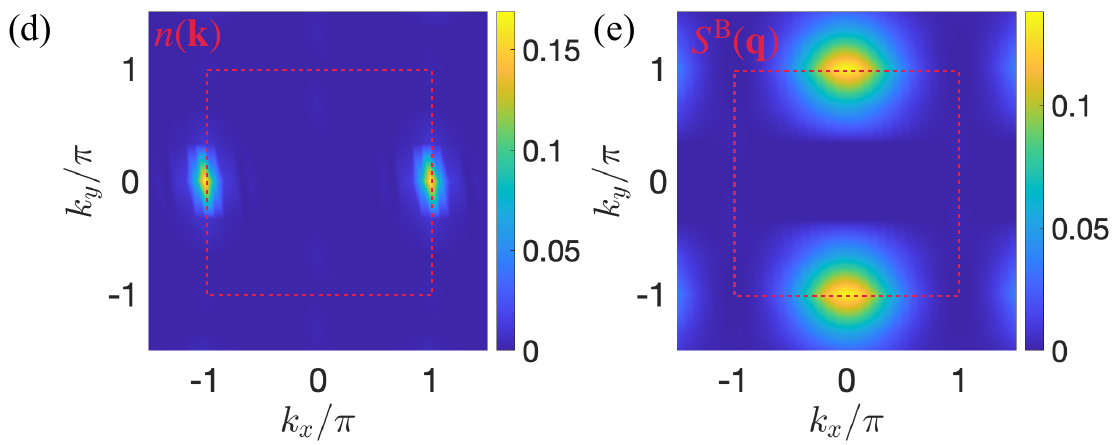}
	\caption{ \textbf{Ground-state properties of FQAHS.} 
		(a) Charge pumping 
		from DMRG and $\Delta Q\approx2/3$. 
		(b) The momentum sectors of the 6-fold (quasi)degenerate ground states on tori of sizes $N=3\times6\times2$ and $N=3\times4\times2$.
		(c) Energy spectrum flow of FQAHS ground states at $V_3=1$ using a $3\times6\times2$ torus with twist boundary condition along ${\bf a_1}$ direction.
		(d) DMRG result of Fourier transformation of the real-space density distribution $n(\mathbf{k})$ and (e) structure factor $S^\mathrm{B}(\mathbf{q})$ in a $N_y=6$ cylinder at $V_3=1$. To minimize the background noise of (d) and (e), we apply a Gaussian window in the Fourier transformation.
	}
	\label{fig_fqahs_gs}
\end{figure}

In this paper, we study  intertwined charge and topological orders using the archetypal correlated topological flat-band model on a checkerboard lattice~\cite{KSun2011_model,DNSheng2011_fci, HYL2023_thermoFQAH}. In addition to conventional numerical methods, density matrix renormalization group (DMRG)~\cite{White1992_dmrg, dmrg_rmp2005} and exact diagonalization (ED)~\cite{SandvikEDNote, LauchliBookChapter}, to study the thermodynamic properties,  we also utilize the state-of-the-art tensor network technique -- the exponential tensor renormalization group (XTRG)~\cite{BBChen2018_XTRG}.  Our attention is focused on the uncharted $\nu=2/3$ filling, where we observed highly intriguing interplays between smectic charge orders (unidirectional charge stripes) and topological order. Our discoveries can be summarized in the following four points:
\begin{enumerate}
\item As the interaction strength is increased, the Fermi liquid (FL) state first transitions into a FQAH smectic (FQAHS) state, and subsequently to a polar smectic metal (PSM) state [Fig.~\ref{fig_phasediagram}(c)].
\item The FQAHS state is incompressible and has a fractional Hall conductivity of $\sigma_{xy}=\nu=2/3$. Its smectic ordering wavevector, either $(\pi,0)$ or $(0,\pi)$, spontaneously breaks the four-fold rotational symmetry down to two-fold. This charge order also breaks the lattice translational symmetry along the direction of the wavevector. On a torus, this state displays a 12-fold ground state degeneracy, with a factor of 3 from topological degeneracy and a factor of 4 from the rotational and translational symmetry breaking [Figs.~\ref{fig_phasediagram}(c), \ref{fig_spectra_order} (b), and \ref{fig_fqahs_gs}].
\item The thermodynamics of the FQAHS state reveal distinct temperature/energy scales and an intriguing finite-temperature phase diagram: (1) the onset temperature of the fractionalized Hall plateau and non-polar smectic order $T^\ast$; (2) the critical temperature of the polar smectic order $T_c$; (3) the charge gap $T_\mathrm{cg}$.
XTRG simulations suggest $T^\ast \ll T_c \ll T_\mathrm{cg}$. 
The principal fluctuations around $T_c$ are from the polar smectic order, while the dominant fluctuations around $T^\ast$ are charge-neutral magnetorotons whose wavevector differs significantly from the smectic order, and the smectic order also becomes non-polar below $T^\ast$. 
The thermal fluctuations and proliferation of charge-neutral modes lead to charged excitations and the system becomes compressible around $T^\ast$. 
Therefore, it is the incompressible FQAHS state below $T^\ast$, while a compressible PSM state exists between $T^\ast$ and $T_c$ [Fig.~\ref{fig_fqahs_thermal}].
\item The polar smectic metal (PSM) is a non-Fermi liquid. This smectic order shares the same ordering wave vector with FQAHS, but it breaks an additional symmetry (two-fold rotation), having a ferroelectric order, making the ground state degeneracy (arising from spontaneous symmetry breaking) 8-fold [Fig.~\ref{fig_phasediagram}(c)].
The inter-stripe tunneling of the PSM is irrelevant and insulating while it is metallic only inside each stripe [Fig.~\ref{fig_psm}].
\end{enumerate}

In classical liquid crystals, a comparable state to PSM is known as the uniaxial ferroelectric smectic A phase (SmA$_F$), recently identified in polar molecule systems~\cite{Chen2022smeticaf}. The PSM we find here serves as a quantum counterpart of that state. Unlike classical systems, our polar smectic state develops in a system devoid of any polar building blocks. 
Interestingly, the quantum melting of this polar smectic order (upon reducing interaction strength) is highly nontrivial. Instead of directly transitioning into the disordered phase (the FL in our phase diagram), the system first turns into a regular smectic state, thereby restoring part of the broken symmetries (two-fold rotation). This two-step transition process strongly echoes the phenomena of vestigial order~\cite{quenchedNie2014,interwinedFernandes2019,wangVestigial2021,sunDimensionality2023,franciniSpin2023}.
More intriguingly, the thermal melting of the FQAHS phase leads to an intermediate PSM phase at finite temperature. 
This scenario, absent in literature, further stresses the significance of thermodynamic properties and finite-temperature phase diagrams to understand the exotic intertwined orders, just like in other systems such as the high-temperature superconductors~\cite{Fradkin2015interwine,dai2015sc, Keimer2015highTc}.  
We also notice similar physics could happen in the superconductors coexisting with other symmetry-broken orders, where such orders could still exist in the intermediate-temperature phases while the superconductivity being melted~\cite{Babaev2004metal,Bojesen2014metal,Babaev2021metal}.


\vspace{0.1cm}
\section{Model and Phase Diagram}\label{sec_phasediagram}

We consider a two-band spinless fermion model on the checkerboard lattice,
\begin{equation}
	\begin{aligned}
		H =&-\sum_{\langle i,j\rangle}te^{\mathrm{i}\phi_{ij}}(c_i^\dagger c^{\ }_j+h.c.)-\sum_{\langle\hskip-.5mm\langle i,j \rangle\hskip-.5mm\rangle}t'_{ij}(c_i^\dagger c^{\ }_j+h.c.)\\
		&-\sum_{\langle\hskip-.5mm\langle\hskip-.5mm\langle i,j \rangle\hskip-.5mm\rangle\hskip-.5mm\rangle} t''(c_i^\dagger c^{\ }_j+h.c.)+\sum_{\langle\hskip-.5mm\langle\hskip-.5mm\langle i,j \rangle\hskip-.5mm\rangle\hskip-.5mm\rangle}V_3(n_i-\frac{1}{2})(n_j-\frac{1}{2})
	\end{aligned}
	\label{eq:eq1}
\end{equation}
with nearest-neighbor (NN, $t$), next-nearest-neighbor (NNN, $t'$), and next to next nearest-neighbor (NNNN, $t''$) hoppings, and NNNN repulsive interaction ($V_3$), as shown in Fig.~\ref{fig_phasediagram} (a). The tight-binding parameters are: $t=1$ (as the energy unit), $t'_{ij}=\pm 1/(2+\sqrt{2})$ with alternating sign in edge-sharing plaquettes, $t''=-1/(2+2\sqrt{2})$ and $\phi_{ij}=\frac{\pi}{4}$ along the direction of the arrows,
such that the relationship between the flat-band width $W$, the gap between the flat and remote band $\Delta$ are $W(=0.08)\ll\Delta(=2.34)$, as shown in Fig.\ref{fig_phasediagram}(b). And this tight-binding model acquires opposite Chern number $C=\pm 1$ for the flat and remote bands~\cite{KSun2011_model,DNSheng2011_fci}.

Previous research on this model has confirmed the existence of FQAH states at $\nu=1/3$ and $\nu=1/5$. No competing CDW order was noted, even when the interaction surpassed the band gap $\Delta$~\cite{DNSheng2011_fci}. 
However, the phenomena at fillings of $1/2<\nu<1$ have largely been left unexplored. In contrast to the first Landau level, where fillings $\nu<1/2$ and $\nu>1/2$ are simply connected by the particle-hole symmetry, a general Chern band does not display such symmetry. Consequently, the repulsive interaction and the existence of a remote band result in significantly different physics for $\nu>1/2$ compared to the $\nu<1/2$ regime. As will be demonstrated below, we observe a highly nontrivial interplay between charge order and topological order for $\nu>1/2$, which was absent in previous studies.

\begin{figure}[t!]
	\centering		
	\includegraphics[width=\columnwidth]{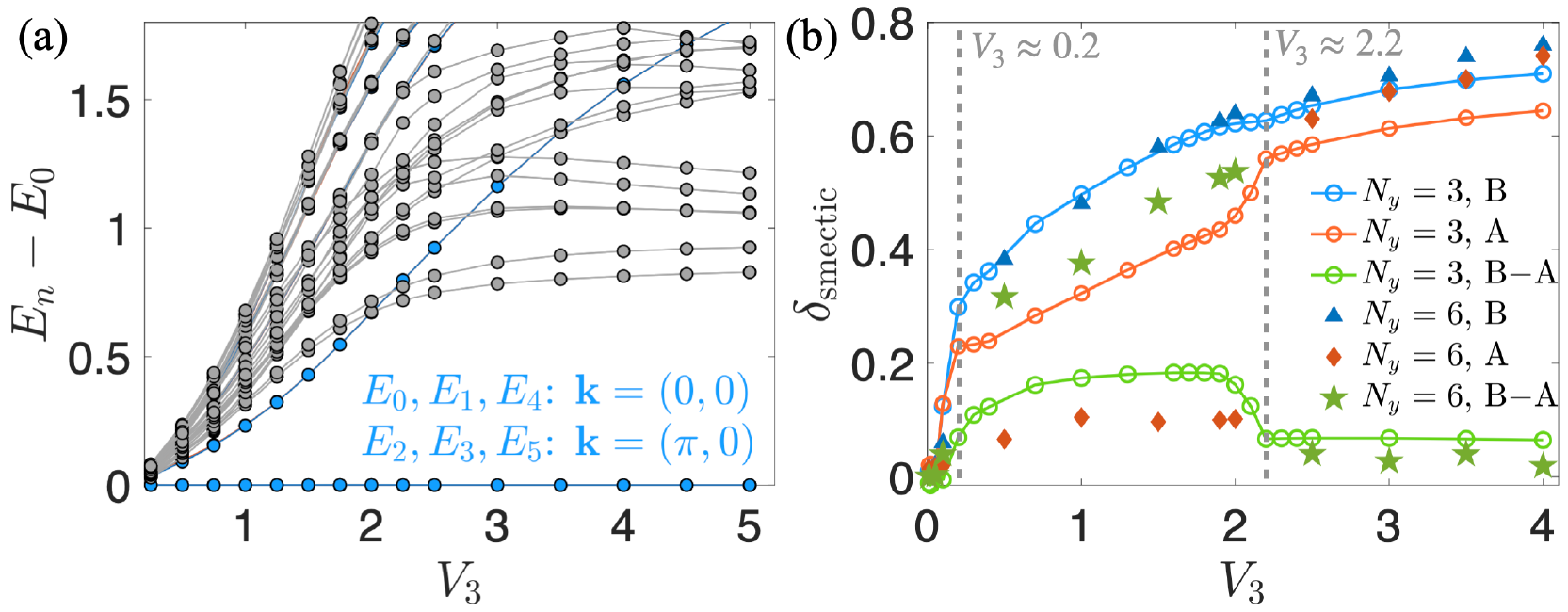}
	\caption{
		\textbf{Determination of phase boundaries via ED and DMRG.}
		(a) $3\times6\times2$ ED spectra with changing $V_3$. The blue lines/dots mark the 6-fold (quasi)degenerate ground states in FQAHS phase for $0.2<V_3<2.2$. 
	(b) The charge-stripe order parameters of A and B sublattices measured via DMRG for cylinders with width $N_y=3$ and $N_y=6$. The difference between $\delta^\mathrm{A}_\mathrm{smectic}$ and $\delta^\mathrm{B}_\mathrm{smectic}$ is also plotted, and the two grey dashed lines label the phase boundaries. 
	}
	\label{fig_spectra_order}
\end{figure}

We focus on $\nu=2/3$ with the NNNN interaction $V_3$, leaving the global phase diagram with NN ($V_1$), NNN ($V_2$) and $V_3$ interactions for future study. 
At the strong coupling limit, $V_3 \to \infty$, the minimization of potential energy leads to a unidirectional stripe order as shown in Fig.~\ref{fig_phasediagram}(c). It is adiabatically connected to the polar smectic metal (PSM) phase at large $V_3$, where sites represented by circles always remain empty and particles  only occupy sites of filled disks. In this charge configuration, particles never occupy any pair of NNNN sites and thus minimize the potential energy. 
This state is an electronic smectic state~\cite{kivelson1998electronic,Fradkin1999_Hall_crystal}. Remarkably, this smectic state is of a unique kind. In direct contrast to the typical electronic smectic state, which is invariant under $C_2$ rotation along the direction perpendicular to the $x$-$y$ plane, this smectic spontaneously breaks this two-fold rotational symmetry. 
This additional symmetry breaking increases the number of degenerate ground state charge configurations by a factor of 2. More importantly,  it implies that this charge ordered state has a spontaneously generated in-plane electric dipole (perpendicular to the stripes), i.e., it is a ferroelectric state~\cite{Niori1996ferro, Chen2022smeticaf}. 
To highlight this ferroelectric order, we call this charge order the polar smectic order. 
As for the physical properties of this PSM state, because the sites of filled disks are only partially occupied with an average density $\nu=2/3$, in principle electrons can move along the stripes and tunnel between the stripes. Such a system can be characterized as coupled Luttinger liquids. Depending on the Luttinger parameters and the inter-stripe couplings, various phases might be stabilized, such as smectic superconductor, smectic crystal (insulator), smectic metal (non-Fermi liquid) and Fermi liquid~\cite{Emery2000}.  We will show in Sec.\ref{sec_psm} that the PSM, observed at $V_3>2.2$, is a non-Fermi-liquid smectic metal phase with exotic thermodynamic properties, and the inter-stripe tunneling is irrelevant while it is metallic only inside the stripes.

In Fig.~\ref{fig_spectra_order}(a), we plot the energy spectra of a $3\times6\times2$ torus obtained from ED [other system sizes are shown in Supplementary Information (SI)~\cite{suppl}, and the smectic order parameter is calculated using DMRG on cylinders of width of $N_y=3$ and $N_y=6$ [Fig.~\ref{fig_spectra_order} (b)].
Here we define two smectic order parameters, for A- and B- sublattices respectively, $\delta^\mathrm{A/B}_\mathrm{smectic}=\tfrac{2}{N'}\sum'_i (-1)^{x_i} n^\mathrm{A/B}_{\mathbf{r}_i}$ with summation over a few unit cells $i$'s in the bulk and $N'$ being the number of such sites. The integer $x_i$ is the $x$ coordinate of the $i$th unit cell (along the cylinder). In the polar smectic phase, both the two order parameters (for A- and B- sublattices) shall take nonzero expectation values, and at $V_3\to\infty$, their values saturate to $\nu=2/3$ as expected. 
It is worthwhile noting that in our DMRG simulations of $N_y=3$ and $N_y=6$, the stripe pattern is found to be along the $y$ axis for the cylindrical geometry, but in the thermodynamic limit, the orientation of the stripes can be either along $x$ or $y$, determined by spontaneous symmetry breaking.

As we reduce $V_3$, quantum fluctuations start to melt the polar smectic order. However, instead of a direct transition to a homogeneous phase, we find an intermediate phase for $0.2<V_3<2.2$. As shown in Fig.~\ref{fig_spectra_order}(b), $\delta^\mathrm{B}$ remains nonzero and large, while $\delta^\mathrm{A}$ becomes very small. 
More importantly, the value of $\delta^\mathrm{A}$ reduces drastically as we increase the system size (from $N_y=3$ to $6$), indicating that the small nonzero value of $\delta^\mathrm{A}$ is a finite size effect, which shall vanish in the thermodynamic limit. 
This phase of $\delta^\mathrm{B}\ne 0$ and $\delta^\mathrm{A}=0$ is a non-polar smectic order, fundamentally different from the polar smectic order. Although it shares the same ordering wavevector $(\pi,0)$ with the polar smectic order, the two-fold rotational symmetry is recovered and thus the in-plane electric dipole moment reduces to zero. 
In other words, this smectic phase at $V<2.2$ doesn't exhibit ferroelectric order. More importantly, this intermediate smectic phase has a nontrivial topological order. It is a gapped FQAH state with Hall conductivity $\sigma_{xy}=\nu=2/3$, thus it is the FQAHS state.

Upon further reduction $V_3$, this nonpolar smectic order is eventually melted for $V_3<0.2$, where the smectic FQAH state gives way to a homogeneous and isotropic Fermi liquid phase. The changes in charge order parameters are discontinuous around transition points.

The ground-state phase digram is one of the key result of this study, which is summarized in Fig.~\ref{fig_phasediagram} (c). To the best of our knowledge, the coexistence of the FQAH effect and smectic order in the FQAHS state has not been observed in microscopic models, even regardless of the fact that the Hall conductivity is $\sigma_{xy}=\nu$. Equally important, the competition between an FQAHS state and non-Fermi-liquid PSM state has not been observed before. In the following sections, we will present more numerical results to show the non-trivial ground-state and thermodynamic characteristics of these intertwined quantum phases, as well as their broad experimental implications.

\section{Fractional quantum anomalous Hall smectic (FQAHS) state} \label{sec_fqahs}

In this section, we set $V_3=1$ and scrutinize the topological properties and thermodynamics of the FQAHS phase. The Hall conductivity is directly measured through charge pumping in DMRG simulations. As we adiabatically introduce a $2 \pi$ magnetic flux ($c_i^\dagger c_j^{\ }+h.c. \rightarrow c_i^\dagger c_j^{\ }e^{\mathrm{i}\theta}+h.c.$) for hopping across the periodic boundary in a cylinder of width $N_y=6$, we find two thirds of an electron charge being pumped from one edge of the cylinder to the opposite one, signifying a fractional Hall conductivity of $\sigma_{xy}=2/3$ [Fig.~\ref{fig_fqahs_gs}(a)].

ED simulations provide further corroboration for this conclusion, revealing six-fold (quasi)degenerate ground states as depicted in Fig.~\ref{fig_fqahs_gs}(b) and (c), as well as in the SI~\cite{suppl}. With twisted boundary conditions, we find that each ground state possesses a fractional Hall conductivity $\sigma_{xy}=2/3$. 
For an $N=3\times6\times2$ torus, three ground states are located at $(0,0)$ and the remainder at $(\pi,0)$. For an $N=3\times4\times2$ cluster, these states can be found at $(0,2 m \pi/3)$ and $(\pi, 2 m \pi/3)$ with $m=-1,0,1$. 
This ground state degeneracy and corresponding momentum sectors are in full alignment with the coexistence states of FQAH and non-polar smectic order. The six-fold ground states can be attributed to the combined effects of translational-symmetry-breaking leading to a 2-fold degeneracy for the ordering wavevector $\mathbf{Q}=(\pi,0)$ and 3-fold topological degeneracy for a $\nu=2/3$ FQAH state on a torus. 
Notably, because the geometry of these ED clusters is incompatible with horizontal stripes, stripes observed here are only along the $y$ direction. In the thermodynamic limit,  stripes along $x$ would further double the ground state degeneracy by a factor of $2$.  We show more ED results to support the smectic order in SI~\cite{suppl}.
Also, it's noteworthy to mention that the observed six-fold ground state degeneracy implies a non-polar smectic order, as a polar smectic would yield a 12-fold degeneracy due to the four degenerate charge patterns for stripes along $y$ as illustrated in SI~\cite{suppl}, confirming the ground state charge pattern previously discussed based on order parameter measurements using DMRG.

For various ED clusters, we observe that the momentum sectors of ground states consistently display this structure: three ground states are located at momentum $(K^{(i)}_x,K_y^{(i)})$ with $i=1,2,3$, in accordance with the anticipated momentum sectors of FQAH states without charge order, while the remaining three have momentum $(K^{(i)}_x,K_y^{(i)})+(\pi,0)$. 
This observation further affirms the charge pattern and its coexistence with topological order. For vertical stripes in the thermodynamic limit, any FQAH ground state $\psi_\mathrm{FQAH}$ coexists with a degenerate state, $T_x \psi_\mathrm{FQAH}$, where $T_x$ is a translation operator shifting the system along the x-axis by one lattice constant. For ED simulations on a finite-sized torus, these two degenerate ground states hybridize and their superpositions result in two nearly degenerate states with total momentum $(K_x,K_y)$ and $(K_x,K_y)+(\pi,0)$ respectively.

Besides the coexistence of charge and topological orders, this FQAHS state also exhibits nontrivial quantum fluctuations. As shown in Fig.~\ref{fig_fqahs_gs}(d), 
the Fourier transformation of the real-space charge density, $n({\bf k}) = \sum_{\bf r} e^{\mathrm{i}\mathbf{k\cdot r}} (n({\bf r})-\bar n)/N$, 
acquired from DMRG for a $N_y=6$ cylinder, reveals a sharp peak at $(\pi,0)$, thereby verifying the smectic order. 
Within the same simulation, a peak in the density-density correlation function, $S^\mathrm{A/B}(\mathbf{q})=\sum_j e^{-\mathrm{i}\mathbf{q}(\mathbf{r_0}-\mathbf{r_j})}( \langle n^\mathrm{A/B}_0n^\mathrm{A/B}_j\rangle-\langle n^\mathrm{A/B}_0 \rangle\langle n^\mathrm{A/B}_j \rangle)$, is noted at $(0,\pi)$ [See Fig.~\ref{fig_fqahs_gs}(e)]. 
This correlation function peak does not arise from the smectic order as it is situated at a completely different $k$ point. Instead, it suggests that low-energy charge-neutral fluctuations are dominated by excitations with a finite momentum $\mathbf{q}\sim (0,\pi)$, also referred to as magnetorotons, analogous to similar excitations observed in FQAH states without charge orders~\cite{HYL2023_thermoFQAH}.

\begin{figure}[htp!]
	\centering		
	\includegraphics[width=\columnwidth]{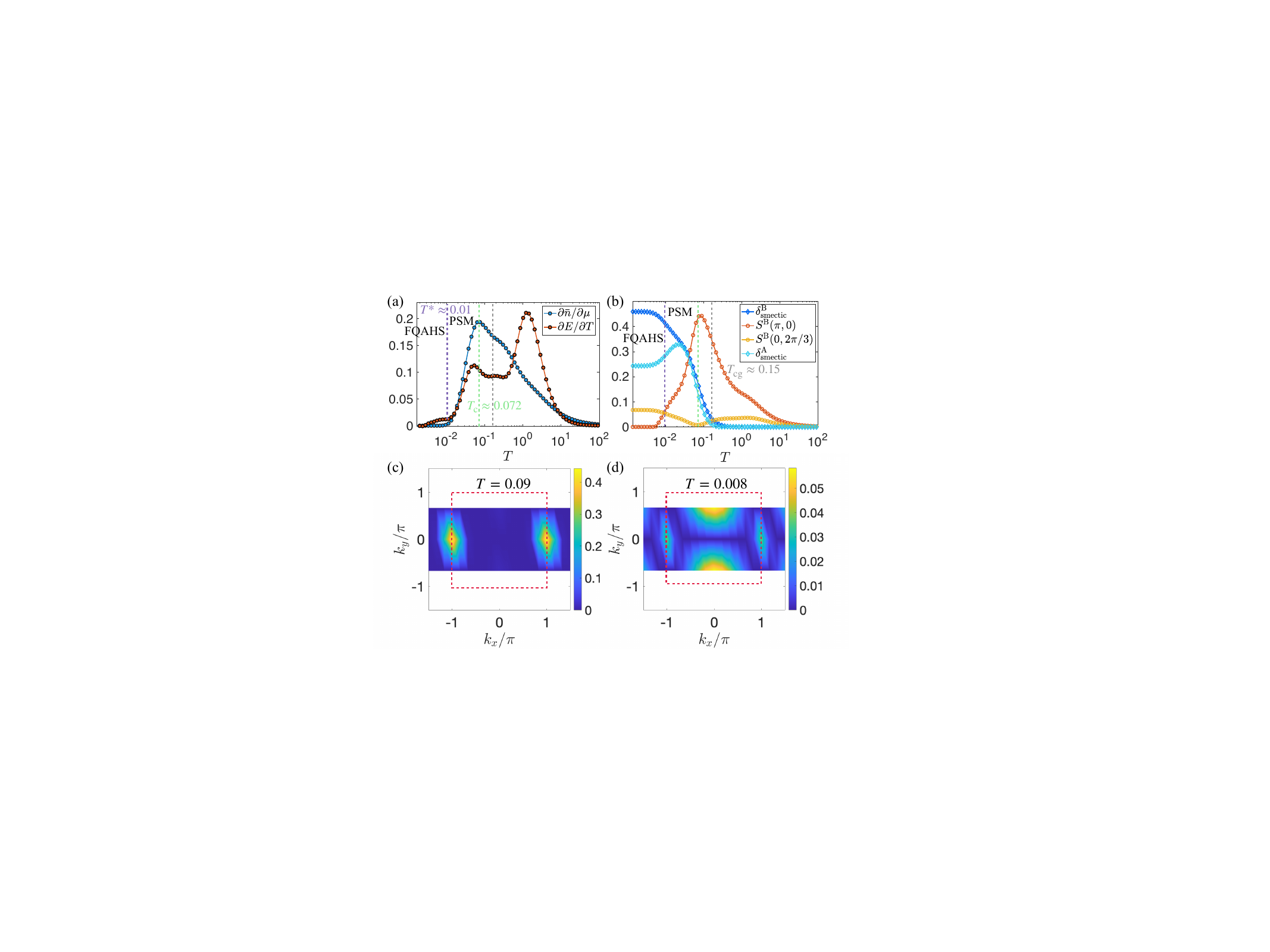}
	\caption{ \textbf{Thermodynamics of the FQAHS state at $V_3=1$.} (a) Specific heat and compressibility, and (b) structure factors of the B sublattice and charge-smectic order parameters of both sublattices versus temperature. 
	(c,d) Structure factors plotted at $T=0.09$ (one sees the smectic fluctuation at $(\pi,0)$) and $T=0.008$ (supports the magneto-roton minimum at $(0,\pi)$), respectively. The dashed lines in (a) and (b) represent $T^\ast\approx0.01$, $T_c\approx0.072$ and $T_\mathrm{cg}\approx0.15$, with colors in purple, green, and gray respectively. Below $T^\ast$ is the incompressible FQAHS phase and the compressible PSM phase exists at the intermediate $T^\ast<T<T_c$.
	}
	\label{fig_fqahs_thermal}
\end{figure}

Furthermore, we study the thermodynamics via XTRG of the FQAHS state with $V_3=1$ in a $3\times12\times2$ cylinder and we show the specific heat,  compressibility, and smectic order parameter of both sublattices and structure factor of B sublattice as an example in Fig.~\ref{fig_fqahs_thermal} (a,b), with $S^\mathrm{B}(\mathbf{q})$ at different tempertures in Fig.~\ref{fig_fqahs_thermal} (c,d). 
Here, we find three different temperature scales with two thermal phase transitions.
The gap of charge excitations $T_\mathrm{cg}\approx 0.15$ is estimated from the width of $\bar n-\mu$ plateau with details in SI~\cite{suppl}, where the specific heat shows a small hump around $T_\mathrm{cg}$.
The intermediate temperature scale $T_c\approx0.072$ is the critical temperature of the translational symmetry breaking and the onset of the polar smectic order, where the smectic order parameters of both sublattices establish and the smectic density fluctuation, denoted by $S^\mathrm{B}(\pi,0)$, reaches the maximal value in Fig.~\ref{fig_fqahs_thermal} (b). 
For $T<T_c$, the formation of smectic order leads to a decrease in smectic fluctuation ($S^\mathrm{B}(\pi,0)$). 
This estimation of $T_c$ is consistent with a small specific heat peak observed around this temperature in Fig.~\ref{fig_fqahs_thermal} (a).

We note that, in Fig.~\ref{fig_fqahs_gs} (d) and (e), while DMRG simulations peak at $\mathbf{q}=(\pi,0)$ smectic pattern, 
there exists strong charge fluctuation in the other direction, i.e. $(0,\pi)$, which belongs to the magnetoroton excitation of the FQAH state.
In our $N_y=3$ cylinder geometry, the closest allowed momentum to the broad roton peak is $(0,\pm2\pi/3)$.
As shown in Fig.~\ref{fig_fqahs_thermal} (b), while $S^\mathrm{B}(\pi,0)$ approaches 0 with decreasing temperature, the density fluctuation at $(0,2\pi/3)$ increases and goes to the highest value around $T^\ast\approx0.01$, which is the third temperature scale of magneto-roton where the specific heat also shows a shoulder. 
For better demonstration, we also show the finite-temperature $S^\mathrm{B}(\pi,0)$ around $T_c$ and $T^\ast$ in Fig.\ref{fig_fqahs_thermal}(c,d) respectively, with more results at different temperatures in SI~\cite{suppl}.
Besides the three important temperature scales, we also note the highest peak of specific heat around $T\sim2$. This is related to the single-particle band gap of the two Chern bands, which is almost universal for different ground states in this model.

The behavior of compressibility is rather interesting. Below $T^\ast$,  compressibility converges to 0 and thus $T^\ast$ (also the scale of the magneto-roton mode) is the onset temperature of quantized Hall plateau of the incompressible FQAHS liquid phase. 
Above $T^\ast$, the compressibility increases with temperature and comes to the maximal value at $T_c$, however, $T^\ast$ and $T_c$ are both much lower than $T_\mathrm{cg}$.
This exotic phenomenon is consistent with our previous thermal study of an isotropic $\nu=1/3$ FQAH state where one of the main conclusions is: the charge-neutral gap is the lowest energy scale of FQAH states which could be much lower than the charge gap, and the thermal fluctuations together with the proliferation of charge-neutral excitations could lead to charged excitations at temperature above the roton gap but much lower than the charge gap~\cite{HYL2023_thermoFQAH}. 
Since the thermal gas of neutral excitations weakens the FQAHS state and the associated charge gap, the system becomes compressible above the temperature scale of magneto-roton gap ($T^\ast$).
Besides, with the higher density of the neutral excitations, such effect could be more obvious, and the peak of compressibility in Fig.~\ref{fig_fqahs_thermal} (a) at $T_c$ coincides with the peak of $S^\mathrm{B}(\pi,0)$ in Fig.~\ref{fig_fqahs_thermal} (b), which suggests the proliferation of smectic fluctuations around intermediate $T_c$. 

Due to the coexistence of smectic order, the thermodynamics of the FQAHS state is very intriguing. At intermediate temperature $T^\ast<T<T_c$, due to the compressible nature and established polar smectic order, the system is inside a finite temperature PSM phase that connects to the PSM found in the $T=0$ phase diagram [Fig.\ref{fig_phasediagram} (c)] at $V_3>2.2$. 
Therefore, $T_c$ also refers to the thermal transition between PSM and the higher-temperature isotropic normal phase.
In the finite-temperature PSM phase, when approaching $T^\ast$, while the smectic order parameter of B sublattice $\delta^\mathrm{B}_\mathrm{smectic}$ continues increasing until convergence, the smectic order parameter of A sublattice $\delta^\mathrm{A}_\mathrm{smectic}$ decreases to a much smaller value, as shown in Fig.~\ref{fig_fqahs_thermal} (b). 
The difference of the magnitude of order parameters of the sublattices at the finite $N_y=3$ system is consistent with the ground-state DMRG simulations in Fig.~\ref{fig_spectra_order} (b), which supports that the smectic order of the FQAHS phase is non-polar.
Consequently, $T^\ast$ as the onset of incompressibility is also the transition point between FQAHS and PSM, which coincides with the scale of magneto-roton gap.

\begin{figure}[htp!]
	\centering		
	\includegraphics[width=\columnwidth]{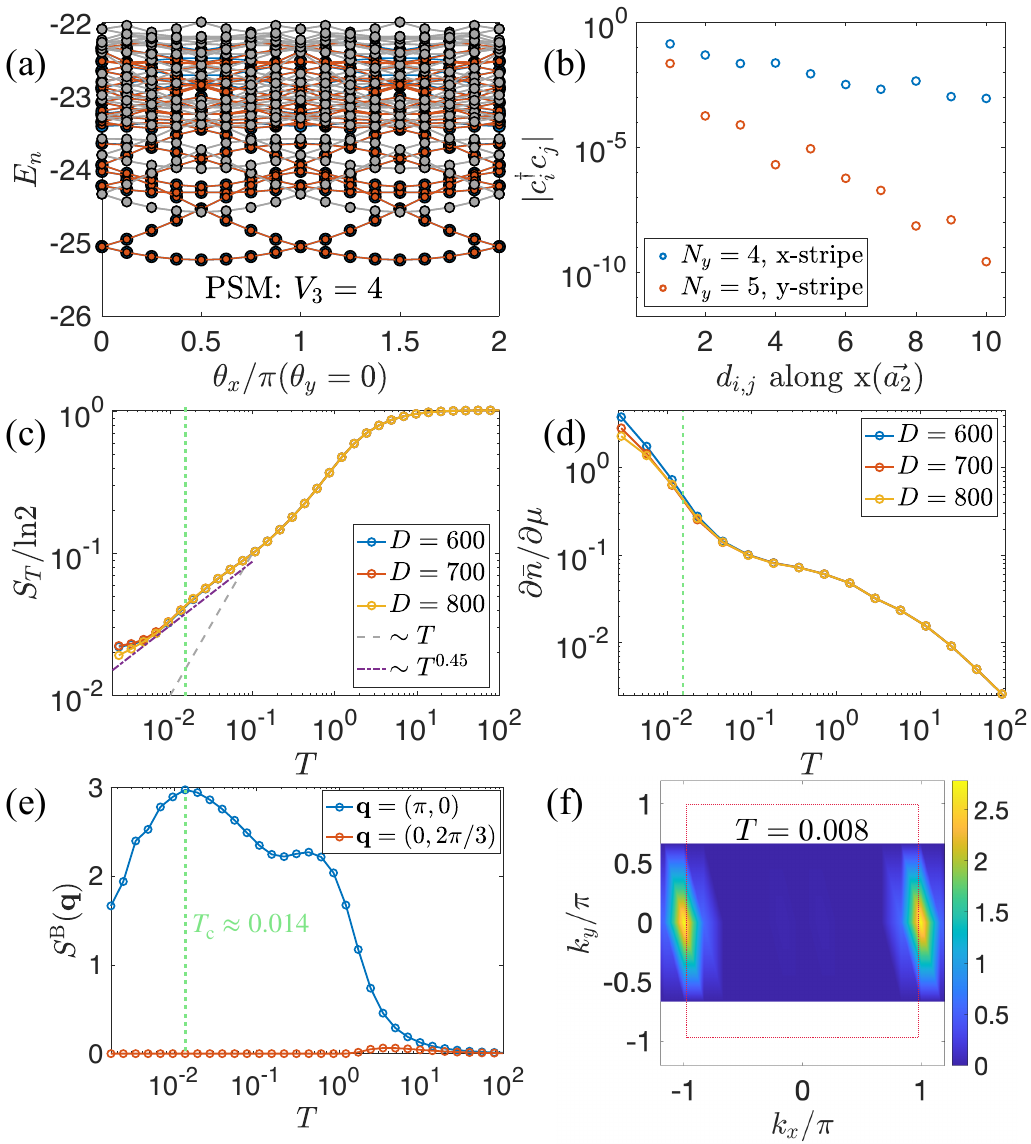}
	\caption{ \textbf{Non-Fermi-liquid PSM state at $V_3=4$.}
		(a) Gapless spectrum in a $3\times4\times2$ torus in ED with twisted boundary conditions. 
		(b) Correlation function along x ($\vec{a_2}$) direction for $4\times24\times2$ (x-stripe with ($0,\pi$) smectic order) and $5\times24\times2$ (y-stripe with ($\pi,0$) smectic order) from DMRG results.  We take a reference site $i$ in the centre of cylinder and $d_{i,j}$ refers to the distance between the two lattice sites.
		(c) Thermodynamic entropy $S_T$ and (d) compressibility $\partial\bar n/\partial\mu$ with bond dimensions up to $D=800$.
		(e) Change of structure factors $S^\mathrm{B}(\mathbf{q})$ versus $T$ for different $\mathbf{q}$.  (f) Structure factor $S^\mathrm{B}(\mathbf{q})$ at $T=0.008$. 
		(c-f) are from XTRG results of a $3\times12\times2$ cylinder, and the green dashed line in (c-e) is the critical temperature $T_c\approx0.014$ of translational symmetry breaking, obtained from the peak of smectic fluctuations in (e).
	}
	\label{fig_psm}
\end{figure}

Although the $T=0 $ quantum phase transition between FQAHS and PSM is first-order from the discontinuous change of order parameters in Fig.~\ref{fig_spectra_order} (b), this FQAHS-PSM transition is highly possible to be continuous at finite temperature from the temperature dependence of order parameters, which is interesting for further verification from thermodynamic simulations of larger system sizes and possible finite-size criticality analysis.
In addition to the current results, it is meaningful for future investigations of the full $T-V$ phase diagram, especially around the $T=0$ phase transition point between FQAHS and PSM, to figure out, for example, the evolution of the critical behavior, especially considering that the PSM state is non-Fermi-liquid, as we will elaborate in the following section.


\section{Polar smectic metal (PSM) state} \label{sec_psm}

As shown in earlier sections, for $V_3> 2.2$, the smectic order further breaks the two-fold rotational symmetry, 
resulting in a spontaneously generated in-plane electric dipole moment, but the nature of this state has not been resolved.
In general, charge stripe states can often be treated as coupled Luttinger liquids. Depending on microscopic details and values of control parameters, various phases have been proposed, such as smectic superconductor, smectic crystal (insulator), smectic metal (non-Fermi liquid), and Fermi liquid~\cite{Emery2000}.

For a deep dive into this polar smectic state, we take $V_3=4$ for example. The gapless spectrum under twisted boundary conditions from ED simulations is shown in Fig.\ref{fig_psm}(a), and the gapless/compressible nature of this state is also supported by the large compressibility at low temperature [Fig.~\ref{fig_psm}(d)].
In this sense, this state is possible to be either a smectic metal or a Fermi liquid.
The smectic metal is a non-Fermi liquid with anisotropic transport and quasi-1D Fermi surface from theoretical studies such as coupled-wire models~\cite{Wen1990NFL,Sondhi2001NFL,Vishwanath2001NFL,Mukhopadhyay2001NFL}.
These two scenarios are also characterized by different exponents in thermodynamic quantities, where the smectic metal exhibits anomalous dimensions deviated from the Fermi liquid theory. 
In our DMRG and XTRG simulations, due to the limited length of circumference and considering the periodic conditions along this direction ($N_y$), it is hard to verify the anisotropy by directly comparing the couplings along two directions of the cylinders.
However, we can use a trick. In a $N_y=5$ system, the smectic order must be ($\pi,0$) (stripe along the $y$ direction) instead of ($0,\pi$). While for $N_y=4$, the energies of  ($\pi,0$) and ($0,\pi$) orders are close, so we can take a very small pinning field such that the DMRG simulations would pick the ($0,\pi$) order (stripe along the $x$ direction).
Then we fix a reference lattice site $i$ and measure the magnitude of the correlation function $\langle c_i^\dagger c_j \rangle$ along the $x$ direction, which would be the interstripe correlation for $N_y=5$ with ($\pi,0$) order and intrastripe correlation for $N_y=4$ with ($0,\pi$) order. 
As shown in Fig.~\ref{fig_psm}(b), the interstripe correlation decays much faster with distance than that of the intrastripe correlations, manifesting the anisotropic nature that this state is almost insulting across the stripes while it is metallic inside each stripe, which is consistent with the theories of smectic metal that interstripe tunneling is irrelevant while the transport would be large inside each stripe~\cite{Emery2000}.
Therefore, we name this state as polar smectic metal (PSM).
This observation of the PSM state at large $V_3$ without topological order, also supports the early proposal that FQH effect can coexist with partially polarized stripe state, but the FQH effect vanishes when the CDW order becomes strong~\cite{Halperin1989_QHC}.

The evidence of two dimensional anisotropic Luttinger liquids has recently been reported in a moir\'e superlattice made of twisted bilayer tungsten ditelluride at millikelvin temperatures~\cite{Wang2022LL, Yu2023LL}, where the temperature-dependence of the interchain conductivity is shown to be power-law, as theoretically predicted~\cite{Wen1990NFL}.  
Different from the typical electronic smectic state, which is invariant under $C_2$ rotation along the direction perpendicular to the $x$-$y$ plane, the PSM state in our work additionally and spontaneously breaks this two-fold rotational symmetry, having a spontaneously generated in-plane electric dipole (perpendicular to the stripes), i.e., a ferroelectric order.
We expect that it will be highly interesting to probe the ferroelectric order apart from the anisotropy, induced by the PSM phase in future transport measurements.

The thermodynamic results of this PSM state are interesting as well. The critical temperature of the smectic order $T_c\approx0.014$ is obtained from the peak of smectic fluctuations $S^\mathrm{B}(\pi,0)$, as shown in Fig.~\ref{fig_psm} (e). Below $T_c$, due to the established smectic order, such smectic fluctuation decreases (approaching the $T=0$ polarized stripes).
It is different from the FQAHS where there exist other dominating collective excitations below $T_c$ such as the magnetoroton modes. In the PSM state, we observe no other density fluctuations by showing the temperature-dependent structure along $k_y$ in Fig.~\ref{fig_psm} (e) and the plot of structure factors at low temperature $T=0.008$.
Furthermore, we show the log-log plot of thermodynamic entropy versus temperature of this non-Fermi-liquid PSM state in Fig.~\ref{fig_psm} (c).
With the increasing bond dimension in XTRG simulations, we observe that the low-$T$ thermal entropy is approaching a power-law scaling, but still largely deviates from the linear dependence of the ordinary Fermi liquid phase. 
However, whether there exists any correction to the linear dependence and what would be the exact correction to the scaling of thermodynamic entropy might still need more accurate simulations for some reasons, including the limitation from finite-size effect and the fact that the exact low-temperature dependence of entropy might be detected at even lower temperatures than the simulated temperature region in our work.
This is similar for the results of compressibility in Fig.~\ref{fig_psm} (d), which is still increasing when temperature goes down even in a log-log plot. However, the speed of increase gets slowed with the enhanced bond dimension of XTRG simulations. Therefore, whether or how the compressibility will converge needs more accurate simulations down to lower temperatures and maybe larger system sizes.
These are meaningful for future work with even more efficient thermodynamic algorithms.


\section{Discussions} \label{sec_discussions}
The FQAHS states identified in this research show unique characteristics that set them apart from other coexisting states of charge order and topological order. For instance, most CDW orders in anomalous Hall crystals are found to be commensurate with the fractional filling, which leads to effective integer filling of folded Chern bands and the integer Hall conductance~\cite{Sheng2024QAHC, pan2022ahc} . 
And, in topological pinball fluids, a portion of the electrons form a charge-ordered crystal, while the remaining contribute to topological states. Consequently, the Hall conductivity strays from the filling factor $\sigma_{xy}\ne \nu$~\cite{Kourtis2014,stefanos2018fqahc}.  
However, in the FQAHS states that we report, $\sigma_{xy}=\nu$ indicating that all electrons participate in forming the stripe order and, simultaneously, contribute to the FQAH effect. 
One conceptual way to understand these FQAHS states starts with an FQAH state without any charge order with $\sigma_{xy}=\nu$, and then perturbatively turning on the charge order. Because the FQAH effect and $\sigma_{xy}$ remains robust against any perturbations, here we obtain a FQAHS state with $\sigma_{xy}=\nu$. 
The FQAHS state observed in our numerical simulations should be adiabatically connected to the ground state of this perturbative picture. 
To verify this conjecture numerically, innovative strategies, analogous to the adiabatic path demonstrated in Ref.~\cite{Wu2012Adiabatic}, can be helpful. 
And the direct transition from FQAHS phase to an FQAH phase at the same filling without charge order in the global phase diagram of our model can be intriguing and helpful for deeper understanding. 
This perturbative picture is rather interesting since the perturbation could arise from the roton instability, which is similar to some formations of supersolids from the roton instability of superfluids, considering the composite-boson picture.
Considering the recent experimental progress in realizing FQH states in cold atom systems or using  circuit
quantum electrodynamics techniques~\cite{Julian2023photonFQH, Wang2024photonFQH}, the roton-instability-triggered scenario (which is universal for either bosonic or fermionic FQH states at different fillings) might provide hints on how to experimentally realize translational-symmetry-broken FQH states by (quasi)adiabatic evolution from isotropic FQH states, similar to the preparation of roton-triggered supersolids~\cite{Chomaz2018roton, Tanzi2019supersolid, Fabien2019supersolid,Chomaz2019supersolid, zhang2019supersolid, Alana2023supersolid}, which would be a significant step forward for quantum simulations.

Theoretically, it is worth noting that for FQAH nematics, a composite-fermion description, based on a lattice Chern-Simon’s gauge theory~\cite{Fradkin1989CS, ELIEZER1992CS, Kumar2014CS,KSun2015CS}, has been achieved \cite{Sohal2018CF}. How to expand such descriptions to FQAHS could be an interesting subject.

Because our FQAHS state shares the same Hall conductivity as conventional FQAH states (without charge order), current experimental studies of FQAH states, mainly focusing on directly or indirectly measuring $\sigma_{xy}$, cannot differentiate these two types of states. Thus, it is not totally impossible that some of the reported FQAH states might actually fall under the FQAHS category or something similar. 
Two experimental probes could provide significant insights to distinguish FQAHS states from FQAH states. The first is longitudinal transport. Since smectic order breaks the rotational symmetry spontaneously, it leads to anisotropy in longitudinal conductivity, thus yielding a nonzero expectation value for the nematic order parameter $(\sigma_{xx}-\sigma_{yy})/(\sigma_{xx}+\sigma_{yy})$. 
At the ideal $T\to 0$ limit, this quantity is undefined due to $\sigma_{xx}=\sigma_{yy}=0$. 
However, at finite temperature, this ratio provides hints about the existence or absence of smectic order. Because this order parameter is unaffected by the breaking of translational symmetry, it cannot distinguish between FQAH nematic and FQAH smectic states. 
The definitive proof of FQAHS order should involve measurements capable of probing the ordering wavevector, such as X-ray scatterings, and/or real-space imaging methods, such as scanning tunneling microscopy (STM).

In our ground-state phase diagram, the interaction-driven PSM state is also intriguing, since this is a non-Fermi liquid with anisotropic transport that it is metallic only inside each stripe while the interstripe coupling is irrelevant. 
Different from ordinary smectic metals, this PSM state has an additional ferroelectric order. As its classical counterpart has been identified, we expect this ferroelectric order of PSM phase could be probed in future experiments of quantum systems.
Besides, our work also provides nontrivial thermodynamic results of this PSM state, considering the limited numerical knowledge, although more accurate and larger-scale simulations are still needed for the exact scalings of the thermodynamic quantities.

When studying the quantum states with intertwined orders, the finite-temperature properties are rather interesting and provide more and even deeper understanding beyond the ground-state analysis.
This is true for the FQAHS state in our work.
The onset temperature $T^\ast$ of incompressibility of this FQAHS is determined by the magneto-roton gap, which coincides with the onset of non-polar smectic order. 
More intriguingly, there exists an intermediate PSM phase at temperature $T^\ast<T<T_c$, which means the thermal fluctuations of this FQAHS phase do not directly melt the Hall plateau and smectic charge order together. 
Although the $T=0$ FQAHS-PSM transition is first-order, this transition at finite temperature is highly possible to be continuous, and thus the full $T-V_3$ phase diagram would also be interesting.
The finite-temperature transition between the non-Fermi-liquid PSM phase and intertwined FQAHS is rather interesting as another exotic metal-insulator transition, as in other cases, the evolution of Fermi surface (anisotropic here) would also be intriguing for future work~\cite{Senthil2008mott}.
Moreover, another open question is whether there is a universal jump of both longitudinal and Hall resistivities of order $\frac{h}{e^2}$ at the finite-temperature critical point, like the proposed critical theory of quantum phase transition between composite Fermi liquid and Fermi liquid phases~\cite{Song2024transition}, and similarly in the continuous Mott transition~\cite{Senthil2008mott}.
Considering these exotic features and open questions, it would be meaningful for the experimental realization of such $\sigma_{xy}=\nu$ intertwined FQAH states.
Beyond this, we also note that the understanding of the thermodynamic properties of other kinds of  intertwined states is rather limited. For example, in the integer quantum Hall crystals at fractional filling of Chern bands, since these could be treated using the mean-field band folding, whether the thermal fluctuations melt the Hall plateaus and CDW orders together or there exist any interesting intermediate phases, is not answered in the previous works~\cite{pan2022ahc, Sheng2024QAHC}.
We believe our work paves the way for better understanding of both the ground-state and thermodynamic properties of intertwined charge and topological order.

\section{Method} \label{sec_Method}
For the ground state calculations, we employ exact diagonalization (ED)~\cite{SandvikEDNote, LauchliBookChapter} and density matrix renormalization group (DMRG)~\cite{White1992_dmrg, dmrg_rmp2005} with the particle number fixed. We have used lattice translational symmetry and parallel computing to accelerate the ED calculations with the tori up to 36 lattice sites. In the DMRG calculations, we use cylinders from $N_y=3$ to $N_y=6$ for ground-state simulations, where $N_y$ / $N_x$ the number of unit cells along the ${\bf a_1}$/${\bf a_2}$ direction. Then we denote the total number of lattice sites as $N=N_y\times N_x \times 2$, the average density as $\bar n$ and the filling of the lower flat band $\nu$ as $\bar n=N_e/N = \nu/2$. For DMRG, we have considered different open-boundary conditions, including that in Fig.\ref{fig_phasediagram}(a) and an inversion-symmetric case (same sublattice in the two open boundaries).
For finite-temperature simulations, we work in the grand canonical ensemble in exponential tensor renormalization groups (XTRG)~\cite{BBChen2018_XTRG} by including the chemical potential term $H_\mu=\mu\sum_i (\hat n_i-\frac{1}{2})$ to tune the particle number $N_e=\sum_i \langle \hat n_i\rangle_\beta $ (here, $\langle\cdot\rangle_\beta$ denotes the ensemble average at inverse temperature $\beta\equiv\tfrac{1}{T}$). The charge gap $\Delta_{\mathrm{cg}}$ is estimated by the change of $H_\mu$ to add or subtract an electron in the system. For thermal simulations, we mainly use $N_y=3$ cylinder in XTRG calculations. The DMRG and XTRG simulations are based on the QSpace framework\cite{AW2012_QSpace} with U(1) symmetry and complex numbers, and the bond dimensions are up to $D=2048$ and $D=800$ respectively.

\begin{acknowledgments}
	We thank Cenke Xu, Bo Yang, Wang Yao and Todadri Senthil for helpful discussions.
	HYL, BBC and ZYM acknowledge the support from the Research Grants Council (RGC) of Hong Kong Special Administrative Region of China (Project Nos. AoE/P-701/20, 17301721, 17309822, HKU C7037-22GF, 17302223), the ANR/RGC Joint Research Scheme sponsored by RGC of Hong Kong and French National Research Agency (Project No. A\_HKU703/22) and the HKU Seed Funding for Strategic Interdisciplinary Research. We thank HPC2021 system under the Information Technology Services and the Blackbody HPC system at the Department of Physics, University of Hong Kong, as well as the Beijng PARATERA Tech CO.,Ltd. (URL: https://cloud.paratera.com) for providing HPC resources that have contributed to the research results reported within this paper. H.Q. Wu acknowledge the support from GuangDong Basic and Applied Basic Research Foundation (No. 2023B1515120013) and Youth S$\&$T Talent Support Programme of Guangdong Provincial Association for Science and Technology (GDSTA) (No. SKXRC202404). The ED calculations reported were performed on resources provided by the Guangdong Provincial Key Laboratory of Magnetoelectric Physics and Devices (No. 2022B1212010008).
\end{acknowledgments}

\bibliographystyle{apsrev4-1}

\begin{thebibliography}{111}%
	\makeatletter
	\providecommand \@ifxundefined [1]{%
		\@ifx{#1\undefined}
	}%
	\providecommand \@ifnum [1]{%
		\ifnum #1\expandafter \@firstoftwo
		\else \expandafter \@secondoftwo
		\fi
	}%
	\providecommand \@ifx [1]{%
		\ifx #1\expandafter \@firstoftwo
		\else \expandafter \@secondoftwo
		\fi
	}%
	\providecommand \natexlab [1]{#1}%
	\providecommand \enquote  [1]{``#1''}%
	\providecommand \bibnamefont  [1]{#1}%
	\providecommand \bibfnamefont [1]{#1}%
	\providecommand \citenamefont [1]{#1}%
	\providecommand \href@noop [0]{\@secondoftwo}%
	\providecommand \href [0]{\begingroup \@sanitize@url \@href}%
	\providecommand \@href[1]{\@@startlink{#1}\@@href}%
	\providecommand \@@href[1]{\endgroup#1\@@endlink}%
	\providecommand \@sanitize@url [0]{\catcode `\\12\catcode `\$12\catcode
		`\&12\catcode `\#12\catcode `\^12\catcode `\_12\catcode `\%12\relax}%
	\providecommand \@@startlink[1]{}%
	\providecommand \@@endlink[0]{}%
	\providecommand \url  [0]{\begingroup\@sanitize@url \@url }%
	\providecommand \@url [1]{\endgroup\@href {#1}{\urlprefix }}%
	\providecommand \urlprefix  [0]{URL }%
	\providecommand \Eprint [0]{\href }%
	\providecommand \doibase [0]{http://dx.doi.org/}%
	\providecommand \selectlanguage [0]{\@gobble}%
	\providecommand \bibinfo  [0]{\@secondoftwo}%
	\providecommand \bibfield  [0]{\@secondoftwo}%
	\providecommand \translation [1]{[#1]}%
	\providecommand \BibitemOpen [0]{}%
	\providecommand \bibitemStop [0]{}%
	\providecommand \bibitemNoStop [0]{.\EOS\space}%
	\providecommand \EOS [0]{\spacefactor3000\relax}%
	\providecommand \BibitemShut  [1]{\csname bibitem#1\endcsname}%
	\let\auto@bib@innerbib\@empty
	\bibitem [{\citenamefont {Fradkin}\ \emph {et~al.}(2015)\citenamefont
		{Fradkin}, \citenamefont {Kivelson},\ and\ \citenamefont
		{Tranquada}}]{Fradkin2015interwine}%
	\BibitemOpen
	\bibfield  {author} {\bibinfo {author} {\bibfnamefont {E.}~\bibnamefont
			{Fradkin}}, \bibinfo {author} {\bibfnamefont {S.~A.}\ \bibnamefont
			{Kivelson}}, \ and\ \bibinfo {author} {\bibfnamefont {J.~M.}\ \bibnamefont
			{Tranquada}},\ }\href {\doibase 10.1103/RevModPhys.87.457} {\bibfield
		{journal} {\bibinfo  {journal} {Rev. Mod. Phys.}\ }\textbf {\bibinfo {volume}
			{87}},\ \bibinfo {pages} {457} (\bibinfo {year} {2015})}\BibitemShut
	{NoStop}%
	\bibitem [{\citenamefont {Dai}(2015)}]{dai2015sc}%
	\BibitemOpen
	\bibfield  {author} {\bibinfo {author} {\bibfnamefont {P.}~\bibnamefont
			{Dai}},\ }\href {\doibase 10.1103/RevModPhys.87.855} {\bibfield  {journal}
		{\bibinfo  {journal} {Rev. Mod. Phys.}\ }\textbf {\bibinfo {volume} {87}},\
		\bibinfo {pages} {855} (\bibinfo {year} {2015})}\BibitemShut {NoStop}%
	\bibitem [{\citenamefont {Keimer}\ \emph {et~al.}(2015)\citenamefont {Keimer},
		\citenamefont {Kivelson}, \citenamefont {Norman}, \citenamefont {Uchida},\
		and\ \citenamefont {Zaanen}}]{Keimer2015highTc}%
	\BibitemOpen
	\bibfield  {author} {\bibinfo {author} {\bibfnamefont {B.}~\bibnamefont
			{Keimer}}, \bibinfo {author} {\bibfnamefont {S.~A.}\ \bibnamefont
			{Kivelson}}, \bibinfo {author} {\bibfnamefont {M.~R.}\ \bibnamefont
			{Norman}}, \bibinfo {author} {\bibfnamefont {S.}~\bibnamefont {Uchida}}, \
		and\ \bibinfo {author} {\bibfnamefont {J.}~\bibnamefont {Zaanen}},\ }\href
	{\doibase 10.1038/nature14165} {\bibfield  {journal} {\bibinfo  {journal}
			{Nature}\ }\textbf {\bibinfo {volume} {518}},\ \bibinfo {pages} {179}
		(\bibinfo {year} {2015})}\BibitemShut {NoStop}%
	\bibitem [{\citenamefont {Wang}\ \emph {et~al.}(2015)\citenamefont {Wang},
		\citenamefont {Agterberg},\ and\ \citenamefont {Chubukov}}]{Wang2015coexist}%
	\BibitemOpen
	\bibfield  {author} {\bibinfo {author} {\bibfnamefont {Y.}~\bibnamefont
			{Wang}}, \bibinfo {author} {\bibfnamefont {D.~F.}\ \bibnamefont {Agterberg}},
		\ and\ \bibinfo {author} {\bibfnamefont {A.}~\bibnamefont {Chubukov}},\
	}\href {\doibase 10.1103/PhysRevLett.114.197001} {\bibfield  {journal}
		{\bibinfo  {journal} {Phys. Rev. Lett.}\ }\textbf {\bibinfo {volume} {114}},\
		\bibinfo {pages} {197001} (\bibinfo {year} {2015})}\BibitemShut {NoStop}%
	\bibitem [{\citenamefont {Vishik}(2018)}]{Vishik2018SC_CDW}%
	\BibitemOpen
	\bibfield  {author} {\bibinfo {author} {\bibfnamefont {I.~M.}\ \bibnamefont
			{Vishik}},\ }\href {\doibase 10.1088/1361-6633/aaba96} {\bibfield  {journal}
		{\bibinfo  {journal} {Reports on Progress in Physics}\ }\textbf {\bibinfo
			{volume} {81}},\ \bibinfo {pages} {062501} (\bibinfo {year}
		{2018})}\BibitemShut {NoStop}%
	\bibitem [{\citenamefont {Ortiz}\ \emph {et~al.}(2020)\citenamefont {Ortiz},
		\citenamefont {Teicher}, \citenamefont {Hu}, \citenamefont {Zuo},
		\citenamefont {Sarte}, \citenamefont {Schueller}, \citenamefont {Abeykoon},
		\citenamefont {Krogstad}, \citenamefont {Rosenkranz}, \citenamefont {Osborn},
		\citenamefont {Seshadri}, \citenamefont {Balents}, \citenamefont {He},\ and\
		\citenamefont {Wilson}}]{Ortiz2020SC}%
	\BibitemOpen
	\bibfield  {author} {\bibinfo {author} {\bibfnamefont {B.~R.}\ \bibnamefont
			{Ortiz}}, \bibinfo {author} {\bibfnamefont {S.~M.~L.}\ \bibnamefont
			{Teicher}}, \bibinfo {author} {\bibfnamefont {Y.}~\bibnamefont {Hu}},
		\bibinfo {author} {\bibfnamefont {J.~L.}\ \bibnamefont {Zuo}}, \bibinfo
		{author} {\bibfnamefont {P.~M.}\ \bibnamefont {Sarte}}, \bibinfo {author}
		{\bibfnamefont {E.~C.}\ \bibnamefont {Schueller}}, \bibinfo {author}
		{\bibfnamefont {A.~M.~M.}\ \bibnamefont {Abeykoon}}, \bibinfo {author}
		{\bibfnamefont {M.~J.}\ \bibnamefont {Krogstad}}, \bibinfo {author}
		{\bibfnamefont {S.}~\bibnamefont {Rosenkranz}}, \bibinfo {author}
		{\bibfnamefont {R.}~\bibnamefont {Osborn}}, \bibinfo {author} {\bibfnamefont
			{R.}~\bibnamefont {Seshadri}}, \bibinfo {author} {\bibfnamefont
			{L.}~\bibnamefont {Balents}}, \bibinfo {author} {\bibfnamefont
			{J.}~\bibnamefont {He}}, \ and\ \bibinfo {author} {\bibfnamefont {S.~D.}\
			\bibnamefont {Wilson}},\ }\href {\doibase 10.1103/PhysRevLett.125.247002}
	{\bibfield  {journal} {\bibinfo  {journal} {Phys. Rev. Lett.}\ }\textbf
		{\bibinfo {volume} {125}},\ \bibinfo {pages} {247002} (\bibinfo {year}
		{2020})}\BibitemShut {NoStop}%
	\bibitem [{\citenamefont {Xu}\ \emph {et~al.}(2024)\citenamefont {Xu},
		\citenamefont {Chung}, \citenamefont {Qin}, \citenamefont {Schollwöck},
		\citenamefont {White},\ and\ \citenamefont {Zhang}}]{Xu2024hubbard}%
	\BibitemOpen
	\bibfield  {author} {\bibinfo {author} {\bibfnamefont {H.}~\bibnamefont
			{Xu}}, \bibinfo {author} {\bibfnamefont {C.-M.}\ \bibnamefont {Chung}},
		\bibinfo {author} {\bibfnamefont {M.}~\bibnamefont {Qin}}, \bibinfo {author}
		{\bibfnamefont {U.}~\bibnamefont {Schollwöck}}, \bibinfo {author}
		{\bibfnamefont {S.~R.}\ \bibnamefont {White}}, \ and\ \bibinfo {author}
		{\bibfnamefont {S.}~\bibnamefont {Zhang}},\ }\href {\doibase
		10.1126/science.adh7691} {\bibfield  {journal} {\bibinfo  {journal}
			{Science}\ }\textbf {\bibinfo {volume} {384}},\ \bibinfo {pages} {eadh7691}
		(\bibinfo {year} {2024})}\BibitemShut {NoStop}%
	\bibitem [{\citenamefont {Kivelson}\ \emph {et~al.}(1986)\citenamefont
		{Kivelson}, \citenamefont {Kallin}, \citenamefont {Arovas},\ and\
		\citenamefont {Schrieffer}}]{Kivelson1986}%
	\BibitemOpen
	\bibfield  {author} {\bibinfo {author} {\bibfnamefont {S.}~\bibnamefont
			{Kivelson}}, \bibinfo {author} {\bibfnamefont {C.}~\bibnamefont {Kallin}},
		\bibinfo {author} {\bibfnamefont {D.~P.}\ \bibnamefont {Arovas}}, \ and\
		\bibinfo {author} {\bibfnamefont {J.~R.}\ \bibnamefont {Schrieffer}},\ }\href
	{\doibase 10.1103/PhysRevLett.56.873} {\bibfield  {journal} {\bibinfo
			{journal} {Phys. Rev. Lett.}\ }\textbf {\bibinfo {volume} {56}},\ \bibinfo
		{pages} {873} (\bibinfo {year} {1986})}\BibitemShut {NoStop}%
	\bibitem [{\citenamefont {Halperin}\ \emph {et~al.}(1986)\citenamefont
		{Halperin}, \citenamefont {Te\ifmmode \check{s}\else
			\v{s}\fi{}anovi\ifmmode~\acute{c}\else \'{c}\fi{}},\ and\ \citenamefont
		{Axel}}]{Halperin1986_QHC}%
	\BibitemOpen
	\bibfield  {author} {\bibinfo {author} {\bibfnamefont {B.~I.}\ \bibnamefont
			{Halperin}}, \bibinfo {author} {\bibfnamefont {Z.}~\bibnamefont {Te\ifmmode
				\check{s}\else \v{s}\fi{}anovi\ifmmode~\acute{c}\else \'{c}\fi{}}}, \ and\
		\bibinfo {author} {\bibfnamefont {F.}~\bibnamefont {Axel}},\ }\href {\doibase
		10.1103/PhysRevLett.57.922} {\bibfield  {journal} {\bibinfo  {journal} {Phys.
				Rev. Lett.}\ }\textbf {\bibinfo {volume} {57}},\ \bibinfo {pages} {922}
		(\bibinfo {year} {1986})}\BibitemShut {NoStop}%
	\bibitem [{\citenamefont {Te\ifmmode \check{s}\else
			\v{s}\fi{}anovi\ifmmode~\acute{c}\else \'{c}\fi{}}\ \emph
		{et~al.}(1989)\citenamefont {Te\ifmmode \check{s}\else
			\v{s}\fi{}anovi\ifmmode~\acute{c}\else \'{c}\fi{}}, \citenamefont {Axel},\
		and\ \citenamefont {Halperin}}]{Halperin1989_QHC}%
	\BibitemOpen
	\bibfield  {author} {\bibinfo {author} {\bibfnamefont {Z.}~\bibnamefont
			{Te\ifmmode \check{s}\else \v{s}\fi{}anovi\ifmmode~\acute{c}\else
				\'{c}\fi{}}}, \bibinfo {author} {\bibfnamefont {F.~m.~c.}\ \bibnamefont
			{Axel}}, \ and\ \bibinfo {author} {\bibfnamefont {B.~I.}\ \bibnamefont
			{Halperin}},\ }\href {\doibase 10.1103/PhysRevB.39.8525} {\bibfield
		{journal} {\bibinfo  {journal} {Phys. Rev. B}\ }\textbf {\bibinfo {volume}
			{39}},\ \bibinfo {pages} {8525} (\bibinfo {year} {1989})}\BibitemShut
	{NoStop}%
	\bibitem [{\citenamefont {Balents}(1996)}]{Balents1996_FQHC}%
	\BibitemOpen
	\bibfield  {author} {\bibinfo {author} {\bibfnamefont {L.}~\bibnamefont
			{Balents}},\ }\href {\doibase 10.1209/epl/i1996-00335-x} {\bibfield
		{journal} {\bibinfo  {journal} {Europhysics Letters}\ }\textbf {\bibinfo
			{volume} {33}},\ \bibinfo {pages} {291} (\bibinfo {year} {1996})}\BibitemShut
	{NoStop}%
	\bibitem [{\citenamefont {Murthy}(2000{\natexlab{a}})}]{Murthy2000_QHC}%
	\BibitemOpen
	\bibfield  {author} {\bibinfo {author} {\bibfnamefont {G.}~\bibnamefont
			{Murthy}},\ }\href {\doibase 10.1103/PhysRevLett.85.1954} {\bibfield
		{journal} {\bibinfo  {journal} {Phys. Rev. Lett.}\ }\textbf {\bibinfo
			{volume} {85}},\ \bibinfo {pages} {1954} (\bibinfo {year}
		{2000}{\natexlab{a}})}\BibitemShut {NoStop}%
	\bibitem [{\citenamefont {Fradkin}\ and\ \citenamefont
		{Kivelson}(1999)}]{Fradkin1999_Hall_crystal}%
	\BibitemOpen
	\bibfield  {author} {\bibinfo {author} {\bibfnamefont {E.}~\bibnamefont
			{Fradkin}}\ and\ \bibinfo {author} {\bibfnamefont {S.~A.}\ \bibnamefont
			{Kivelson}},\ }\href {\doibase 10.1103/PhysRevB.59.8065} {\bibfield
		{journal} {\bibinfo  {journal} {Phys. Rev. B}\ }\textbf {\bibinfo {volume}
			{59}},\ \bibinfo {pages} {8065} (\bibinfo {year} {1999})}\BibitemShut
	{NoStop}%
	\bibitem [{\citenamefont {Barci}\ \emph {et~al.}(2002)\citenamefont {Barci},
		\citenamefont {Fradkin}, \citenamefont {Kivelson},\ and\ \citenamefont
		{Oganesyan}}]{Fradkin2002_smecticHall1}%
	\BibitemOpen
	\bibfield  {author} {\bibinfo {author} {\bibfnamefont {D.~G.}\ \bibnamefont
			{Barci}}, \bibinfo {author} {\bibfnamefont {E.}~\bibnamefont {Fradkin}},
		\bibinfo {author} {\bibfnamefont {S.~A.}\ \bibnamefont {Kivelson}}, \ and\
		\bibinfo {author} {\bibfnamefont {V.}~\bibnamefont {Oganesyan}},\ }\href
	{\doibase 10.1103/PhysRevB.65.245319} {\bibfield  {journal} {\bibinfo
			{journal} {Phys. Rev. B}\ }\textbf {\bibinfo {volume} {65}},\ \bibinfo
		{pages} {245319} (\bibinfo {year} {2002})}\BibitemShut {NoStop}%
	\bibitem [{\citenamefont {Barci}\ and\ \citenamefont
		{Fradkin}(2002)}]{Fradkin2002_smecticHall2}%
	\BibitemOpen
	\bibfield  {author} {\bibinfo {author} {\bibfnamefont {D.~G.}\ \bibnamefont
			{Barci}}\ and\ \bibinfo {author} {\bibfnamefont {E.}~\bibnamefont
			{Fradkin}},\ }\href {\doibase 10.1103/PhysRevB.65.245320} {\bibfield
		{journal} {\bibinfo  {journal} {Phys. Rev. B}\ }\textbf {\bibinfo {volume}
			{65}},\ \bibinfo {pages} {245320} (\bibinfo {year} {2002})}\BibitemShut
	{NoStop}%
	\bibitem [{\citenamefont {Murthy}(2000{\natexlab{b}})}]{Murthy2000_CFC}%
	\BibitemOpen
	\bibfield  {author} {\bibinfo {author} {\bibfnamefont {G.}~\bibnamefont
			{Murthy}},\ }\href {\doibase 10.1103/PhysRevLett.84.350} {\bibfield
		{journal} {\bibinfo  {journal} {Phys. Rev. Lett.}\ }\textbf {\bibinfo
			{volume} {84}},\ \bibinfo {pages} {350} (\bibinfo {year}
		{2000}{\natexlab{b}})}\BibitemShut {NoStop}%
	\bibitem [{\citenamefont {Chang}\ \emph {et~al.}(2006)\citenamefont {Chang},
		\citenamefont {T\"oke}, \citenamefont {Jeon},\ and\ \citenamefont
		{Jain}}]{Jain2006_CFC}%
	\BibitemOpen
	\bibfield  {author} {\bibinfo {author} {\bibfnamefont {C.-C.}\ \bibnamefont
			{Chang}}, \bibinfo {author} {\bibfnamefont {C.}~\bibnamefont {T\"oke}},
		\bibinfo {author} {\bibfnamefont {G.~S.}\ \bibnamefont {Jeon}}, \ and\
		\bibinfo {author} {\bibfnamefont {J.~K.}\ \bibnamefont {Jain}},\ }\href
	{\doibase 10.1103/PhysRevB.73.155323} {\bibfield  {journal} {\bibinfo
			{journal} {Phys. Rev. B}\ }\textbf {\bibinfo {volume} {73}},\ \bibinfo
		{pages} {155323} (\bibinfo {year} {2006})}\BibitemShut {NoStop}%
	\bibitem [{\citenamefont {You}\ \emph {et~al.}(2014)\citenamefont {You},
		\citenamefont {Cho},\ and\ \citenamefont {Fradkin}}]{You2014nematic}%
	\BibitemOpen
	\bibfield  {author} {\bibinfo {author} {\bibfnamefont {Y.}~\bibnamefont
			{You}}, \bibinfo {author} {\bibfnamefont {G.~Y.}\ \bibnamefont {Cho}}, \ and\
		\bibinfo {author} {\bibfnamefont {E.}~\bibnamefont {Fradkin}},\ }\href
	{\doibase 10.1103/PhysRevX.4.041050} {\bibfield  {journal} {\bibinfo
			{journal} {Phys. Rev. X}\ }\textbf {\bibinfo {volume} {4}},\ \bibinfo {pages}
		{041050} (\bibinfo {year} {2014})}\BibitemShut {NoStop}%
	\bibitem [{\citenamefont {Zuo}\ \emph {et~al.}(2020)\citenamefont {Zuo},
		\citenamefont {Balram}, \citenamefont {Pu}, \citenamefont {Zhao},
		\citenamefont {Jolicoeur}, \citenamefont {W\'ojs},\ and\ \citenamefont
		{Jain}}]{Jain2020_QHC}%
	\BibitemOpen
	\bibfield  {author} {\bibinfo {author} {\bibfnamefont {Z.-W.}\ \bibnamefont
			{Zuo}}, \bibinfo {author} {\bibfnamefont {A.~C.}\ \bibnamefont {Balram}},
		\bibinfo {author} {\bibfnamefont {S.}~\bibnamefont {Pu}}, \bibinfo {author}
		{\bibfnamefont {J.}~\bibnamefont {Zhao}}, \bibinfo {author} {\bibfnamefont
			{T.}~\bibnamefont {Jolicoeur}}, \bibinfo {author} {\bibfnamefont
			{A.}~\bibnamefont {W\'ojs}}, \ and\ \bibinfo {author} {\bibfnamefont {J.~K.}\
			\bibnamefont {Jain}},\ }\href {\doibase 10.1103/PhysRevB.102.075307}
	{\bibfield  {journal} {\bibinfo  {journal} {Phys. Rev. B}\ }\textbf {\bibinfo
			{volume} {102}},\ \bibinfo {pages} {075307} (\bibinfo {year}
		{2020})}\BibitemShut {NoStop}%
	\bibitem [{\citenamefont {Trung}\ and\ \citenamefont
		{Yang}(2021)}]{trungFractionalization2021}%
	\BibitemOpen
	\bibfield  {author} {\bibinfo {author} {\bibfnamefont {H.~Q.}\ \bibnamefont
			{Trung}}\ and\ \bibinfo {author} {\bibfnamefont {B.}~\bibnamefont {Yang}},\
	}\href {\doibase 10.1103/PhysRevLett.127.046402} {\bibfield  {journal}
		{\bibinfo  {journal} {Phys. Rev. Lett.}\ }\textbf {\bibinfo {volume} {127}},\
		\bibinfo {pages} {046402} (\bibinfo {year} {2021})}\BibitemShut {NoStop}%
	\bibitem [{\citenamefont {Cs\'athy}\ \emph {et~al.}(2004)\citenamefont
		{Cs\'athy}, \citenamefont {Tsui}, \citenamefont {Pfeiffer},\ and\
		\citenamefont {West}}]{West2004_FQHC}%
	\BibitemOpen
	\bibfield  {author} {\bibinfo {author} {\bibfnamefont {G.~A.}\ \bibnamefont
			{Cs\'athy}}, \bibinfo {author} {\bibfnamefont {D.~C.}\ \bibnamefont {Tsui}},
		\bibinfo {author} {\bibfnamefont {L.~N.}\ \bibnamefont {Pfeiffer}}, \ and\
		\bibinfo {author} {\bibfnamefont {K.~W.}\ \bibnamefont {West}},\ }\href
	{\doibase 10.1103/PhysRevLett.92.256804} {\bibfield  {journal} {\bibinfo
			{journal} {Phys. Rev. Lett.}\ }\textbf {\bibinfo {volume} {92}},\ \bibinfo
		{pages} {256804} (\bibinfo {year} {2004})}\BibitemShut {NoStop}%
	\bibitem [{\citenamefont {Zhu}\ \emph {et~al.}(2010)\citenamefont {Zhu},
		\citenamefont {Chen}, \citenamefont {Jiang}, \citenamefont {Engel},
		\citenamefont {Tsui}, \citenamefont {Pfeiffer},\ and\ \citenamefont
		{West}}]{West2010_FQHC}%
	\BibitemOpen
	\bibfield  {author} {\bibinfo {author} {\bibfnamefont {H.}~\bibnamefont
			{Zhu}}, \bibinfo {author} {\bibfnamefont {Y.~P.}\ \bibnamefont {Chen}},
		\bibinfo {author} {\bibfnamefont {P.}~\bibnamefont {Jiang}}, \bibinfo
		{author} {\bibfnamefont {L.~W.}\ \bibnamefont {Engel}}, \bibinfo {author}
		{\bibfnamefont {D.~C.}\ \bibnamefont {Tsui}}, \bibinfo {author}
		{\bibfnamefont {L.~N.}\ \bibnamefont {Pfeiffer}}, \ and\ \bibinfo {author}
		{\bibfnamefont {K.~W.}\ \bibnamefont {West}},\ }\href {\doibase
		10.1103/PhysRevLett.105.126803} {\bibfield  {journal} {\bibinfo  {journal}
			{Phys. Rev. Lett.}\ }\textbf {\bibinfo {volume} {105}},\ \bibinfo {pages}
		{126803} (\bibinfo {year} {2010})}\BibitemShut {NoStop}%
	\bibitem [{\citenamefont {Xia}\ \emph {et~al.}(2011)\citenamefont {Xia},
		\citenamefont {Eisenstein}, \citenamefont {Pfeiffer},\ and\ \citenamefont
		{West}}]{xiaEvidence2011}%
	\BibitemOpen
	\bibfield  {author} {\bibinfo {author} {\bibfnamefont {J.}~\bibnamefont
			{Xia}}, \bibinfo {author} {\bibfnamefont {J.~P.}\ \bibnamefont {Eisenstein}},
		\bibinfo {author} {\bibfnamefont {L.~N.}\ \bibnamefont {Pfeiffer}}, \ and\
		\bibinfo {author} {\bibfnamefont {K.~W.}\ \bibnamefont {West}},\ }\href
	{\doibase 10.1038/nphys2118} {\bibfield  {journal} {\bibinfo  {journal}
			{Nature Physics}\ }\textbf {\bibinfo {volume} {7}},\ \bibinfo {pages} {845 }
		(\bibinfo {year} {2011})}\BibitemShut {NoStop}%
	\bibitem [{\citenamefont {Du}\ \emph {et~al.}(2019)\citenamefont {Du},
		\citenamefont {Wurstbauer}, \citenamefont {West}, \citenamefont {Pfeiffer},
		\citenamefont {Fallahi}, \citenamefont {Gardner}, \citenamefont {Manfra},\
		and\ \citenamefont {Pinczuk}}]{duObservation2019}%
	\BibitemOpen
	\bibfield  {author} {\bibinfo {author} {\bibfnamefont {L.}~\bibnamefont
			{Du}}, \bibinfo {author} {\bibfnamefont {U.}~\bibnamefont {Wurstbauer}},
		\bibinfo {author} {\bibfnamefont {K.~W.}\ \bibnamefont {West}}, \bibinfo
		{author} {\bibfnamefont {L.~N.}\ \bibnamefont {Pfeiffer}}, \bibinfo {author}
		{\bibfnamefont {S.}~\bibnamefont {Fallahi}}, \bibinfo {author} {\bibfnamefont
			{G.~C.}\ \bibnamefont {Gardner}}, \bibinfo {author} {\bibfnamefont {M.~J.}\
			\bibnamefont {Manfra}}, \ and\ \bibinfo {author} {\bibfnamefont
			{A.}~\bibnamefont {Pinczuk}},\ }\href {\doibase 10.1126/sciadv.aav3407}
	{\bibfield  {journal} {\bibinfo  {journal} {Science Advances}\ }\textbf
		{\bibinfo {volume} {5}},\ \bibinfo {pages} {eaav3407} (\bibinfo {year}
		{2019})}\BibitemShut {NoStop}%
	\bibitem [{\citenamefont {Fu}\ \emph {et~al.}(2020)\citenamefont {Fu},
		\citenamefont {Shi}, \citenamefont {Zudov}, \citenamefont {Gardner},
		\citenamefont {Watson}, \citenamefont {Manfra}, \citenamefont {Baldwin},
		\citenamefont {Pfeiffer},\ and\ \citenamefont {West}}]{xuAnomalous2020}%
	\BibitemOpen
	\bibfield  {author} {\bibinfo {author} {\bibfnamefont {X.}~\bibnamefont
			{Fu}}, \bibinfo {author} {\bibfnamefont {Q.}~\bibnamefont {Shi}}, \bibinfo
		{author} {\bibfnamefont {M.~A.}\ \bibnamefont {Zudov}}, \bibinfo {author}
		{\bibfnamefont {G.~C.}\ \bibnamefont {Gardner}}, \bibinfo {author}
		{\bibfnamefont {J.~D.}\ \bibnamefont {Watson}}, \bibinfo {author}
		{\bibfnamefont {M.~J.}\ \bibnamefont {Manfra}}, \bibinfo {author}
		{\bibfnamefont {K.~W.}\ \bibnamefont {Baldwin}}, \bibinfo {author}
		{\bibfnamefont {L.~N.}\ \bibnamefont {Pfeiffer}}, \ and\ \bibinfo {author}
		{\bibfnamefont {K.~W.}\ \bibnamefont {West}},\ }\href {\doibase
		10.1103/PhysRevLett.124.067601} {\bibfield  {journal} {\bibinfo  {journal}
			{Phys. Rev. Lett.}\ }\textbf {\bibinfo {volume} {124}},\ \bibinfo {pages}
		{067601} (\bibinfo {year} {2020})}\BibitemShut {NoStop}%
	\bibitem [{\citenamefont {Shingla}\ \emph {et~al.}(2023)\citenamefont
		{Shingla}, \citenamefont {Huang}, \citenamefont {Kumar}, \citenamefont
		{Pfeiffer}, \citenamefont {West}, \citenamefont {Baldwin},\ and\
		\citenamefont {Csáthy}}]{Shingla2023_bubble}%
	\BibitemOpen
	\bibfield  {author} {\bibinfo {author} {\bibfnamefont {V.}~\bibnamefont
			{Shingla}}, \bibinfo {author} {\bibfnamefont {H.}~\bibnamefont {Huang}},
		\bibinfo {author} {\bibfnamefont {A.}~\bibnamefont {Kumar}}, \bibinfo
		{author} {\bibfnamefont {L.~N.}\ \bibnamefont {Pfeiffer}}, \bibinfo {author}
		{\bibfnamefont {K.~W.}\ \bibnamefont {West}}, \bibinfo {author}
		{\bibfnamefont {K.~W.}\ \bibnamefont {Baldwin}}, \ and\ \bibinfo {author}
		{\bibfnamefont {G.~A.}\ \bibnamefont {Csáthy}},\ }\href {\doibase
		10.1038/s41567-023-01939-2} {\bibfield  {journal} {\bibinfo  {journal}
			{Nature Physics}\ }\textbf {\bibinfo {volume} {19}},\ \bibinfo {pages} {689}
		(\bibinfo {year} {2023})}\BibitemShut {NoStop}%
	\bibitem [{\citenamefont {Feldman}\ \emph {et~al.}(2016)\citenamefont
		{Feldman}, \citenamefont {Randeria}, \citenamefont {Gyenis}, \citenamefont
		{Wu}, \citenamefont {Ji}, \citenamefont {Cava}, \citenamefont {MacDonald},\
		and\ \citenamefont {Yazdani}}]{Benjamin2016nematic}%
	\BibitemOpen
	\bibfield  {author} {\bibinfo {author} {\bibfnamefont {B.~E.}\ \bibnamefont
			{Feldman}}, \bibinfo {author} {\bibfnamefont {M.~T.}\ \bibnamefont
			{Randeria}}, \bibinfo {author} {\bibfnamefont {A.}~\bibnamefont {Gyenis}},
		\bibinfo {author} {\bibfnamefont {F.}~\bibnamefont {Wu}}, \bibinfo {author}
		{\bibfnamefont {H.}~\bibnamefont {Ji}}, \bibinfo {author} {\bibfnamefont
			{R.~J.}\ \bibnamefont {Cava}}, \bibinfo {author} {\bibfnamefont {A.~H.}\
			\bibnamefont {MacDonald}}, \ and\ \bibinfo {author} {\bibfnamefont
			{A.}~\bibnamefont {Yazdani}},\ }\href {\doibase 10.1126/science.aag1715}
	{\bibfield  {journal} {\bibinfo  {journal} {Science}\ }\textbf {\bibinfo
			{volume} {354}},\ \bibinfo {pages} {316} (\bibinfo {year}
		{2016})}\BibitemShut {NoStop}%
	\bibitem [{\citenamefont {Samkharadze}\ \emph {et~al.}(2016)\citenamefont
		{Samkharadze}, \citenamefont {Schreiber}, \citenamefont {Gardner},
		\citenamefont {Manfra}, \citenamefont {Fradkin},\ and\ \citenamefont
		{Csáthy}}]{Samkharadze2016_nematicHall}%
	\BibitemOpen
	\bibfield  {author} {\bibinfo {author} {\bibfnamefont {N.}~\bibnamefont
			{Samkharadze}}, \bibinfo {author} {\bibfnamefont {K.~A.}\ \bibnamefont
			{Schreiber}}, \bibinfo {author} {\bibfnamefont {G.~C.}\ \bibnamefont
			{Gardner}}, \bibinfo {author} {\bibfnamefont {M.~J.}\ \bibnamefont {Manfra}},
		\bibinfo {author} {\bibfnamefont {E.}~\bibnamefont {Fradkin}}, \ and\
		\bibinfo {author} {\bibfnamefont {G.~A.}\ \bibnamefont {Csáthy}},\ }\href
	{\doibase 10.1038/nphys3523} {\bibfield  {journal} {\bibinfo  {journal}
			{Nature Physics}\ }\textbf {\bibinfo {volume} {12}},\ \bibinfo {pages} {191}
		(\bibinfo {year} {2016})}\BibitemShut {NoStop}%
	\bibitem [{\citenamefont {Yang}(2020)}]{yangMicroscopic2020}%
	\BibitemOpen
	\bibfield  {author} {\bibinfo {author} {\bibfnamefont {B.}~\bibnamefont
			{Yang}},\ }\href {\doibase 10.1103/PhysRevResearch.2.033362} {\bibfield
		{journal} {\bibinfo  {journal} {Phys. Rev. Res.}\ }\textbf {\bibinfo {volume}
			{2}},\ \bibinfo {pages} {033362} (\bibinfo {year} {2020})}\BibitemShut
	{NoStop}%
	\bibitem [{\citenamefont {Pu}\ \emph {et~al.}(2024)\citenamefont {Pu},
		\citenamefont {Balram}, \citenamefont {Taylor}, \citenamefont {Fradkin},\
		and\ \citenamefont {Papi\ifmmode~\acute{c}\else \'{c}\fi{}}}]{Pu2024nematic}%
	\BibitemOpen
	\bibfield  {author} {\bibinfo {author} {\bibfnamefont {S.}~\bibnamefont
			{Pu}}, \bibinfo {author} {\bibfnamefont {A.~C.}\ \bibnamefont {Balram}},
		\bibinfo {author} {\bibfnamefont {J.}~\bibnamefont {Taylor}}, \bibinfo
		{author} {\bibfnamefont {E.}~\bibnamefont {Fradkin}}, \ and\ \bibinfo
		{author} {\bibfnamefont {Z.}~\bibnamefont {Papi\ifmmode~\acute{c}\else
				\'{c}\fi{}}},\ }\href {\doibase 10.1103/PhysRevLett.132.236503} {\bibfield
		{journal} {\bibinfo  {journal} {Phys. Rev. Lett.}\ }\textbf {\bibinfo
			{volume} {132}},\ \bibinfo {pages} {236503} (\bibinfo {year}
		{2024})}\BibitemShut {NoStop}%
	\bibitem [{\citenamefont {Haldane}(1988)}]{Haldane1988_QAH}%
	\BibitemOpen
	\bibfield  {author} {\bibinfo {author} {\bibfnamefont {F.~D.~M.}\
			\bibnamefont {Haldane}},\ }\href {\doibase 10.1103/PhysRevLett.61.2015}
	{\bibfield  {journal} {\bibinfo  {journal} {Phys. Rev. Lett.}\ }\textbf
		{\bibinfo {volume} {61}},\ \bibinfo {pages} {2015} (\bibinfo {year}
		{1988})}\BibitemShut {NoStop}%
	\bibitem [{\citenamefont {Sun}\ \emph {et~al.}(2011)\citenamefont {Sun},
		\citenamefont {Gu}, \citenamefont {Katsura},\ and\ \citenamefont
		{Das~Sarma}}]{KSun2011_model}%
	\BibitemOpen
	\bibfield  {author} {\bibinfo {author} {\bibfnamefont {K.}~\bibnamefont
			{Sun}}, \bibinfo {author} {\bibfnamefont {Z.}~\bibnamefont {Gu}}, \bibinfo
		{author} {\bibfnamefont {H.}~\bibnamefont {Katsura}}, \ and\ \bibinfo
		{author} {\bibfnamefont {S.}~\bibnamefont {Das~Sarma}},\ }\href {\doibase
		10.1103/PhysRevLett.106.236803} {\bibfield  {journal} {\bibinfo  {journal}
			{Phys. Rev. Lett.}\ }\textbf {\bibinfo {volume} {106}},\ \bibinfo {pages}
		{236803} (\bibinfo {year} {2011})}\BibitemShut {NoStop}%
	\bibitem [{\citenamefont {Sheng}\ \emph {et~al.}(2011)\citenamefont {Sheng},
		\citenamefont {Gu}, \citenamefont {Sun},\ and\ \citenamefont
		{Sheng}}]{DNSheng2011_fci}%
	\BibitemOpen
	\bibfield  {author} {\bibinfo {author} {\bibfnamefont {D.}~\bibnamefont
			{Sheng}}, \bibinfo {author} {\bibfnamefont {Z.-C.}\ \bibnamefont {Gu}},
		\bibinfo {author} {\bibfnamefont {K.}~\bibnamefont {Sun}}, \ and\ \bibinfo
		{author} {\bibfnamefont {L.}~\bibnamefont {Sheng}},\ }\href {\doibase
		10.1038/ncomms1380} {\bibfield  {journal} {\bibinfo  {journal} {Nature
				Communications}\ }\textbf {\bibinfo {volume} {2}},\ \bibinfo {pages} {389}
		(\bibinfo {year} {2011})}\BibitemShut {NoStop}%
	\bibitem [{\citenamefont {Neupert}\ \emph {et~al.}(2011)\citenamefont
		{Neupert}, \citenamefont {Santos}, \citenamefont {Chamon},\ and\
		\citenamefont {Mudry}}]{Neupert2011_fci}%
	\BibitemOpen
	\bibfield  {author} {\bibinfo {author} {\bibfnamefont {T.}~\bibnamefont
			{Neupert}}, \bibinfo {author} {\bibfnamefont {L.}~\bibnamefont {Santos}},
		\bibinfo {author} {\bibfnamefont {C.}~\bibnamefont {Chamon}}, \ and\ \bibinfo
		{author} {\bibfnamefont {C.}~\bibnamefont {Mudry}},\ }\href {\doibase
		10.1103/PhysRevLett.106.236804} {\bibfield  {journal} {\bibinfo  {journal}
			{Phys. Rev. Lett.}\ }\textbf {\bibinfo {volume} {106}},\ \bibinfo {pages}
		{236804} (\bibinfo {year} {2011})}\BibitemShut {NoStop}%
	\bibitem [{\citenamefont {Regnault}\ and\ \citenamefont
		{Bernevig}(2011)}]{Bernevig2011fci}%
	\BibitemOpen
	\bibfield  {author} {\bibinfo {author} {\bibfnamefont {N.}~\bibnamefont
			{Regnault}}\ and\ \bibinfo {author} {\bibfnamefont {B.~A.}\ \bibnamefont
			{Bernevig}},\ }\href {\doibase 10.1103/PhysRevX.1.021014} {\bibfield
		{journal} {\bibinfo  {journal} {Phys. Rev. X}\ }\textbf {\bibinfo {volume}
			{1}},\ \bibinfo {pages} {021014} (\bibinfo {year} {2011})}\BibitemShut
	{NoStop}%
	\bibitem [{\citenamefont {Tang}\ \emph {et~al.}(2011)\citenamefont {Tang},
		\citenamefont {Mei},\ and\ \citenamefont {Wen}}]{XGWen2011_fci}%
	\BibitemOpen
	\bibfield  {author} {\bibinfo {author} {\bibfnamefont {E.}~\bibnamefont
			{Tang}}, \bibinfo {author} {\bibfnamefont {J.-W.}\ \bibnamefont {Mei}}, \
		and\ \bibinfo {author} {\bibfnamefont {X.-G.}\ \bibnamefont {Wen}},\ }\href
	{\doibase 10.1103/PhysRevLett.106.236802} {\bibfield  {journal} {\bibinfo
			{journal} {Phys. Rev. Lett.}\ }\textbf {\bibinfo {volume} {106}},\ \bibinfo
		{pages} {236802} (\bibinfo {year} {2011})}\BibitemShut {NoStop}%
	\bibitem [{\citenamefont {Chang}\ \emph {et~al.}(2013)\citenamefont {Chang},
		\citenamefont {Zhang}, \citenamefont {Feng}, \citenamefont {Shen},
		\citenamefont {Zhang}, \citenamefont {Guo}, \citenamefont {Li}, \citenamefont
		{Ou}, \citenamefont {Wei}, \citenamefont {Wang}, \citenamefont {Ji},
		\citenamefont {Feng}, \citenamefont {Ji}, \citenamefont {Chen}, \citenamefont
		{Jia}, \citenamefont {Dai}, \citenamefont {Fang}, \citenamefont {Zhang},
		\citenamefont {He}, \citenamefont {Wang}, \citenamefont {Lu}, \citenamefont
		{Ma},\ and\ \citenamefont {Xue}}]{CZChang2013_QAH}%
	\BibitemOpen
	\bibfield  {author} {\bibinfo {author} {\bibfnamefont {C.-Z.}\ \bibnamefont
			{Chang}}, \bibinfo {author} {\bibfnamefont {J.}~\bibnamefont {Zhang}},
		\bibinfo {author} {\bibfnamefont {X.}~\bibnamefont {Feng}}, \bibinfo {author}
		{\bibfnamefont {J.}~\bibnamefont {Shen}}, \bibinfo {author} {\bibfnamefont
			{Z.}~\bibnamefont {Zhang}}, \bibinfo {author} {\bibfnamefont
			{M.}~\bibnamefont {Guo}}, \bibinfo {author} {\bibfnamefont {K.}~\bibnamefont
			{Li}}, \bibinfo {author} {\bibfnamefont {Y.}~\bibnamefont {Ou}}, \bibinfo
		{author} {\bibfnamefont {P.}~\bibnamefont {Wei}}, \bibinfo {author}
		{\bibfnamefont {L.-L.}\ \bibnamefont {Wang}}, \bibinfo {author}
		{\bibfnamefont {Z.-Q.}\ \bibnamefont {Ji}}, \bibinfo {author} {\bibfnamefont
			{Y.}~\bibnamefont {Feng}}, \bibinfo {author} {\bibfnamefont {S.}~\bibnamefont
			{Ji}}, \bibinfo {author} {\bibfnamefont {X.}~\bibnamefont {Chen}}, \bibinfo
		{author} {\bibfnamefont {J.}~\bibnamefont {Jia}}, \bibinfo {author}
		{\bibfnamefont {X.}~\bibnamefont {Dai}}, \bibinfo {author} {\bibfnamefont
			{Z.}~\bibnamefont {Fang}}, \bibinfo {author} {\bibfnamefont {S.-C.}\
			\bibnamefont {Zhang}}, \bibinfo {author} {\bibfnamefont {K.}~\bibnamefont
			{He}}, \bibinfo {author} {\bibfnamefont {Y.}~\bibnamefont {Wang}}, \bibinfo
		{author} {\bibfnamefont {L.}~\bibnamefont {Lu}}, \bibinfo {author}
		{\bibfnamefont {X.-C.}\ \bibnamefont {Ma}}, \ and\ \bibinfo {author}
		{\bibfnamefont {Q.-K.}\ \bibnamefont {Xue}},\ }\href {\doibase
		10.1126/science.1234414} {\bibfield  {journal} {\bibinfo  {journal}
			{Science}\ }\textbf {\bibinfo {volume} {340}},\ \bibinfo {pages} {167}
		(\bibinfo {year} {2013})}\BibitemShut {NoStop}%
	\bibitem [{\citenamefont {Cai}\ \emph {et~al.}(2023)\citenamefont {Cai},
		\citenamefont {Anderson}, \citenamefont {Wang}, \citenamefont {Zhang},
		\citenamefont {Liu}, \citenamefont {Holtzmann}, \citenamefont {Zhang},
		\citenamefont {Fan}, \citenamefont {Taniguchi}, \citenamefont {Watanabe},
		\citenamefont {Ran}, \citenamefont {Cao}, \citenamefont {Fu}, \citenamefont
		{Xiao}, \citenamefont {Yao},\ and\ \citenamefont {Xu}}]{caiSignature2023}%
	\BibitemOpen
	\bibfield  {author} {\bibinfo {author} {\bibfnamefont {J.}~\bibnamefont
			{Cai}}, \bibinfo {author} {\bibfnamefont {E.}~\bibnamefont {Anderson}},
		\bibinfo {author} {\bibfnamefont {C.}~\bibnamefont {Wang}}, \bibinfo {author}
		{\bibfnamefont {X.}~\bibnamefont {Zhang}}, \bibinfo {author} {\bibfnamefont
			{X.}~\bibnamefont {Liu}}, \bibinfo {author} {\bibfnamefont {W.}~\bibnamefont
			{Holtzmann}}, \bibinfo {author} {\bibfnamefont {Y.}~\bibnamefont {Zhang}},
		\bibinfo {author} {\bibfnamefont {F.}~\bibnamefont {Fan}}, \bibinfo {author}
		{\bibfnamefont {T.}~\bibnamefont {Taniguchi}}, \bibinfo {author}
		{\bibfnamefont {K.}~\bibnamefont {Watanabe}}, \bibinfo {author}
		{\bibfnamefont {Y.}~\bibnamefont {Ran}}, \bibinfo {author} {\bibfnamefont
			{T.}~\bibnamefont {Cao}}, \bibinfo {author} {\bibfnamefont {L.}~\bibnamefont
			{Fu}}, \bibinfo {author} {\bibfnamefont {D.}~\bibnamefont {Xiao}}, \bibinfo
		{author} {\bibfnamefont {W.}~\bibnamefont {Yao}}, \ and\ \bibinfo {author}
		{\bibfnamefont {X.}~\bibnamefont {Xu}},\ }\href {\doibase
		10.1038/s41586-023-06289-w} {\bibfield  {journal} {\bibinfo  {journal}
			{Nature}\ }\textbf {\bibinfo {volume} {622}},\ \bibinfo {pages} {63 }
		(\bibinfo {year} {2023})}\BibitemShut {NoStop}%
	\bibitem [{\citenamefont {Park}\ \emph {et~al.}(2023)\citenamefont {Park},
		\citenamefont {Cai}, \citenamefont {Anderson}, \citenamefont {Zhang},
		\citenamefont {Zhu}, \citenamefont {Liu}, \citenamefont {Wang}, \citenamefont
		{Holtzmann}, \citenamefont {Hu}, \citenamefont {Liu}, \citenamefont
		{Taniguchi}, \citenamefont {Watanabe}, \citenamefont {Chu}, \citenamefont
		{Cao}, \citenamefont {Fu}, \citenamefont {Yao}, \citenamefont {Chang},
		\citenamefont {Cobden}, \citenamefont {Xiao},\ and\ \citenamefont
		{Xu}}]{park2023_fqah}%
	\BibitemOpen
	\bibfield  {author} {\bibinfo {author} {\bibfnamefont {H.}~\bibnamefont
			{Park}}, \bibinfo {author} {\bibfnamefont {J.}~\bibnamefont {Cai}}, \bibinfo
		{author} {\bibfnamefont {E.}~\bibnamefont {Anderson}}, \bibinfo {author}
		{\bibfnamefont {Y.}~\bibnamefont {Zhang}}, \bibinfo {author} {\bibfnamefont
			{J.}~\bibnamefont {Zhu}}, \bibinfo {author} {\bibfnamefont {X.}~\bibnamefont
			{Liu}}, \bibinfo {author} {\bibfnamefont {C.}~\bibnamefont {Wang}}, \bibinfo
		{author} {\bibfnamefont {W.}~\bibnamefont {Holtzmann}}, \bibinfo {author}
		{\bibfnamefont {C.}~\bibnamefont {Hu}}, \bibinfo {author} {\bibfnamefont
			{Z.}~\bibnamefont {Liu}}, \bibinfo {author} {\bibfnamefont {T.}~\bibnamefont
			{Taniguchi}}, \bibinfo {author} {\bibfnamefont {K.}~\bibnamefont {Watanabe}},
		\bibinfo {author} {\bibfnamefont {J.-H.}\ \bibnamefont {Chu}}, \bibinfo
		{author} {\bibfnamefont {T.}~\bibnamefont {Cao}}, \bibinfo {author}
		{\bibfnamefont {L.}~\bibnamefont {Fu}}, \bibinfo {author} {\bibfnamefont
			{W.}~\bibnamefont {Yao}}, \bibinfo {author} {\bibfnamefont {C.-Z.}\
			\bibnamefont {Chang}}, \bibinfo {author} {\bibfnamefont {D.}~\bibnamefont
			{Cobden}}, \bibinfo {author} {\bibfnamefont {D.}~\bibnamefont {Xiao}}, \ and\
		\bibinfo {author} {\bibfnamefont {X.}~\bibnamefont {Xu}},\ }\href {\doibase
		10.1038/s41586-023-06536-0} {\bibfield  {journal} {\bibinfo  {journal}
			{Nature}\ }\textbf {\bibinfo {volume} {622}},\ \bibinfo {pages} {74}
		(\bibinfo {year} {2023})}\BibitemShut {NoStop}%
	\bibitem [{\citenamefont {Zeng}\ \emph {et~al.}(2023)\citenamefont {Zeng},
		\citenamefont {Xia}, \citenamefont {Kang}, \citenamefont {Zhu}, \citenamefont
		{Kn\"uppel}, \citenamefont {Vaswani}, \citenamefont {Watanabe}, \citenamefont
		{Taniguchi}, \citenamefont {Mak},\ and\ \citenamefont
		{Shan}}]{zengThermodynamic2023}%
	\BibitemOpen
	\bibfield  {author} {\bibinfo {author} {\bibfnamefont {Y.}~\bibnamefont
			{Zeng}}, \bibinfo {author} {\bibfnamefont {Z.}~\bibnamefont {Xia}}, \bibinfo
		{author} {\bibfnamefont {K.}~\bibnamefont {Kang}}, \bibinfo {author}
		{\bibfnamefont {J.}~\bibnamefont {Zhu}}, \bibinfo {author} {\bibfnamefont
			{P.}~\bibnamefont {Kn\"uppel}}, \bibinfo {author} {\bibfnamefont
			{C.}~\bibnamefont {Vaswani}}, \bibinfo {author} {\bibfnamefont
			{K.}~\bibnamefont {Watanabe}}, \bibinfo {author} {\bibfnamefont
			{T.}~\bibnamefont {Taniguchi}}, \bibinfo {author} {\bibfnamefont {K.~F.}\
			\bibnamefont {Mak}}, \ and\ \bibinfo {author} {\bibfnamefont
			{J.}~\bibnamefont {Shan}},\ }\href {\doibase 10.1038/s41586-023-06452-3}
	{\bibfield  {journal} {\bibinfo  {journal} {Nature}\ }\textbf {\bibinfo
			{volume} {622}},\ \bibinfo {pages} {69 } (\bibinfo {year}
		{2023})}\BibitemShut {NoStop}%
	\bibitem [{\citenamefont {Xu}\ \emph {et~al.}(2023)\citenamefont {Xu},
		\citenamefont {Sun}, \citenamefont {Jia}, \citenamefont {Liu}, \citenamefont
		{Xu}, \citenamefont {Li}, \citenamefont {Gu}, \citenamefont {Watanabe},
		\citenamefont {Taniguchi}, \citenamefont {Tong}, \citenamefont {Jia},
		\citenamefont {Shi}, \citenamefont {Jiang}, \citenamefont {Zhang},
		\citenamefont {Liu},\ and\ \citenamefont {Li}}]{xu2023_fci}%
	\BibitemOpen
	\bibfield  {author} {\bibinfo {author} {\bibfnamefont {F.}~\bibnamefont
			{Xu}}, \bibinfo {author} {\bibfnamefont {Z.}~\bibnamefont {Sun}}, \bibinfo
		{author} {\bibfnamefont {T.}~\bibnamefont {Jia}}, \bibinfo {author}
		{\bibfnamefont {C.}~\bibnamefont {Liu}}, \bibinfo {author} {\bibfnamefont
			{C.}~\bibnamefont {Xu}}, \bibinfo {author} {\bibfnamefont {C.}~\bibnamefont
			{Li}}, \bibinfo {author} {\bibfnamefont {Y.}~\bibnamefont {Gu}}, \bibinfo
		{author} {\bibfnamefont {K.}~\bibnamefont {Watanabe}}, \bibinfo {author}
		{\bibfnamefont {T.}~\bibnamefont {Taniguchi}}, \bibinfo {author}
		{\bibfnamefont {B.}~\bibnamefont {Tong}}, \bibinfo {author} {\bibfnamefont
			{J.}~\bibnamefont {Jia}}, \bibinfo {author} {\bibfnamefont {Z.}~\bibnamefont
			{Shi}}, \bibinfo {author} {\bibfnamefont {S.}~\bibnamefont {Jiang}}, \bibinfo
		{author} {\bibfnamefont {Y.}~\bibnamefont {Zhang}}, \bibinfo {author}
		{\bibfnamefont {X.}~\bibnamefont {Liu}}, \ and\ \bibinfo {author}
		{\bibfnamefont {T.}~\bibnamefont {Li}},\ }\href {\doibase
		10.1103/PhysRevX.13.031037} {\bibfield  {journal} {\bibinfo  {journal} {Phys.
				Rev. X}\ }\textbf {\bibinfo {volume} {13}},\ \bibinfo {pages} {031037}
		(\bibinfo {year} {2023})}\BibitemShut {NoStop}%
	\bibitem [{\citenamefont {Lu}\ \emph {et~al.}(2024{\natexlab{a}})\citenamefont
		{Lu}, \citenamefont {Han}, \citenamefont {Yao}, \citenamefont {Reddy},
		\citenamefont {Yang}, \citenamefont {Seo}, \citenamefont {Watanabe},
		\citenamefont {Taniguchi}, \citenamefont {Fu},\ and\ \citenamefont
		{Ju}}]{multilayer_graphene_fqah}%
	\BibitemOpen
	\bibfield  {author} {\bibinfo {author} {\bibfnamefont {Z.}~\bibnamefont
			{Lu}}, \bibinfo {author} {\bibfnamefont {T.}~\bibnamefont {Han}}, \bibinfo
		{author} {\bibfnamefont {Y.}~\bibnamefont {Yao}}, \bibinfo {author}
		{\bibfnamefont {A.~P.}\ \bibnamefont {Reddy}}, \bibinfo {author}
		{\bibfnamefont {J.}~\bibnamefont {Yang}}, \bibinfo {author} {\bibfnamefont
			{J.}~\bibnamefont {Seo}}, \bibinfo {author} {\bibfnamefont {K.}~\bibnamefont
			{Watanabe}}, \bibinfo {author} {\bibfnamefont {T.}~\bibnamefont {Taniguchi}},
		\bibinfo {author} {\bibfnamefont {L.}~\bibnamefont {Fu}}, \ and\ \bibinfo
		{author} {\bibfnamefont {L.}~\bibnamefont {Ju}},\ }\href {\doibase
		10.1038/s41586-023-07010-7} {\bibfield  {journal} {\bibinfo  {journal}
			{Nature}\ }\textbf {\bibinfo {volume} {626}},\ \bibinfo {pages} {759}
		(\bibinfo {year} {2024}{\natexlab{a}})}\BibitemShut {NoStop}%
	\bibitem [{\citenamefont {Kang}\ \emph {et~al.}(2024)\citenamefont {Kang},
		\citenamefont {Shen}, \citenamefont {Qiu}, \citenamefont {Zeng},
		\citenamefont {Xia}, \citenamefont {Watanabe}, \citenamefont {Taniguchi},
		\citenamefont {Shan},\ and\ \citenamefont {Mak}}]{Kang2024fqsh}%
	\BibitemOpen
	\bibfield  {author} {\bibinfo {author} {\bibfnamefont {K.}~\bibnamefont
			{Kang}}, \bibinfo {author} {\bibfnamefont {B.}~\bibnamefont {Shen}}, \bibinfo
		{author} {\bibfnamefont {Y.}~\bibnamefont {Qiu}}, \bibinfo {author}
		{\bibfnamefont {Y.}~\bibnamefont {Zeng}}, \bibinfo {author} {\bibfnamefont
			{Z.}~\bibnamefont {Xia}}, \bibinfo {author} {\bibfnamefont {K.}~\bibnamefont
			{Watanabe}}, \bibinfo {author} {\bibfnamefont {T.}~\bibnamefont {Taniguchi}},
		\bibinfo {author} {\bibfnamefont {J.}~\bibnamefont {Shan}}, \ and\ \bibinfo
		{author} {\bibfnamefont {K.~F.}\ \bibnamefont {Mak}},\ }\href {\doibase
		10.1038/s41586-024-07214-5} {\bibfield  {journal} {\bibinfo  {journal}
			{Nature}\ }\textbf {\bibinfo {volume} {628}},\ \bibinfo {pages} {522}
		(\bibinfo {year} {2024})}\BibitemShut {NoStop}%
	\bibitem [{\citenamefont {Song}\ \emph
		{et~al.}(2024{\natexlab{a}})\citenamefont {Song}, \citenamefont {Jian},
		\citenamefont {Fu},\ and\ \citenamefont {Xu}}]{XYSong2023fqahc}%
	\BibitemOpen
	\bibfield  {author} {\bibinfo {author} {\bibfnamefont {X.-Y.}\ \bibnamefont
			{Song}}, \bibinfo {author} {\bibfnamefont {C.-M.}\ \bibnamefont {Jian}},
		\bibinfo {author} {\bibfnamefont {L.}~\bibnamefont {Fu}}, \ and\ \bibinfo
		{author} {\bibfnamefont {C.}~\bibnamefont {Xu}},\ }\href {\doibase
		10.1103/PhysRevB.109.115116} {\bibfield  {journal} {\bibinfo  {journal}
			{Phys. Rev. B}\ }\textbf {\bibinfo {volume} {109}},\ \bibinfo {pages}
		{115116} (\bibinfo {year} {2024}{\natexlab{a}})}\BibitemShut {NoStop}%
	\bibitem [{\citenamefont {Dong}\ \emph
		{et~al.}(2023{\natexlab{a}})\citenamefont {Dong}, \citenamefont {Patri},\
		and\ \citenamefont {Senthil}}]{dong2023graphene}%
	\BibitemOpen
	\bibfield  {author} {\bibinfo {author} {\bibfnamefont {Z.}~\bibnamefont
			{Dong}}, \bibinfo {author} {\bibfnamefont {A.~S.}\ \bibnamefont {Patri}}, \
		and\ \bibinfo {author} {\bibfnamefont {T.}~\bibnamefont {Senthil}},\
	}\href@noop {} {\  (\bibinfo {year} {2023}{\natexlab{a}})},\ \Eprint
	{http://arxiv.org/abs/2311.03445} {arXiv:2311.03445 [cond-mat.str-el]}
	\BibitemShut {NoStop}%
	\bibitem [{\citenamefont {Zhou}\ \emph {et~al.}(2023)\citenamefont {Zhou},
		\citenamefont {Yang},\ and\ \citenamefont {Zhang}}]{BRZhou2023_AHC}%
	\BibitemOpen
	\bibfield  {author} {\bibinfo {author} {\bibfnamefont {B.}~\bibnamefont
			{Zhou}}, \bibinfo {author} {\bibfnamefont {H.}~\bibnamefont {Yang}}, \ and\
		\bibinfo {author} {\bibfnamefont {Y.-H.}\ \bibnamefont {Zhang}},\ }\href@noop
	{} {\  (\bibinfo {year} {2023})},\ \Eprint {http://arxiv.org/abs/2311.04217}
	{arXiv:2311.04217 [cond-mat.str-el]} \BibitemShut {NoStop}%
	\bibitem [{\citenamefont {Dong}\ \emph
		{et~al.}(2023{\natexlab{b}})\citenamefont {Dong}, \citenamefont {Wang},
		\citenamefont {Wang}, \citenamefont {Soejima}, \citenamefont {Zaletel},
		\citenamefont {Vishwanath},\ and\ \citenamefont {Parker}}]{Dong2023_AHC}%
	\BibitemOpen
	\bibfield  {author} {\bibinfo {author} {\bibfnamefont {J.}~\bibnamefont
			{Dong}}, \bibinfo {author} {\bibfnamefont {T.}~\bibnamefont {Wang}}, \bibinfo
		{author} {\bibfnamefont {T.}~\bibnamefont {Wang}}, \bibinfo {author}
		{\bibfnamefont {T.}~\bibnamefont {Soejima}}, \bibinfo {author} {\bibfnamefont
			{M.~P.}\ \bibnamefont {Zaletel}}, \bibinfo {author} {\bibfnamefont
			{A.}~\bibnamefont {Vishwanath}}, \ and\ \bibinfo {author} {\bibfnamefont
			{D.~E.}\ \bibnamefont {Parker}},\ }\href@noop {} {\  (\bibinfo {year}
		{2023}{\natexlab{b}})},\ \Eprint {http://arxiv.org/abs/2311.05568}
	{arXiv:2311.05568 [cond-mat.str-el]} \BibitemShut {NoStop}%
	\bibitem [{\citenamefont {{Kwan}}\ \emph {et~al.}(2023)\citenamefont {{Kwan}},
		\citenamefont {{Yu}}, \citenamefont {{Herzog-Arbeitman}}, \citenamefont
		{{Efetov}}, \citenamefont {{Regnault}},\ and\ \citenamefont
		{{Bernevig}}}]{kwanMoire2023}%
	\BibitemOpen
	\bibfield  {author} {\bibinfo {author} {\bibfnamefont {Y.~H.}\ \bibnamefont
			{{Kwan}}}, \bibinfo {author} {\bibfnamefont {J.}~\bibnamefont {{Yu}}},
		\bibinfo {author} {\bibfnamefont {J.}~\bibnamefont {{Herzog-Arbeitman}}},
		\bibinfo {author} {\bibfnamefont {D.~K.}\ \bibnamefont {{Efetov}}}, \bibinfo
		{author} {\bibfnamefont {N.}~\bibnamefont {{Regnault}}}, \ and\ \bibinfo
		{author} {\bibfnamefont {B.~A.}\ \bibnamefont {{Bernevig}}},\ }\href
	{\doibase 10.48550/arXiv.2312.11617} {\bibfield  {journal} {\bibinfo
			{journal} {arXiv e-prints}\ ,\ \bibinfo {eid} {arXiv:2312.11617}} (\bibinfo
		{year} {2023})},\ \Eprint {http://arxiv.org/abs/2312.11617} {arXiv:2312.11617
		[cond-mat.str-el]} \BibitemShut {NoStop}%
	\bibitem [{\citenamefont {Soejima}\ \emph {et~al.}(2024)\citenamefont
		{Soejima}, \citenamefont {Dong}, \citenamefont {Wang}, \citenamefont {Wang},
		\citenamefont {Zaletel}, \citenamefont {Vishwanath},\ and\ \citenamefont
		{Parker}}]{ashvin2024ahc2}%
	\BibitemOpen
	\bibfield  {author} {\bibinfo {author} {\bibfnamefont {T.}~\bibnamefont
			{Soejima}}, \bibinfo {author} {\bibfnamefont {J.}~\bibnamefont {Dong}},
		\bibinfo {author} {\bibfnamefont {T.}~\bibnamefont {Wang}}, \bibinfo {author}
		{\bibfnamefont {T.}~\bibnamefont {Wang}}, \bibinfo {author} {\bibfnamefont
			{M.~P.}\ \bibnamefont {Zaletel}}, \bibinfo {author} {\bibfnamefont
			{A.}~\bibnamefont {Vishwanath}}, \ and\ \bibinfo {author} {\bibfnamefont
			{D.~E.}\ \bibnamefont {Parker}},\ }\href@noop {} {\  (\bibinfo {year}
		{2024})},\ \Eprint {http://arxiv.org/abs/2403.05522} {arXiv:2403.05522
		[cond-mat.str-el]} \BibitemShut {NoStop}%
	\bibitem [{\citenamefont {Dong}\ \emph {et~al.}(2024)\citenamefont {Dong},
		\citenamefont {Patri},\ and\ \citenamefont {Senthil}}]{dong2024ahc}%
	\BibitemOpen
	\bibfield  {author} {\bibinfo {author} {\bibfnamefont {Z.}~\bibnamefont
			{Dong}}, \bibinfo {author} {\bibfnamefont {A.~S.}\ \bibnamefont {Patri}}, \
		and\ \bibinfo {author} {\bibfnamefont {T.}~\bibnamefont {Senthil}},\
	}\href@noop {} {\  (\bibinfo {year} {2024})},\ \Eprint
	{http://arxiv.org/abs/2403.07873} {arXiv:2403.07873 [cond-mat.str-el]}
	\BibitemShut {NoStop}%
	\bibitem [{\citenamefont {Tan}\ and\ \citenamefont
		{Devakul}(2024)}]{tan2024ahc}%
	\BibitemOpen
	\bibfield  {author} {\bibinfo {author} {\bibfnamefont {T.}~\bibnamefont
			{Tan}}\ and\ \bibinfo {author} {\bibfnamefont {T.}~\bibnamefont {Devakul}},\
	}\href@noop {} {\  (\bibinfo {year} {2024})},\ \Eprint
	{http://arxiv.org/abs/2403.04196} {arXiv:2403.04196 [cond-mat.mes-hall]}
	\BibitemShut {NoStop}%
	\bibitem [{\citenamefont {{Sheng}}\ \emph {et~al.}(2024)\citenamefont
		{{Sheng}}, \citenamefont {{Reddy}}, \citenamefont {{Abouelkomsan}},
		\citenamefont {{Bergholtz}},\ and\ \citenamefont {{Fu}}}]{Sheng2024QAHC}%
	\BibitemOpen
	\bibfield  {author} {\bibinfo {author} {\bibfnamefont {D.~N.}\ \bibnamefont
			{{Sheng}}}, \bibinfo {author} {\bibfnamefont {A.~P.}\ \bibnamefont
			{{Reddy}}}, \bibinfo {author} {\bibfnamefont {A.}~\bibnamefont
			{{Abouelkomsan}}}, \bibinfo {author} {\bibfnamefont {E.~J.}\ \bibnamefont
			{{Bergholtz}}}, \ and\ \bibinfo {author} {\bibfnamefont {L.}~\bibnamefont
			{{Fu}}},\ }\href {\doibase 10.48550/arXiv.2402.17832} {\bibfield  {journal}
		{\bibinfo  {journal} {arXiv e-prints}\ ,\ \bibinfo {eid} {arXiv:2402.17832}}
		(\bibinfo {year} {2024})},\ \Eprint {http://arxiv.org/abs/2402.17832}
	{arXiv:2402.17832 [cond-mat.mes-hall]} \BibitemShut {NoStop}%
	\bibitem [{\citenamefont {Pan}\ \emph {et~al.}(2022)\citenamefont {Pan},
		\citenamefont {Xie}, \citenamefont {Wu},\ and\ \citenamefont
		{Das~Sarma}}]{pan2022ahc}%
	\BibitemOpen
	\bibfield  {author} {\bibinfo {author} {\bibfnamefont {H.}~\bibnamefont
			{Pan}}, \bibinfo {author} {\bibfnamefont {M.}~\bibnamefont {Xie}}, \bibinfo
		{author} {\bibfnamefont {F.}~\bibnamefont {Wu}}, \ and\ \bibinfo {author}
		{\bibfnamefont {S.}~\bibnamefont {Das~Sarma}},\ }\href {\doibase
		10.1103/PhysRevLett.129.056804} {\bibfield  {journal} {\bibinfo  {journal}
			{Phys. Rev. Lett.}\ }\textbf {\bibinfo {volume} {129}},\ \bibinfo {pages}
		{056804} (\bibinfo {year} {2022})}\BibitemShut {NoStop}%
	\bibitem [{\citenamefont {Xie}\ \emph {et~al.}(2021)\citenamefont {Xie},
		\citenamefont {Pierce}, \citenamefont {Park}, \citenamefont {Parker},
		\citenamefont {Khalaf}, \citenamefont {Ledwith}, \citenamefont {Cao},
		\citenamefont {Lee}, \citenamefont {Chen}, \citenamefont {Forrester},
		\citenamefont {Watanabe}, \citenamefont {Taniguchi}, \citenamefont
		{Vishwanath}, \citenamefont {Jarillo-Herrero},\ and\ \citenamefont
		{Yacoby}}]{xie2021fci}%
	\BibitemOpen
	\bibfield  {author} {\bibinfo {author} {\bibfnamefont {Y.}~\bibnamefont
			{Xie}}, \bibinfo {author} {\bibfnamefont {A.~T.}\ \bibnamefont {Pierce}},
		\bibinfo {author} {\bibfnamefont {J.~M.}\ \bibnamefont {Park}}, \bibinfo
		{author} {\bibfnamefont {D.~E.}\ \bibnamefont {Parker}}, \bibinfo {author}
		{\bibfnamefont {E.}~\bibnamefont {Khalaf}}, \bibinfo {author} {\bibfnamefont
			{P.}~\bibnamefont {Ledwith}}, \bibinfo {author} {\bibfnamefont
			{Y.}~\bibnamefont {Cao}}, \bibinfo {author} {\bibfnamefont {S.~H.}\
			\bibnamefont {Lee}}, \bibinfo {author} {\bibfnamefont {S.}~\bibnamefont
			{Chen}}, \bibinfo {author} {\bibfnamefont {P.~R.}\ \bibnamefont {Forrester}},
		\bibinfo {author} {\bibfnamefont {K.}~\bibnamefont {Watanabe}}, \bibinfo
		{author} {\bibfnamefont {T.}~\bibnamefont {Taniguchi}}, \bibinfo {author}
		{\bibfnamefont {A.}~\bibnamefont {Vishwanath}}, \bibinfo {author}
		{\bibfnamefont {P.}~\bibnamefont {Jarillo-Herrero}}, \ and\ \bibinfo {author}
		{\bibfnamefont {A.}~\bibnamefont {Yacoby}},\ }\href {\doibase
		10.1038/s41586-021-04002-3} {\bibfield  {journal} {\bibinfo  {journal}
			{Nature}\ }\textbf {\bibinfo {volume} {600}},\ \bibinfo {pages} {439}
		(\bibinfo {year} {2021})}\BibitemShut {NoStop}%
	\bibitem [{\citenamefont {Polshyn}\ \emph {et~al.}(2022)\citenamefont
		{Polshyn}, \citenamefont {Zhang}, \citenamefont {Kumar}, \citenamefont
		{Soejima}, \citenamefont {Ledwith}, \citenamefont {Watanabe}, \citenamefont
		{Taniguchi}, \citenamefont {Vishwanath}, \citenamefont {Zaletel},\ and\
		\citenamefont {Young}}]{polshyn2022tcdw}%
	\BibitemOpen
	\bibfield  {author} {\bibinfo {author} {\bibfnamefont {H.}~\bibnamefont
			{Polshyn}}, \bibinfo {author} {\bibfnamefont {Y.}~\bibnamefont {Zhang}},
		\bibinfo {author} {\bibfnamefont {M.~A.}\ \bibnamefont {Kumar}}, \bibinfo
		{author} {\bibfnamefont {T.}~\bibnamefont {Soejima}}, \bibinfo {author}
		{\bibfnamefont {P.}~\bibnamefont {Ledwith}}, \bibinfo {author} {\bibfnamefont
			{K.}~\bibnamefont {Watanabe}}, \bibinfo {author} {\bibfnamefont
			{T.}~\bibnamefont {Taniguchi}}, \bibinfo {author} {\bibfnamefont
			{A.}~\bibnamefont {Vishwanath}}, \bibinfo {author} {\bibfnamefont {M.~P.}\
			\bibnamefont {Zaletel}}, \ and\ \bibinfo {author} {\bibfnamefont {A.~F.}\
			\bibnamefont {Young}},\ }\href {\doibase 10.1038/s41567-021-01418-6}
	{\bibfield  {journal} {\bibinfo  {journal} {Nature Physics}\ }\textbf
		{\bibinfo {volume} {18}},\ \bibinfo {pages} {42} (\bibinfo {year}
		{2022})}\BibitemShut {NoStop}%
	\bibitem [{\citenamefont {Kourtis}\ and\ \citenamefont
		{Daghofer}(2014)}]{Kourtis2014}%
	\BibitemOpen
	\bibfield  {author} {\bibinfo {author} {\bibfnamefont {S.}~\bibnamefont
			{Kourtis}}\ and\ \bibinfo {author} {\bibfnamefont {M.}~\bibnamefont
			{Daghofer}},\ }\href {\doibase 10.1103/PhysRevLett.113.216404} {\bibfield
		{journal} {\bibinfo  {journal} {Phys. Rev. Lett.}\ }\textbf {\bibinfo
			{volume} {113}},\ \bibinfo {pages} {216404} (\bibinfo {year}
		{2014})}\BibitemShut {NoStop}%
	\bibitem [{\citenamefont {Kourtis}(2018)}]{stefanos2018fqahc}%
	\BibitemOpen
	\bibfield  {author} {\bibinfo {author} {\bibfnamefont {S.}~\bibnamefont
			{Kourtis}},\ }\href {\doibase 10.1103/PhysRevB.97.085108} {\bibfield
		{journal} {\bibinfo  {journal} {Phys. Rev. B}\ }\textbf {\bibinfo {volume}
			{97}},\ \bibinfo {pages} {085108} (\bibinfo {year} {2018})}\BibitemShut
	{NoStop}%
	\bibitem [{\citenamefont {Girvin}\ \emph {et~al.}(1985)\citenamefont {Girvin},
		\citenamefont {MacDonald},\ and\ \citenamefont {Platzman}}]{GMP1985}%
	\BibitemOpen
	\bibfield  {author} {\bibinfo {author} {\bibfnamefont {S.~M.}\ \bibnamefont
			{Girvin}}, \bibinfo {author} {\bibfnamefont {A.~H.}\ \bibnamefont
			{MacDonald}}, \ and\ \bibinfo {author} {\bibfnamefont {P.~M.}\ \bibnamefont
			{Platzman}},\ }\href {\doibase 10.1103/PhysRevLett.54.581} {\bibfield
		{journal} {\bibinfo  {journal} {Phys. Rev. Lett.}\ }\textbf {\bibinfo
			{volume} {54}},\ \bibinfo {pages} {581} (\bibinfo {year} {1985})}\BibitemShut
	{NoStop}%
	\bibitem [{\citenamefont {Girvin}\ \emph {et~al.}(1986)\citenamefont {Girvin},
		\citenamefont {MacDonald},\ and\ \citenamefont {Platzman}}]{GMP1986}%
	\BibitemOpen
	\bibfield  {author} {\bibinfo {author} {\bibfnamefont {S.~M.}\ \bibnamefont
			{Girvin}}, \bibinfo {author} {\bibfnamefont {A.~H.}\ \bibnamefont
			{MacDonald}}, \ and\ \bibinfo {author} {\bibfnamefont {P.~M.}\ \bibnamefont
			{Platzman}},\ }\href {\doibase 10.1103/PhysRevB.33.2481} {\bibfield
		{journal} {\bibinfo  {journal} {Phys. Rev. B}\ }\textbf {\bibinfo {volume}
			{33}},\ \bibinfo {pages} {2481} (\bibinfo {year} {1986})}\BibitemShut
	{NoStop}%
	\bibitem [{\citenamefont {Zhang}\ \emph {et~al.}(1989)\citenamefont {Zhang},
		\citenamefont {Hansson},\ and\ \citenamefont
		{Kivelson}}]{Zhang1989composite}%
	\BibitemOpen
	\bibfield  {author} {\bibinfo {author} {\bibfnamefont {S.~C.}\ \bibnamefont
			{Zhang}}, \bibinfo {author} {\bibfnamefont {T.~H.}\ \bibnamefont {Hansson}},
		\ and\ \bibinfo {author} {\bibfnamefont {S.}~\bibnamefont {Kivelson}},\
	}\href {\doibase 10.1103/PhysRevLett.62.82} {\bibfield  {journal} {\bibinfo
			{journal} {Phys. Rev. Lett.}\ }\textbf {\bibinfo {volume} {62}},\ \bibinfo
		{pages} {82} (\bibinfo {year} {1989})}\BibitemShut {NoStop}%
	\bibitem [{\citenamefont {Sengupta}\ \emph {et~al.}(2005)\citenamefont
		{Sengupta}, \citenamefont {Pryadko}, \citenamefont {Alet}, \citenamefont
		{Troyer},\ and\ \citenamefont {Schmid}}]{Sengupta2005supersolid}%
	\BibitemOpen
	\bibfield  {author} {\bibinfo {author} {\bibfnamefont {P.}~\bibnamefont
			{Sengupta}}, \bibinfo {author} {\bibfnamefont {L.~P.}\ \bibnamefont
			{Pryadko}}, \bibinfo {author} {\bibfnamefont {F.}~\bibnamefont {Alet}},
		\bibinfo {author} {\bibfnamefont {M.}~\bibnamefont {Troyer}}, \ and\ \bibinfo
		{author} {\bibfnamefont {G.}~\bibnamefont {Schmid}},\ }\href {\doibase
		10.1103/PhysRevLett.94.207202} {\bibfield  {journal} {\bibinfo  {journal}
			{Phys. Rev. Lett.}\ }\textbf {\bibinfo {volume} {94}},\ \bibinfo {pages}
		{207202} (\bibinfo {year} {2005})}\BibitemShut {NoStop}%
	\bibitem [{\citenamefont {Raczkowski}\ and\ \citenamefont
		{Poilblanc}(2009)}]{Raczkowski2009supersolid}%
	\BibitemOpen
	\bibfield  {author} {\bibinfo {author} {\bibfnamefont {M.}~\bibnamefont
			{Raczkowski}}\ and\ \bibinfo {author} {\bibfnamefont {D.}~\bibnamefont
			{Poilblanc}},\ }\href {\doibase 10.1103/PhysRevLett.103.027001} {\bibfield
		{journal} {\bibinfo  {journal} {Phys. Rev. Lett.}\ }\textbf {\bibinfo
			{volume} {103}},\ \bibinfo {pages} {027001} (\bibinfo {year}
		{2009})}\BibitemShut {NoStop}%
	\bibitem [{\citenamefont {Xi}\ \emph {et~al.}(2011)\citenamefont {Xi},
		\citenamefont {Ye}, \citenamefont {Chen}, \citenamefont {Zhang},\ and\
		\citenamefont {Su}}]{Xi2011supersolid}%
	\BibitemOpen
	\bibfield  {author} {\bibinfo {author} {\bibfnamefont {B.}~\bibnamefont
			{Xi}}, \bibinfo {author} {\bibfnamefont {F.}~\bibnamefont {Ye}}, \bibinfo
		{author} {\bibfnamefont {W.}~\bibnamefont {Chen}}, \bibinfo {author}
		{\bibfnamefont {F.}~\bibnamefont {Zhang}}, \ and\ \bibinfo {author}
		{\bibfnamefont {G.}~\bibnamefont {Su}},\ }\href {\doibase
		10.1103/PhysRevB.84.054512} {\bibfield  {journal} {\bibinfo  {journal} {Phys.
				Rev. B}\ }\textbf {\bibinfo {volume} {84}},\ \bibinfo {pages} {054512}
		(\bibinfo {year} {2011})}\BibitemShut {NoStop}%
	\bibitem [{\citenamefont {Chomaz}\ \emph {et~al.}(2018)\citenamefont {Chomaz},
		\citenamefont {van Bijnen}, \citenamefont {Petter}, \citenamefont {Faraoni},
		\citenamefont {Baier}, \citenamefont {Becher}, \citenamefont {Mark},
		\citenamefont {Wächtler}, \citenamefont {Santos},\ and\ \citenamefont
		{Ferlaino}}]{Chomaz2018roton}%
	\BibitemOpen
	\bibfield  {author} {\bibinfo {author} {\bibfnamefont {L.}~\bibnamefont
			{Chomaz}}, \bibinfo {author} {\bibfnamefont {R.~M.~W.}\ \bibnamefont {van
				Bijnen}}, \bibinfo {author} {\bibfnamefont {D.}~\bibnamefont {Petter}},
		\bibinfo {author} {\bibfnamefont {G.}~\bibnamefont {Faraoni}}, \bibinfo
		{author} {\bibfnamefont {S.}~\bibnamefont {Baier}}, \bibinfo {author}
		{\bibfnamefont {J.~H.}\ \bibnamefont {Becher}}, \bibinfo {author}
		{\bibfnamefont {M.~J.}\ \bibnamefont {Mark}}, \bibinfo {author}
		{\bibfnamefont {F.}~\bibnamefont {Wächtler}}, \bibinfo {author}
		{\bibfnamefont {L.}~\bibnamefont {Santos}}, \ and\ \bibinfo {author}
		{\bibfnamefont {F.}~\bibnamefont {Ferlaino}},\ }\href {\doibase
		10.1038/s41567-018-0054-7} {\bibfield  {journal} {\bibinfo  {journal} {Nature
				Physics}\ }\textbf {\bibinfo {volume} {14}},\ \bibinfo {pages} {442}
		(\bibinfo {year} {2018})}\BibitemShut {NoStop}%
	\bibitem [{\citenamefont {Tanzi}\ \emph {et~al.}(2019)\citenamefont {Tanzi},
		\citenamefont {Lucioni}, \citenamefont {Fam\`a}, \citenamefont {Catani},
		\citenamefont {Fioretti}, \citenamefont {Gabbanini}, \citenamefont {Bisset},
		\citenamefont {Santos},\ and\ \citenamefont {Modugno}}]{Tanzi2019supersolid}%
	\BibitemOpen
	\bibfield  {author} {\bibinfo {author} {\bibfnamefont {L.}~\bibnamefont
			{Tanzi}}, \bibinfo {author} {\bibfnamefont {E.}~\bibnamefont {Lucioni}},
		\bibinfo {author} {\bibfnamefont {F.}~\bibnamefont {Fam\`a}}, \bibinfo
		{author} {\bibfnamefont {J.}~\bibnamefont {Catani}}, \bibinfo {author}
		{\bibfnamefont {A.}~\bibnamefont {Fioretti}}, \bibinfo {author}
		{\bibfnamefont {C.}~\bibnamefont {Gabbanini}}, \bibinfo {author}
		{\bibfnamefont {R.~N.}\ \bibnamefont {Bisset}}, \bibinfo {author}
		{\bibfnamefont {L.}~\bibnamefont {Santos}}, \ and\ \bibinfo {author}
		{\bibfnamefont {G.}~\bibnamefont {Modugno}},\ }\href {\doibase
		10.1103/PhysRevLett.122.130405} {\bibfield  {journal} {\bibinfo  {journal}
			{Phys. Rev. Lett.}\ }\textbf {\bibinfo {volume} {122}},\ \bibinfo {pages}
		{130405} (\bibinfo {year} {2019})}\BibitemShut {NoStop}%
	\bibitem [{\citenamefont {B\"ottcher}\ \emph {et~al.}(2019)\citenamefont
		{B\"ottcher}, \citenamefont {Schmidt}, \citenamefont {Wenzel}, \citenamefont
		{Hertkorn}, \citenamefont {Guo}, \citenamefont {Langen},\ and\ \citenamefont
		{Pfau}}]{Fabien2019supersolid}%
	\BibitemOpen
	\bibfield  {author} {\bibinfo {author} {\bibfnamefont {F.}~\bibnamefont
			{B\"ottcher}}, \bibinfo {author} {\bibfnamefont {J.-N.}\ \bibnamefont
			{Schmidt}}, \bibinfo {author} {\bibfnamefont {M.}~\bibnamefont {Wenzel}},
		\bibinfo {author} {\bibfnamefont {J.}~\bibnamefont {Hertkorn}}, \bibinfo
		{author} {\bibfnamefont {M.}~\bibnamefont {Guo}}, \bibinfo {author}
		{\bibfnamefont {T.}~\bibnamefont {Langen}}, \ and\ \bibinfo {author}
		{\bibfnamefont {T.}~\bibnamefont {Pfau}},\ }\href {\doibase
		10.1103/PhysRevX.9.011051} {\bibfield  {journal} {\bibinfo  {journal} {Phys.
				Rev. X}\ }\textbf {\bibinfo {volume} {9}},\ \bibinfo {pages} {011051}
		(\bibinfo {year} {2019})}\BibitemShut {NoStop}%
	\bibitem [{\citenamefont {Chomaz}\ \emph {et~al.}(2019)\citenamefont {Chomaz},
		\citenamefont {Petter}, \citenamefont {Ilzh\"ofer}, \citenamefont {Natale},
		\citenamefont {Trautmann}, \citenamefont {Politi}, \citenamefont
		{Durastante}, \citenamefont {van Bijnen}, \citenamefont {Patscheider},
		\citenamefont {Sohmen}, \citenamefont {Mark},\ and\ \citenamefont
		{Ferlaino}}]{Chomaz2019supersolid}%
	\BibitemOpen
	\bibfield  {author} {\bibinfo {author} {\bibfnamefont {L.}~\bibnamefont
			{Chomaz}}, \bibinfo {author} {\bibfnamefont {D.}~\bibnamefont {Petter}},
		\bibinfo {author} {\bibfnamefont {P.}~\bibnamefont {Ilzh\"ofer}}, \bibinfo
		{author} {\bibfnamefont {G.}~\bibnamefont {Natale}}, \bibinfo {author}
		{\bibfnamefont {A.}~\bibnamefont {Trautmann}}, \bibinfo {author}
		{\bibfnamefont {C.}~\bibnamefont {Politi}}, \bibinfo {author} {\bibfnamefont
			{G.}~\bibnamefont {Durastante}}, \bibinfo {author} {\bibfnamefont {R.~M.~W.}\
			\bibnamefont {van Bijnen}}, \bibinfo {author} {\bibfnamefont
			{A.}~\bibnamefont {Patscheider}}, \bibinfo {author} {\bibfnamefont
			{M.}~\bibnamefont {Sohmen}}, \bibinfo {author} {\bibfnamefont {M.~J.}\
			\bibnamefont {Mark}}, \ and\ \bibinfo {author} {\bibfnamefont
			{F.}~\bibnamefont {Ferlaino}},\ }\href {\doibase 10.1103/PhysRevX.9.021012}
	{\bibfield  {journal} {\bibinfo  {journal} {Phys. Rev. X}\ }\textbf {\bibinfo
			{volume} {9}},\ \bibinfo {pages} {021012} (\bibinfo {year}
		{2019})}\BibitemShut {NoStop}%
	\bibitem [{\citenamefont {Zhang}\ \emph {et~al.}(2019)\citenamefont {Zhang},
		\citenamefont {Maucher},\ and\ \citenamefont {Pohl}}]{zhang2019supersolid}%
	\BibitemOpen
	\bibfield  {author} {\bibinfo {author} {\bibfnamefont {Y.-C.}\ \bibnamefont
			{Zhang}}, \bibinfo {author} {\bibfnamefont {F.}~\bibnamefont {Maucher}}, \
		and\ \bibinfo {author} {\bibfnamefont {T.}~\bibnamefont {Pohl}},\ }\href
	{\doibase 10.1103/PhysRevLett.123.015301} {\bibfield  {journal} {\bibinfo
			{journal} {Phys. Rev. Lett.}\ }\textbf {\bibinfo {volume} {123}},\ \bibinfo
		{pages} {015301} (\bibinfo {year} {2019})}\BibitemShut {NoStop}%
	\bibitem [{\citenamefont {Bland}\ \emph {et~al.}(2022)\citenamefont {Bland},
		\citenamefont {Poli}, \citenamefont {Politi}, \citenamefont {Klaus},
		\citenamefont {Norcia}, \citenamefont {Ferlaino}, \citenamefont {Santos},\
		and\ \citenamefont {Bisset}}]{Bland2022supersolid}%
	\BibitemOpen
	\bibfield  {author} {\bibinfo {author} {\bibfnamefont {T.}~\bibnamefont
			{Bland}}, \bibinfo {author} {\bibfnamefont {E.}~\bibnamefont {Poli}},
		\bibinfo {author} {\bibfnamefont {C.}~\bibnamefont {Politi}}, \bibinfo
		{author} {\bibfnamefont {L.}~\bibnamefont {Klaus}}, \bibinfo {author}
		{\bibfnamefont {M.~A.}\ \bibnamefont {Norcia}}, \bibinfo {author}
		{\bibfnamefont {F.}~\bibnamefont {Ferlaino}}, \bibinfo {author}
		{\bibfnamefont {L.}~\bibnamefont {Santos}}, \ and\ \bibinfo {author}
		{\bibfnamefont {R.~N.}\ \bibnamefont {Bisset}},\ }\href {\doibase
		10.1103/PhysRevLett.128.195302} {\bibfield  {journal} {\bibinfo  {journal}
			{Phys. Rev. Lett.}\ }\textbf {\bibinfo {volume} {128}},\ \bibinfo {pages}
		{195302} (\bibinfo {year} {2022})}\BibitemShut {NoStop}%
	\bibitem [{\citenamefont {Ala\~na}\ \emph {et~al.}(2023)\citenamefont
		{Ala\~na}, \citenamefont {Egusquiza},\ and\ \citenamefont
		{Modugno}}]{Alana2023supersolid}%
	\BibitemOpen
	\bibfield  {author} {\bibinfo {author} {\bibfnamefont {A.}~\bibnamefont
			{Ala\~na}}, \bibinfo {author} {\bibfnamefont {I.~n.~L.}\ \bibnamefont
			{Egusquiza}}, \ and\ \bibinfo {author} {\bibfnamefont {M.}~\bibnamefont
			{Modugno}},\ }\href {\doibase 10.1103/PhysRevA.108.033316} {\bibfield
		{journal} {\bibinfo  {journal} {Phys. Rev. A}\ }\textbf {\bibinfo {volume}
			{108}},\ \bibinfo {pages} {033316} (\bibinfo {year} {2023})}\BibitemShut
	{NoStop}%
	\bibitem [{\citenamefont {Lu}\ \emph {et~al.}(2024{\natexlab{b}})\citenamefont
		{Lu}, \citenamefont {Chen}, \citenamefont {Wu}, \citenamefont {Sun},\ and\
		\citenamefont {Meng}}]{HYL2023_thermoFQAH}%
	\BibitemOpen
	\bibfield  {author} {\bibinfo {author} {\bibfnamefont {H.}~\bibnamefont
			{Lu}}, \bibinfo {author} {\bibfnamefont {B.-B.}\ \bibnamefont {Chen}},
		\bibinfo {author} {\bibfnamefont {H.-Q.}\ \bibnamefont {Wu}}, \bibinfo
		{author} {\bibfnamefont {K.}~\bibnamefont {Sun}}, \ and\ \bibinfo {author}
		{\bibfnamefont {Z.~Y.}\ \bibnamefont {Meng}},\ }\href {\doibase
		10.1103/PhysRevLett.132.236502} {\bibfield  {journal} {\bibinfo  {journal}
			{Phys. Rev. Lett.}\ }\textbf {\bibinfo {volume} {132}},\ \bibinfo {pages}
		{236502} (\bibinfo {year} {2024}{\natexlab{b}})}\BibitemShut {NoStop}%
	\bibitem [{\citenamefont {Lin}\ \emph {et~al.}(2022)\citenamefont {Lin},
		\citenamefont {Chen}, \citenamefont {Li}, \citenamefont {Meng},\ and\
		\citenamefont {Shi}}]{linExciton2022}%
	\BibitemOpen
	\bibfield  {author} {\bibinfo {author} {\bibfnamefont {X.}~\bibnamefont
			{Lin}}, \bibinfo {author} {\bibfnamefont {B.-B.}\ \bibnamefont {Chen}},
		\bibinfo {author} {\bibfnamefont {W.}~\bibnamefont {Li}}, \bibinfo {author}
		{\bibfnamefont {Z.~Y.}\ \bibnamefont {Meng}}, \ and\ \bibinfo {author}
		{\bibfnamefont {T.}~\bibnamefont {Shi}},\ }\href {\doibase
		10.1103/PhysRevLett.128.157201} {\bibfield  {journal} {\bibinfo  {journal}
			{Phys. Rev. Lett.}\ }\textbf {\bibinfo {volume} {128}},\ \bibinfo {pages}
		{157201} (\bibinfo {year} {2022})}\BibitemShut {NoStop}%
	\bibitem [{\citenamefont {Pan}\ \emph {et~al.}(2023)\citenamefont {Pan},
		\citenamefont {Zhang}, \citenamefont {Lu}, \citenamefont {Li}, \citenamefont
		{Chen}, \citenamefont {Sun},\ and\ \citenamefont
		{Meng}}]{panThermodynamic2023}%
	\BibitemOpen
	\bibfield  {author} {\bibinfo {author} {\bibfnamefont {G.}~\bibnamefont
			{Pan}}, \bibinfo {author} {\bibfnamefont {X.}~\bibnamefont {Zhang}}, \bibinfo
		{author} {\bibfnamefont {H.}~\bibnamefont {Lu}}, \bibinfo {author}
		{\bibfnamefont {H.}~\bibnamefont {Li}}, \bibinfo {author} {\bibfnamefont
			{B.-B.}\ \bibnamefont {Chen}}, \bibinfo {author} {\bibfnamefont
			{K.}~\bibnamefont {Sun}}, \ and\ \bibinfo {author} {\bibfnamefont {Z.~Y.}\
			\bibnamefont {Meng}},\ }\href {\doibase 10.1103/PhysRevLett.130.016401}
	{\bibfield  {journal} {\bibinfo  {journal} {Phys. Rev. Lett.}\ }\textbf
		{\bibinfo {volume} {130}},\ \bibinfo {pages} {016401} (\bibinfo {year}
		{2023})}\BibitemShut {NoStop}%
	\bibitem [{\citenamefont {White}(1992)}]{White1992_dmrg}%
	\BibitemOpen
	\bibfield  {author} {\bibinfo {author} {\bibfnamefont {S.~R.}\ \bibnamefont
			{White}},\ }\href {\doibase 10.1103/PhysRevLett.69.2863} {\bibfield
		{journal} {\bibinfo  {journal} {Phys. Rev. Lett.}\ }\textbf {\bibinfo
			{volume} {69}},\ \bibinfo {pages} {2863} (\bibinfo {year}
		{1992})}\BibitemShut {NoStop}%
	\bibitem [{\citenamefont {Schollw\"ock}(2005)}]{dmrg_rmp2005}%
	\BibitemOpen
	\bibfield  {author} {\bibinfo {author} {\bibfnamefont {U.}~\bibnamefont
			{Schollw\"ock}},\ }\href {\doibase 10.1103/RevModPhys.77.259} {\bibfield
		{journal} {\bibinfo  {journal} {Rev. Mod. Phys.}\ }\textbf {\bibinfo {volume}
			{77}},\ \bibinfo {pages} {259} (\bibinfo {year} {2005})}\BibitemShut
	{NoStop}%
	\bibitem [{\citenamefont {Sandvik}(2010)}]{SandvikEDNote}%
	\BibitemOpen
	\bibfield  {author} {\bibinfo {author} {\bibfnamefont {A.~W.}\ \bibnamefont
			{Sandvik}},\ }\href {\doibase 10.1063/1.3518900} {\bibfield  {journal}
		{\bibinfo  {journal} {AIP Conference Proceedings}\ }\textbf {\bibinfo
			{volume} {1297}},\ \bibinfo {pages} {135} (\bibinfo {year} {2010})},\ \Eprint
	{http://arxiv.org/abs/https://pubs.aip.org/aip/acp/article-pdf/1297/1/135/11407753/135\_1\_online.pdf}
	{https://pubs.aip.org/aip/acp/article-pdf/1297/1/135/11407753/135\_1\_online.pdf}
	\BibitemShut {NoStop}%
	\bibitem [{\citenamefont {L{\"a}uchli}(2011)}]{LauchliBookChapter}%
	\BibitemOpen
	\bibfield  {author} {\bibinfo {author} {\bibfnamefont {A.~M.}\ \bibnamefont
			{L{\"a}uchli}},\ }\enquote {\bibinfo {title} {Numerical simulations of
			frustrated systems},}\ in\ \href {\doibase 10.1007/978-3-642-10589-0_18}
	{\emph {\bibinfo {booktitle} {Introduction to Frustrated Magnetism:
				Materials, Experiments, Theory}}},\ \bibinfo {editor} {edited by\ \bibinfo
		{editor} {\bibfnamefont {C.}~\bibnamefont {Lacroix}}, \bibinfo {editor}
		{\bibfnamefont {P.}~\bibnamefont {Mendels}}, \ and\ \bibinfo {editor}
		{\bibfnamefont {F.}~\bibnamefont {Mila}}}\ (\bibinfo  {publisher} {Springer
		Berlin Heidelberg},\ \bibinfo {address} {Berlin, Heidelberg},\ \bibinfo
	{year} {2011})\ pp.\ \bibinfo {pages} {481--511}\BibitemShut {NoStop}%
	\bibitem [{\citenamefont {Chen}\ \emph {et~al.}(2018)\citenamefont {Chen},
		\citenamefont {Chen}, \citenamefont {Chen}, \citenamefont {Li},\ and\
		\citenamefont {Weichselbaum}}]{BBChen2018_XTRG}%
	\BibitemOpen
	\bibfield  {author} {\bibinfo {author} {\bibfnamefont {B.-B.}\ \bibnamefont
			{Chen}}, \bibinfo {author} {\bibfnamefont {L.}~\bibnamefont {Chen}}, \bibinfo
		{author} {\bibfnamefont {Z.}~\bibnamefont {Chen}}, \bibinfo {author}
		{\bibfnamefont {W.}~\bibnamefont {Li}}, \ and\ \bibinfo {author}
		{\bibfnamefont {A.}~\bibnamefont {Weichselbaum}},\ }\href {\doibase
		10.1103/PhysRevX.8.031082} {\bibfield  {journal} {\bibinfo  {journal} {Phys.
				Rev. X}\ }\textbf {\bibinfo {volume} {8}},\ \bibinfo {pages} {031082}
		(\bibinfo {year} {2018})}\BibitemShut {NoStop}%
	\bibitem [{\citenamefont {Chen}\ \emph {et~al.}(2022)\citenamefont {Chen},
		\citenamefont {Martinez}, \citenamefont {Nacke}, \citenamefont {Korblova},
		\citenamefont {Manabe}, \citenamefont {Klasen-Memmer}, \citenamefont
		{Freychet}, \citenamefont {Zhernenkov}, \citenamefont {Glaser}, \citenamefont
		{Radzihovsky}, \citenamefont {Maclennan}, \citenamefont {Walba},
		\citenamefont {Bremer}, \citenamefont {Giesselmann},\ and\ \citenamefont
		{Clark}}]{Chen2022smeticaf}%
	\BibitemOpen
	\bibfield  {author} {\bibinfo {author} {\bibfnamefont {X.}~\bibnamefont
			{Chen}}, \bibinfo {author} {\bibfnamefont {V.}~\bibnamefont {Martinez}},
		\bibinfo {author} {\bibfnamefont {P.}~\bibnamefont {Nacke}}, \bibinfo
		{author} {\bibfnamefont {E.}~\bibnamefont {Korblova}}, \bibinfo {author}
		{\bibfnamefont {A.}~\bibnamefont {Manabe}}, \bibinfo {author} {\bibfnamefont
			{M.}~\bibnamefont {Klasen-Memmer}}, \bibinfo {author} {\bibfnamefont
			{G.}~\bibnamefont {Freychet}}, \bibinfo {author} {\bibfnamefont
			{M.}~\bibnamefont {Zhernenkov}}, \bibinfo {author} {\bibfnamefont {M.~A.}\
			\bibnamefont {Glaser}}, \bibinfo {author} {\bibfnamefont {L.}~\bibnamefont
			{Radzihovsky}}, \bibinfo {author} {\bibfnamefont {J.~E.}\ \bibnamefont
			{Maclennan}}, \bibinfo {author} {\bibfnamefont {D.~M.}\ \bibnamefont
			{Walba}}, \bibinfo {author} {\bibfnamefont {M.}~\bibnamefont {Bremer}},
		\bibinfo {author} {\bibfnamefont {F.}~\bibnamefont {Giesselmann}}, \ and\
		\bibinfo {author} {\bibfnamefont {N.~A.}\ \bibnamefont {Clark}},\ }\href
	{\doibase 10.1073/pnas.2210062119} {\bibfield  {journal} {\bibinfo  {journal}
			{Proceedings of the National Academy of Sciences}\ }\textbf {\bibinfo
			{volume} {119}},\ \bibinfo {pages} {e2210062119} (\bibinfo {year}
		{2022})}\BibitemShut {NoStop}%
	\bibitem [{\citenamefont {Nie}\ \emph {et~al.}(2014)\citenamefont {Nie},
		\citenamefont {Tarjus},\ and\ \citenamefont {Kivelson}}]{quenchedNie2014}%
	\BibitemOpen
	\bibfield  {author} {\bibinfo {author} {\bibfnamefont {L.}~\bibnamefont
			{Nie}}, \bibinfo {author} {\bibfnamefont {G.}~\bibnamefont {Tarjus}}, \ and\
		\bibinfo {author} {\bibfnamefont {S.~A.}\ \bibnamefont {Kivelson}},\ }\href
	{\doibase 10.1073/pnas.1406019111} {\bibfield  {journal} {\bibinfo  {journal}
			{Proceedings of the National Academy of Sciences}\ }\textbf {\bibinfo
			{volume} {111}},\ \bibinfo {pages} {7980} (\bibinfo {year}
		{2014})}\BibitemShut {NoStop}%
	\bibitem [{\citenamefont {Fernandes}\ \emph {et~al.}(2019)\citenamefont
		{Fernandes}, \citenamefont {Orth},\ and\ \citenamefont
		{Schmalian}}]{interwinedFernandes2019}%
	\BibitemOpen
	\bibfield  {author} {\bibinfo {author} {\bibfnamefont {R.~M.}\ \bibnamefont
			{Fernandes}}, \bibinfo {author} {\bibfnamefont {P.~P.}\ \bibnamefont {Orth}},
		\ and\ \bibinfo {author} {\bibfnamefont {J.}~\bibnamefont {Schmalian}},\
	}\href {\doibase 10.1146/annurev-conmatphys-031218-013200} {\bibfield
		{journal} {\bibinfo  {journal} {Annual Review of Condensed Matter Physics}\
		}\textbf {\bibinfo {volume} {10}},\ \bibinfo {pages} {133} (\bibinfo {year}
		{2019})}\BibitemShut {NoStop}%
	\bibitem [{\citenamefont {Wang}\ \emph {et~al.}(2021)\citenamefont {Wang},
		\citenamefont {Yan}, \citenamefont {Wang}, \citenamefont {Qi},\ and\
		\citenamefont {Meng}}]{wangVestigial2021}%
	\BibitemOpen
	\bibfield  {author} {\bibinfo {author} {\bibfnamefont {Y.-C.}\ \bibnamefont
			{Wang}}, \bibinfo {author} {\bibfnamefont {Z.}~\bibnamefont {Yan}}, \bibinfo
		{author} {\bibfnamefont {C.}~\bibnamefont {Wang}}, \bibinfo {author}
		{\bibfnamefont {Y.}~\bibnamefont {Qi}}, \ and\ \bibinfo {author}
		{\bibfnamefont {Z.~Y.}\ \bibnamefont {Meng}},\ }\href {\doibase
		10.1103/PhysRevB.103.014408} {\bibfield  {journal} {\bibinfo  {journal}
			{Phys. Rev. B}\ }\textbf {\bibinfo {volume} {103}},\ \bibinfo {pages}
		{014408} (\bibinfo {year} {2021})}\BibitemShut {NoStop}%
	\bibitem [{\citenamefont {{Sun}}\ \emph {et~al.}(2023)\citenamefont {{Sun}},
		\citenamefont {{Ye}}, \citenamefont {{Huang}}, \citenamefont {{Zhou}},
		\citenamefont {{Huang}}, \citenamefont {{Li}}, \citenamefont {{Ye}},
		\citenamefont {{Nnokwe}}, \citenamefont {{Deng}}, \citenamefont {{Mandrus}},
		\citenamefont {{Meng}}, \citenamefont {{Sun}}, \citenamefont {{Du}},
		\citenamefont {{He}},\ and\ \citenamefont {{Zhao}}}]{sunDimensionality2023}%
	\BibitemOpen
	\bibfield  {author} {\bibinfo {author} {\bibfnamefont {Z.}~\bibnamefont
			{{Sun}}}, \bibinfo {author} {\bibfnamefont {G.}~\bibnamefont {{Ye}}},
		\bibinfo {author} {\bibfnamefont {M.}~\bibnamefont {{Huang}}}, \bibinfo
		{author} {\bibfnamefont {C.}~\bibnamefont {{Zhou}}}, \bibinfo {author}
		{\bibfnamefont {N.}~\bibnamefont {{Huang}}}, \bibinfo {author} {\bibfnamefont
			{Q.}~\bibnamefont {{Li}}}, \bibinfo {author} {\bibfnamefont {Z.}~\bibnamefont
			{{Ye}}}, \bibinfo {author} {\bibfnamefont {C.}~\bibnamefont {{Nnokwe}}},
		\bibinfo {author} {\bibfnamefont {H.}~\bibnamefont {{Deng}}}, \bibinfo
		{author} {\bibfnamefont {D.}~\bibnamefont {{Mandrus}}}, \bibinfo {author}
		{\bibfnamefont {Z.~Y.}\ \bibnamefont {{Meng}}}, \bibinfo {author}
		{\bibfnamefont {K.}~\bibnamefont {{Sun}}}, \bibinfo {author} {\bibfnamefont
			{C.}~\bibnamefont {{Du}}}, \bibinfo {author} {\bibfnamefont {R.}~\bibnamefont
			{{He}}}, \ and\ \bibinfo {author} {\bibfnamefont {L.}~\bibnamefont
			{{Zhao}}},\ }\href {\doibase 10.48550/arXiv.2311.03493} {\bibfield  {journal}
		{\bibinfo  {journal} {arXiv e-prints}\ ,\ \bibinfo {eid} {arXiv:2311.03493}}
		(\bibinfo {year} {2023})},\ \Eprint {http://arxiv.org/abs/2311.03493}
	{arXiv:2311.03493 [cond-mat.mtrl-sci]} \BibitemShut {NoStop}%
	\bibitem [{\citenamefont {Francini}\ and\ \citenamefont
		{Janssen}(2024)}]{franciniSpin2023}%
	\BibitemOpen
	\bibfield  {author} {\bibinfo {author} {\bibfnamefont {N.}~\bibnamefont
			{Francini}}\ and\ \bibinfo {author} {\bibfnamefont {L.}~\bibnamefont
			{Janssen}},\ }\href {\doibase 10.1103/PhysRevB.109.075104} {\bibfield
		{journal} {\bibinfo  {journal} {Phys. Rev. B}\ }\textbf {\bibinfo {volume}
			{109}},\ \bibinfo {pages} {075104} (\bibinfo {year} {2024})}\BibitemShut
	{NoStop}%
	\bibitem [{\citenamefont {Babaev}\ \emph {et~al.}(2004)\citenamefont {Babaev},
		\citenamefont {Sudbø},\ and\ \citenamefont {Ashcroft}}]{Babaev2004metal}%
	\BibitemOpen
	\bibfield  {author} {\bibinfo {author} {\bibfnamefont {E.}~\bibnamefont
			{Babaev}}, \bibinfo {author} {\bibfnamefont {A.}~\bibnamefont {Sudbø}}, \
		and\ \bibinfo {author} {\bibfnamefont {N.~W.}\ \bibnamefont {Ashcroft}},\
	}\href {\doibase 10.1038/nature02910} {\bibfield  {journal} {\bibinfo
			{journal} {Nature}\ }\textbf {\bibinfo {volume} {431}},\ \bibinfo {pages}
		{666} (\bibinfo {year} {2004})}\BibitemShut {NoStop}%
	\bibitem [{\citenamefont {Bojesen}\ \emph {et~al.}(2014)\citenamefont
		{Bojesen}, \citenamefont {Babaev},\ and\ \citenamefont
		{Sudb\o{}}}]{Bojesen2014metal}%
	\BibitemOpen
	\bibfield  {author} {\bibinfo {author} {\bibfnamefont {T.~A.}\ \bibnamefont
			{Bojesen}}, \bibinfo {author} {\bibfnamefont {E.}~\bibnamefont {Babaev}}, \
		and\ \bibinfo {author} {\bibfnamefont {A.}~\bibnamefont {Sudb\o{}}},\ }\href
	{\doibase 10.1103/PhysRevB.89.104509} {\bibfield  {journal} {\bibinfo
			{journal} {Phys. Rev. B}\ }\textbf {\bibinfo {volume} {89}},\ \bibinfo
		{pages} {104509} (\bibinfo {year} {2014})}\BibitemShut {NoStop}%
	\bibitem [{\citenamefont {Grinenko}\ \emph {et~al.}(2021)\citenamefont
		{Grinenko}, \citenamefont {Weston}, \citenamefont {Caglieris}, \citenamefont
		{Wuttke}, \citenamefont {Hess}, \citenamefont {Gottschall}, \citenamefont
		{Maccari}, \citenamefont {Gorbunov}, \citenamefont {Zherlitsyn},
		\citenamefont {Wosnitza}, \citenamefont {Rydh}, \citenamefont {Kihou},
		\citenamefont {Lee}, \citenamefont {Sarkar}, \citenamefont {Dengre},
		\citenamefont {Garaud}, \citenamefont {Charnukha}, \citenamefont {Hühne},
		\citenamefont {Nielsch}, \citenamefont {Büchner}, \citenamefont {Klauss},\
		and\ \citenamefont {Babaev}}]{Babaev2021metal}%
	\BibitemOpen
	\bibfield  {author} {\bibinfo {author} {\bibfnamefont {V.}~\bibnamefont
			{Grinenko}}, \bibinfo {author} {\bibfnamefont {D.}~\bibnamefont {Weston}},
		\bibinfo {author} {\bibfnamefont {F.}~\bibnamefont {Caglieris}}, \bibinfo
		{author} {\bibfnamefont {C.}~\bibnamefont {Wuttke}}, \bibinfo {author}
		{\bibfnamefont {C.}~\bibnamefont {Hess}}, \bibinfo {author} {\bibfnamefont
			{T.}~\bibnamefont {Gottschall}}, \bibinfo {author} {\bibfnamefont
			{I.}~\bibnamefont {Maccari}}, \bibinfo {author} {\bibfnamefont
			{D.}~\bibnamefont {Gorbunov}}, \bibinfo {author} {\bibfnamefont
			{S.}~\bibnamefont {Zherlitsyn}}, \bibinfo {author} {\bibfnamefont
			{J.}~\bibnamefont {Wosnitza}}, \bibinfo {author} {\bibfnamefont
			{A.}~\bibnamefont {Rydh}}, \bibinfo {author} {\bibfnamefont {K.}~\bibnamefont
			{Kihou}}, \bibinfo {author} {\bibfnamefont {C.-H.}\ \bibnamefont {Lee}},
		\bibinfo {author} {\bibfnamefont {R.}~\bibnamefont {Sarkar}}, \bibinfo
		{author} {\bibfnamefont {S.}~\bibnamefont {Dengre}}, \bibinfo {author}
		{\bibfnamefont {J.}~\bibnamefont {Garaud}}, \bibinfo {author} {\bibfnamefont
			{A.}~\bibnamefont {Charnukha}}, \bibinfo {author} {\bibfnamefont
			{R.}~\bibnamefont {Hühne}}, \bibinfo {author} {\bibfnamefont
			{K.}~\bibnamefont {Nielsch}}, \bibinfo {author} {\bibfnamefont
			{B.}~\bibnamefont {Büchner}}, \bibinfo {author} {\bibfnamefont {H.-H.}\
			\bibnamefont {Klauss}}, \ and\ \bibinfo {author} {\bibfnamefont
			{E.}~\bibnamefont {Babaev}},\ }\href {\doibase 10.1038/s41567-021-01350-9}
	{\bibfield  {journal} {\bibinfo  {journal} {Nature Physics}\ }\textbf
		{\bibinfo {volume} {17}},\ \bibinfo {pages} {1254} (\bibinfo {year}
		{2021})}\BibitemShut {NoStop}%
	\bibitem [{\citenamefont {Kivelson}\ \emph {et~al.}(1998)\citenamefont
		{Kivelson}, \citenamefont {Fradkin},\ and\ \citenamefont
		{Emery}}]{kivelson1998electronic}%
	\BibitemOpen
	\bibfield  {author} {\bibinfo {author} {\bibfnamefont {S.~A.}\ \bibnamefont
			{Kivelson}}, \bibinfo {author} {\bibfnamefont {E.}~\bibnamefont {Fradkin}}, \
		and\ \bibinfo {author} {\bibfnamefont {V.~J.}\ \bibnamefont {Emery}},\ }\href
	{\doibase 10.1038/31177} {\bibfield  {journal} {\bibinfo  {journal} {Nature}\
		}\textbf {\bibinfo {volume} {393}},\ \bibinfo {pages} {550} (\bibinfo {year}
		{1998})}\BibitemShut {NoStop}%
	\bibitem [{\citenamefont {Niori}\ \emph {et~al.}(1996)\citenamefont {Niori},
		\citenamefont {Sekine}, \citenamefont {Watanabe}, \citenamefont {Furukawa},\
		and\ \citenamefont {Takezoe}}]{Niori1996ferro}%
	\BibitemOpen
	\bibfield  {author} {\bibinfo {author} {\bibfnamefont {T.}~\bibnamefont
			{Niori}}, \bibinfo {author} {\bibfnamefont {T.}~\bibnamefont {Sekine}},
		\bibinfo {author} {\bibfnamefont {J.}~\bibnamefont {Watanabe}}, \bibinfo
		{author} {\bibfnamefont {T.}~\bibnamefont {Furukawa}}, \ and\ \bibinfo
		{author} {\bibfnamefont {H.}~\bibnamefont {Takezoe}},\ }\href {\doibase
		10.1039/JM9960601231} {\bibfield  {journal} {\bibinfo  {journal} {J. Mater.
				Chem.}\ }\textbf {\bibinfo {volume} {6}},\ \bibinfo {pages} {1231} (\bibinfo
		{year} {1996})}\BibitemShut {NoStop}%
	\bibitem [{\citenamefont {Emery}\ \emph {et~al.}(2000)\citenamefont {Emery},
		\citenamefont {Fradkin}, \citenamefont {Kivelson},\ and\ \citenamefont
		{Lubensky}}]{Emery2000}%
	\BibitemOpen
	\bibfield  {author} {\bibinfo {author} {\bibfnamefont {V.~J.}\ \bibnamefont
			{Emery}}, \bibinfo {author} {\bibfnamefont {E.}~\bibnamefont {Fradkin}},
		\bibinfo {author} {\bibfnamefont {S.~A.}\ \bibnamefont {Kivelson}}, \ and\
		\bibinfo {author} {\bibfnamefont {T.~C.}\ \bibnamefont {Lubensky}},\ }\href
	{\doibase 10.1103/PhysRevLett.85.2160} {\bibfield  {journal} {\bibinfo
			{journal} {Phys. Rev. Lett.}\ }\textbf {\bibinfo {volume} {85}},\ \bibinfo
		{pages} {2160} (\bibinfo {year} {2000})}\BibitemShut {NoStop}%
	\bibitem [{sup()}]{suppl}%
	\BibitemOpen
	\href@noop {} {\bibinfo  {journal} {In Section A, we show supplementary ED
			results. In Section B, we provide more detailed finite-temperature structure
			factor of the FQAHS state. Moreover, we show supplementary thermodynamic data
			of FQAHS and PSM phases in Section C and Section D, respectively. The
			degenerate stripe patterns are shown in Section E. In addition, the order
			parameters, symmetry analysis and the Ginzburg-Landau theory of smectic
			phases are shown in Section F.}\ }\BibitemShut {NoStop}%
	\bibitem [{\citenamefont {Wen}(1990)}]{Wen1990NFL}%
	\BibitemOpen
	\bibfield  {journal} {  }\bibfield  {author} {\bibinfo {author} {\bibfnamefont
			{X.~G.}\ \bibnamefont {Wen}},\ }\href {\doibase 10.1103/PhysRevB.42.6623}
	{\bibfield  {journal} {\bibinfo  {journal} {Phys. Rev. B}\ }\textbf {\bibinfo
			{volume} {42}},\ \bibinfo {pages} {6623} (\bibinfo {year}
		{1990})}\BibitemShut {NoStop}%
	\bibitem [{\citenamefont {Sondhi}\ and\ \citenamefont
		{Yang}(2001)}]{Sondhi2001NFL}%
	\BibitemOpen
	\bibfield  {author} {\bibinfo {author} {\bibfnamefont {S.~L.}\ \bibnamefont
			{Sondhi}}\ and\ \bibinfo {author} {\bibfnamefont {K.}~\bibnamefont {Yang}},\
	}\href {\doibase 10.1103/PhysRevB.63.054430} {\bibfield  {journal} {\bibinfo
			{journal} {Phys. Rev. B}\ }\textbf {\bibinfo {volume} {63}},\ \bibinfo
		{pages} {054430} (\bibinfo {year} {2001})}\BibitemShut {NoStop}%
	\bibitem [{\citenamefont {Vishwanath}\ and\ \citenamefont
		{Carpentier}(2001)}]{Vishwanath2001NFL}%
	\BibitemOpen
	\bibfield  {author} {\bibinfo {author} {\bibfnamefont {A.}~\bibnamefont
			{Vishwanath}}\ and\ \bibinfo {author} {\bibfnamefont {D.}~\bibnamefont
			{Carpentier}},\ }\href {\doibase 10.1103/PhysRevLett.86.676} {\bibfield
		{journal} {\bibinfo  {journal} {Phys. Rev. Lett.}\ }\textbf {\bibinfo
			{volume} {86}},\ \bibinfo {pages} {676} (\bibinfo {year} {2001})}\BibitemShut
	{NoStop}%
	\bibitem [{\citenamefont {Mukhopadhyay}\ \emph {et~al.}(2001)\citenamefont
		{Mukhopadhyay}, \citenamefont {Kane},\ and\ \citenamefont
		{Lubensky}}]{Mukhopadhyay2001NFL}%
	\BibitemOpen
	\bibfield  {author} {\bibinfo {author} {\bibfnamefont {R.}~\bibnamefont
			{Mukhopadhyay}}, \bibinfo {author} {\bibfnamefont {C.~L.}\ \bibnamefont
			{Kane}}, \ and\ \bibinfo {author} {\bibfnamefont {T.~C.}\ \bibnamefont
			{Lubensky}},\ }\href {\doibase 10.1103/PhysRevB.64.045120} {\bibfield
		{journal} {\bibinfo  {journal} {Phys. Rev. B}\ }\textbf {\bibinfo {volume}
			{64}},\ \bibinfo {pages} {045120} (\bibinfo {year} {2001})}\BibitemShut
	{NoStop}%
	\bibitem [{\citenamefont {Wang}\ \emph {et~al.}(2022)\citenamefont {Wang},
		\citenamefont {Yu}, \citenamefont {Kwan}, \citenamefont {Jia}, \citenamefont
		{Lei}, \citenamefont {Klemenz}, \citenamefont {Cevallos}, \citenamefont
		{Singha}, \citenamefont {Devakul}, \citenamefont {Watanabe}, \citenamefont
		{Taniguchi}, \citenamefont {Sondhi}, \citenamefont {Cava}, \citenamefont
		{Schoop}, \citenamefont {Parameswaran},\ and\ \citenamefont
		{Wu}}]{Wang2022LL}%
	\BibitemOpen
	\bibfield  {author} {\bibinfo {author} {\bibfnamefont {P.}~\bibnamefont
			{Wang}}, \bibinfo {author} {\bibfnamefont {G.}~\bibnamefont {Yu}}, \bibinfo
		{author} {\bibfnamefont {Y.~H.}\ \bibnamefont {Kwan}}, \bibinfo {author}
		{\bibfnamefont {Y.}~\bibnamefont {Jia}}, \bibinfo {author} {\bibfnamefont
			{S.}~\bibnamefont {Lei}}, \bibinfo {author} {\bibfnamefont {S.}~\bibnamefont
			{Klemenz}}, \bibinfo {author} {\bibfnamefont {F.~A.}\ \bibnamefont
			{Cevallos}}, \bibinfo {author} {\bibfnamefont {R.}~\bibnamefont {Singha}},
		\bibinfo {author} {\bibfnamefont {T.}~\bibnamefont {Devakul}}, \bibinfo
		{author} {\bibfnamefont {K.}~\bibnamefont {Watanabe}}, \bibinfo {author}
		{\bibfnamefont {T.}~\bibnamefont {Taniguchi}}, \bibinfo {author}
		{\bibfnamefont {S.~L.}\ \bibnamefont {Sondhi}}, \bibinfo {author}
		{\bibfnamefont {R.~J.}\ \bibnamefont {Cava}}, \bibinfo {author}
		{\bibfnamefont {L.~M.}\ \bibnamefont {Schoop}}, \bibinfo {author}
		{\bibfnamefont {S.~A.}\ \bibnamefont {Parameswaran}}, \ and\ \bibinfo
		{author} {\bibfnamefont {S.}~\bibnamefont {Wu}},\ }\href {\doibase
		10.1038/s41586-022-04514-6} {\bibfield  {journal} {\bibinfo  {journal}
			{Nature}\ }\textbf {\bibinfo {volume} {605}},\ \bibinfo {pages} {52}
		(\bibinfo {year} {2022})}\BibitemShut {NoStop}%
	\bibitem [{\citenamefont {Yu}\ \emph {et~al.}(2023)\citenamefont {Yu},
		\citenamefont {Wang}, \citenamefont {Uzan-Narovlansky}, \citenamefont {Jia},
		\citenamefont {Onyszczak}, \citenamefont {Singha}, \citenamefont {Gui},
		\citenamefont {Song}, \citenamefont {Tang}, \citenamefont {Watanabe},
		\citenamefont {Taniguchi}, \citenamefont {Cava}, \citenamefont {Schoop},\
		and\ \citenamefont {Wu}}]{Yu2023LL}%
	\BibitemOpen
	\bibfield  {author} {\bibinfo {author} {\bibfnamefont {G.}~\bibnamefont
			{Yu}}, \bibinfo {author} {\bibfnamefont {P.}~\bibnamefont {Wang}}, \bibinfo
		{author} {\bibfnamefont {A.~J.}\ \bibnamefont {Uzan-Narovlansky}}, \bibinfo
		{author} {\bibfnamefont {Y.}~\bibnamefont {Jia}}, \bibinfo {author}
		{\bibfnamefont {M.}~\bibnamefont {Onyszczak}}, \bibinfo {author}
		{\bibfnamefont {R.}~\bibnamefont {Singha}}, \bibinfo {author} {\bibfnamefont
			{X.}~\bibnamefont {Gui}}, \bibinfo {author} {\bibfnamefont {T.}~\bibnamefont
			{Song}}, \bibinfo {author} {\bibfnamefont {Y.}~\bibnamefont {Tang}}, \bibinfo
		{author} {\bibfnamefont {K.}~\bibnamefont {Watanabe}}, \bibinfo {author}
		{\bibfnamefont {T.}~\bibnamefont {Taniguchi}}, \bibinfo {author}
		{\bibfnamefont {R.~J.}\ \bibnamefont {Cava}}, \bibinfo {author}
		{\bibfnamefont {L.~M.}\ \bibnamefont {Schoop}}, \ and\ \bibinfo {author}
		{\bibfnamefont {S.}~\bibnamefont {Wu}},\ }\href {\doibase
		10.1038/s41467-023-42821-2} {\bibfield  {journal} {\bibinfo  {journal}
			{Nature Communications}\ }\textbf {\bibinfo {volume} {14}},\ \bibinfo {pages}
		{7025} (\bibinfo {year} {2023})}\BibitemShut {NoStop}%
	\bibitem [{\citenamefont {Wu}\ \emph {et~al.}(2012)\citenamefont {Wu},
		\citenamefont {Jain},\ and\ \citenamefont {Sun}}]{Wu2012Adiabatic}%
	\BibitemOpen
	\bibfield  {author} {\bibinfo {author} {\bibfnamefont {Y.-H.}\ \bibnamefont
			{Wu}}, \bibinfo {author} {\bibfnamefont {J.~K.}\ \bibnamefont {Jain}}, \ and\
		\bibinfo {author} {\bibfnamefont {K.}~\bibnamefont {Sun}},\ }\href {\doibase
		10.1103/PhysRevB.86.165129} {\bibfield  {journal} {\bibinfo  {journal} {Phys.
				Rev. B}\ }\textbf {\bibinfo {volume} {86}},\ \bibinfo {pages} {165129}
		(\bibinfo {year} {2012})}\BibitemShut {NoStop}%
	\bibitem [{\citenamefont {Léonard}\ \emph {et~al.}(2023)\citenamefont
		{Léonard}, \citenamefont {Kim}, \citenamefont {Kwan}, \citenamefont
		{Segura}, \citenamefont {Grusdt}, \citenamefont {Repellin}, \citenamefont
		{Goldman},\ and\ \citenamefont {Greiner}}]{Julian2023photonFQH}%
	\BibitemOpen
	\bibfield  {author} {\bibinfo {author} {\bibfnamefont {J.}~\bibnamefont
			{Léonard}}, \bibinfo {author} {\bibfnamefont {S.}~\bibnamefont {Kim}},
		\bibinfo {author} {\bibfnamefont {J.}~\bibnamefont {Kwan}}, \bibinfo {author}
		{\bibfnamefont {P.}~\bibnamefont {Segura}}, \bibinfo {author} {\bibfnamefont
			{F.}~\bibnamefont {Grusdt}}, \bibinfo {author} {\bibfnamefont
			{C.}~\bibnamefont {Repellin}}, \bibinfo {author} {\bibfnamefont
			{N.}~\bibnamefont {Goldman}}, \ and\ \bibinfo {author} {\bibfnamefont
			{M.}~\bibnamefont {Greiner}},\ }\href {\doibase 10.1038/s41586-023-06122-4}
	{\bibfield  {journal} {\bibinfo  {journal} {Nature}\ }\textbf {\bibinfo
			{volume} {619}},\ \bibinfo {pages} {495} (\bibinfo {year}
		{2023})}\BibitemShut {NoStop}%
	\bibitem [{\citenamefont {Wang}\ \emph {et~al.}(2024)\citenamefont {Wang},
		\citenamefont {Liu}, \citenamefont {Chen}, \citenamefont {Chen},
		\citenamefont {Zhao}, \citenamefont {Ying}, \citenamefont {Shang},
		\citenamefont {Wang}, \citenamefont {Huo}, \citenamefont {Peng},
		\citenamefont {Zhu}, \citenamefont {Lu},\ and\ \citenamefont
		{Pan}}]{Wang2024photonFQH}%
	\BibitemOpen
	\bibfield  {author} {\bibinfo {author} {\bibfnamefont {C.}~\bibnamefont
			{Wang}}, \bibinfo {author} {\bibfnamefont {F.-M.}\ \bibnamefont {Liu}},
		\bibinfo {author} {\bibfnamefont {M.-C.}\ \bibnamefont {Chen}}, \bibinfo
		{author} {\bibfnamefont {H.}~\bibnamefont {Chen}}, \bibinfo {author}
		{\bibfnamefont {X.-H.}\ \bibnamefont {Zhao}}, \bibinfo {author}
		{\bibfnamefont {C.}~\bibnamefont {Ying}}, \bibinfo {author} {\bibfnamefont
			{Z.-X.}\ \bibnamefont {Shang}}, \bibinfo {author} {\bibfnamefont {J.-W.}\
			\bibnamefont {Wang}}, \bibinfo {author} {\bibfnamefont {Y.-H.}\ \bibnamefont
			{Huo}}, \bibinfo {author} {\bibfnamefont {C.-Z.}\ \bibnamefont {Peng}},
		\bibinfo {author} {\bibfnamefont {X.}~\bibnamefont {Zhu}}, \bibinfo {author}
		{\bibfnamefont {C.-Y.}\ \bibnamefont {Lu}}, \ and\ \bibinfo {author}
		{\bibfnamefont {J.-W.}\ \bibnamefont {Pan}},\ }\href {\doibase
		10.1126/science.ado3912} {\bibfield  {journal} {\bibinfo  {journal}
			{Science}\ }\textbf {\bibinfo {volume} {384}},\ \bibinfo {pages} {579}
		(\bibinfo {year} {2024})}\BibitemShut {NoStop}%
	\bibitem [{\citenamefont {Fradkin}(1989)}]{Fradkin1989CS}%
	\BibitemOpen
	\bibfield  {author} {\bibinfo {author} {\bibfnamefont {E.}~\bibnamefont
			{Fradkin}},\ }\href {\doibase 10.1103/PhysRevLett.63.322} {\bibfield
		{journal} {\bibinfo  {journal} {Phys. Rev. Lett.}\ }\textbf {\bibinfo
			{volume} {63}},\ \bibinfo {pages} {322} (\bibinfo {year} {1989})}\BibitemShut
	{NoStop}%
	\bibitem [{\citenamefont {Eliezer}\ and\ \citenamefont
		{Semenoff}(1992)}]{ELIEZER1992CS}%
	\BibitemOpen
	\bibfield  {author} {\bibinfo {author} {\bibfnamefont {D.}~\bibnamefont
			{Eliezer}}\ and\ \bibinfo {author} {\bibfnamefont {G.}~\bibnamefont
			{Semenoff}},\ }\href {\doibase https://doi.org/10.1016/0003-4916(92)90339-N}
	{\bibfield  {journal} {\bibinfo  {journal} {Annals of Physics}\ }\textbf
		{\bibinfo {volume} {217}},\ \bibinfo {pages} {66} (\bibinfo {year}
		{1992})}\BibitemShut {NoStop}%
	\bibitem [{\citenamefont {Kumar}\ \emph {et~al.}(2014)\citenamefont {Kumar},
		\citenamefont {Sun},\ and\ \citenamefont {Fradkin}}]{Kumar2014CS}%
	\BibitemOpen
	\bibfield  {author} {\bibinfo {author} {\bibfnamefont {K.}~\bibnamefont
			{Kumar}}, \bibinfo {author} {\bibfnamefont {K.}~\bibnamefont {Sun}}, \ and\
		\bibinfo {author} {\bibfnamefont {E.}~\bibnamefont {Fradkin}},\ }\href
	{\doibase 10.1103/PhysRevB.90.174409} {\bibfield  {journal} {\bibinfo
			{journal} {Phys. Rev. B}\ }\textbf {\bibinfo {volume} {90}},\ \bibinfo
		{pages} {174409} (\bibinfo {year} {2014})}\BibitemShut {NoStop}%
	\bibitem [{\citenamefont {Sun}\ \emph {et~al.}(2015)\citenamefont {Sun},
		\citenamefont {Kumar},\ and\ \citenamefont {Fradkin}}]{KSun2015CS}%
	\BibitemOpen
	\bibfield  {author} {\bibinfo {author} {\bibfnamefont {K.}~\bibnamefont
			{Sun}}, \bibinfo {author} {\bibfnamefont {K.}~\bibnamefont {Kumar}}, \ and\
		\bibinfo {author} {\bibfnamefont {E.}~\bibnamefont {Fradkin}},\ }\href
	{\doibase 10.1103/PhysRevB.92.115148} {\bibfield  {journal} {\bibinfo
			{journal} {Phys. Rev. B}\ }\textbf {\bibinfo {volume} {92}},\ \bibinfo
		{pages} {115148} (\bibinfo {year} {2015})}\BibitemShut {NoStop}%
	\bibitem [{\citenamefont {Sohal}\ \emph {et~al.}(2018)\citenamefont {Sohal},
		\citenamefont {Santos},\ and\ \citenamefont {Fradkin}}]{Sohal2018CF}%
	\BibitemOpen
	\bibfield  {author} {\bibinfo {author} {\bibfnamefont {R.}~\bibnamefont
			{Sohal}}, \bibinfo {author} {\bibfnamefont {L.~H.}\ \bibnamefont {Santos}}, \
		and\ \bibinfo {author} {\bibfnamefont {E.}~\bibnamefont {Fradkin}},\ }\href
	{\doibase 10.1103/PhysRevB.97.125131} {\bibfield  {journal} {\bibinfo
			{journal} {Phys. Rev. B}\ }\textbf {\bibinfo {volume} {97}},\ \bibinfo
		{pages} {125131} (\bibinfo {year} {2018})}\BibitemShut {NoStop}%
	\bibitem [{\citenamefont {Senthil}(2008)}]{Senthil2008mott}%
	\BibitemOpen
	\bibfield  {author} {\bibinfo {author} {\bibfnamefont {T.}~\bibnamefont
			{Senthil}},\ }\href {\doibase 10.1103/PhysRevB.78.045109} {\bibfield
		{journal} {\bibinfo  {journal} {Phys. Rev. B}\ }\textbf {\bibinfo {volume}
			{78}},\ \bibinfo {pages} {045109} (\bibinfo {year} {2008})}\BibitemShut
	{NoStop}%
	\bibitem [{\citenamefont {Song}\ \emph
		{et~al.}(2024{\natexlab{b}})\citenamefont {Song}, \citenamefont {Zhang},\
		and\ \citenamefont {Senthil}}]{Song2024transition}%
	\BibitemOpen
	\bibfield  {author} {\bibinfo {author} {\bibfnamefont {X.-Y.}\ \bibnamefont
			{Song}}, \bibinfo {author} {\bibfnamefont {Y.-H.}\ \bibnamefont {Zhang}}, \
		and\ \bibinfo {author} {\bibfnamefont {T.}~\bibnamefont {Senthil}},\ }\href
	{\doibase 10.1103/PhysRevB.109.085143} {\bibfield  {journal} {\bibinfo
			{journal} {Phys. Rev. B}\ }\textbf {\bibinfo {volume} {109}},\ \bibinfo
		{pages} {085143} (\bibinfo {year} {2024}{\natexlab{b}})}\BibitemShut
	{NoStop}%
	\bibitem [{\citenamefont {Weichselbaum}(2012)}]{AW2012_QSpace}%
	\BibitemOpen
	\bibfield  {author} {\bibinfo {author} {\bibfnamefont {A.}~\bibnamefont
			{Weichselbaum}},\ }\href {\doibase https://doi.org/10.1016/j.aop.2012.07.009}
	{\bibfield  {journal} {\bibinfo  {journal} {Annals of Physics}\ }\textbf
		{\bibinfo {volume} {327}},\ \bibinfo {pages} {2972} (\bibinfo {year}
		{2012})}\BibitemShut {NoStop}%
	\bibitem [{\citenamefont {Niu}\ \emph {et~al.}(1985)\citenamefont {Niu},
		\citenamefont {Thouless},\ and\ \citenamefont
		{Wu}}]{Niu1985_hallconductance}%
	\BibitemOpen
	\bibfield  {author} {\bibinfo {author} {\bibfnamefont {Q.}~\bibnamefont
			{Niu}}, \bibinfo {author} {\bibfnamefont {D.~J.}\ \bibnamefont {Thouless}}, \
		and\ \bibinfo {author} {\bibfnamefont {Y.-S.}\ \bibnamefont {Wu}},\ }\href
	{\doibase 10.1103/PhysRevB.31.3372} {\bibfield  {journal} {\bibinfo
			{journal} {Phys. Rev. B}\ }\textbf {\bibinfo {volume} {31}},\ \bibinfo
		{pages} {3372} (\bibinfo {year} {1985})}\BibitemShut {NoStop}%
	\bibitem [{\citenamefont {Varney}\ \emph {et~al.}(2011)\citenamefont {Varney},
		\citenamefont {Sun}, \citenamefont {Rigol},\ and\ \citenamefont
		{Galitski}}]{Varney2011_haldanemodel}%
	\BibitemOpen
	\bibfield  {author} {\bibinfo {author} {\bibfnamefont {C.~N.}\ \bibnamefont
			{Varney}}, \bibinfo {author} {\bibfnamefont {K.}~\bibnamefont {Sun}},
		\bibinfo {author} {\bibfnamefont {M.}~\bibnamefont {Rigol}}, \ and\ \bibinfo
		{author} {\bibfnamefont {V.}~\bibnamefont {Galitski}},\ }\href {\doibase
		10.1103/PhysRevB.84.241105} {\bibfield  {journal} {\bibinfo  {journal} {Phys.
				Rev. B}\ }\textbf {\bibinfo {volume} {84}},\ \bibinfo {pages} {241105}
		(\bibinfo {year} {2011})}\BibitemShut {NoStop}%
	\bibitem [{\citenamefont {Fukui}\ \emph {et~al.}(2005)\citenamefont {Fukui},
		\citenamefont {Hatsugai},\ and\ \citenamefont
		{Suzuki}}]{Fukui2005_hallconductance}%
	\BibitemOpen
	\bibfield  {author} {\bibinfo {author} {\bibfnamefont {T.}~\bibnamefont
			{Fukui}}, \bibinfo {author} {\bibfnamefont {Y.}~\bibnamefont {Hatsugai}}, \
		and\ \bibinfo {author} {\bibfnamefont {H.}~\bibnamefont {Suzuki}},\ }\href
	{\doibase 10.1143/JPSJ.74.1674} {\bibfield  {journal} {\bibinfo  {journal}
			{Journal of the Physical Society of Japan}\ }\textbf {\bibinfo {volume}
			{74}},\ \bibinfo {pages} {1674} (\bibinfo {year} {2005})},\ \Eprint
	{http://arxiv.org/abs/https://doi.org/10.1143/JPSJ.74.1674}
	{https://doi.org/10.1143/JPSJ.74.1674} \BibitemShut {NoStop}%
\end{thebibliography}


\renewcommand{\theequation}{S\arabic{equation}} \renewcommand{\thefigure}{S%
\arabic{figure}} \setcounter{equation}{0} \setcounter{figure}{0}

\newpage
\begin{widetext}
\section*{Supplementary Information for \\[0.5em]
	From Fractional Quantum Anomalous Hall Smectics to Polar Smectic Metals: Nontrivial Interplay Between Electronic Liquid Crystal Order and Topological Order in Correlated Topological Flat Bands}
In Section A, we show supplementary ED results. In Section B, we provide more detailed finite-temperature structure factors of the FQAHS state. Moreover, we show supplementary thermodynamic data of FQAHS and PSM phases in Section C and Section D,
respectively.  The degenerate stripe patterns are shown in Section E. In addition, the order parameters, symmetry analysis and
the Ginzburg-Landau theory of smectic phases are shown in Section F.

\subsection*{A. Supplementary ED results} 

\begin{figure}[htp!]
	\centering		
	\includegraphics[width=0.6\columnwidth]{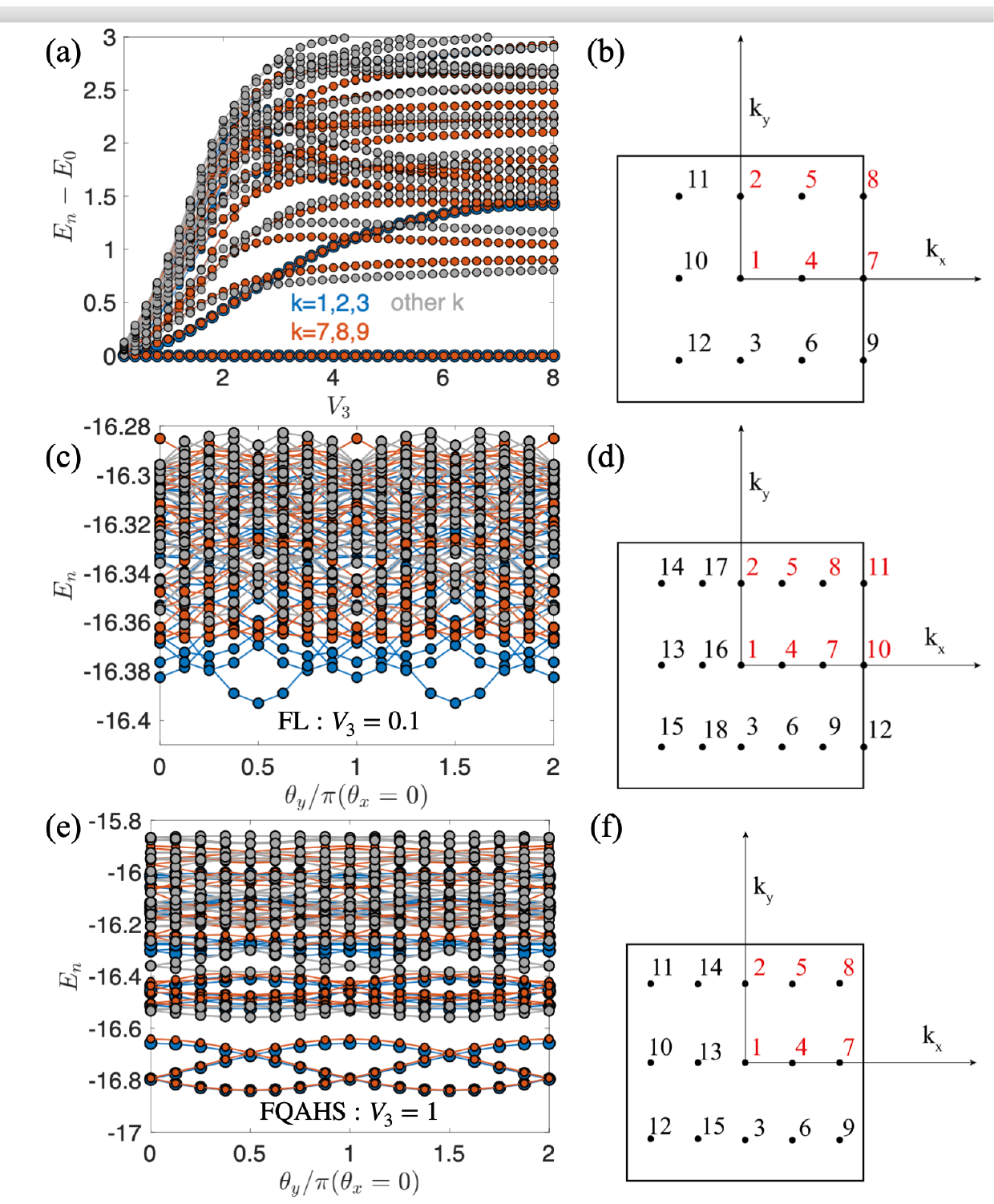}
	\caption{ (a) Energy spectra of a $3\times4\times2$ torus with changing $V_3$. We use four different colors to represent these energy levels, where light blue points represent the energy levels from momentum sectors labeling $k=1,2,3$. The red color points represent the energy levels from $k=7,8,9$, while the energy levels from other $k$ are represented with gray color.  (b), (d) and (f) show the momentum points in the Brillouin zone of the $3\times4\times2$, $3\times6\times2$ and $3\times5\times2$ tori, respectively. The energy spectra of momentum sectors marked by red color are simulated using Lanczos, while the black ones can be obtained by mirror or rotational point-group symmetry.
		Energy spectra of (c) FL, and (e) FQAHS phases are shown with twist boundary conditions using a $3\times4\times2$ torus. The colors are the same as those of (a).}
	\label{figsm_Spectra24}
\end{figure}

Throughout the main text and supplementary information, the ED simulations of Chern numbers are based on the following formula first proposed in Ref.~\cite{Niu1985_hallconductance}. More details on implementation can be found in Ref.~\cite{Varney2011_haldanemodel,Fukui2005_hallconductance}.

\begin{equation}
	\begin{aligned}
		C=\frac{i}{2\pi}\int\int d\phi_1 d\phi_2  [ &\frac{\partial}{\partial \phi_1}\langle \Omega(\phi_1,\phi_2)| \frac{\partial}{\partial \phi_2}| \Omega(\phi_1,\phi_2)\rangle \\
		-&\frac{\partial}{\partial \phi_2}\langle \Omega(\phi_1,\phi_2)| \frac{\partial}{\partial \phi_1}| \Omega(\phi_1,\phi_2)\rangle ] .
	\end{aligned}
	\label{eq_chern}
\end{equation}

In the main text, we have shown the energy spectra on a $3\times 6\times 2$ torus in Fig.\ref{fig_spectra_order}(a), and here we show the spectra on a $3\times 4\times 2$ torus in Fig.\ref{figsm_Spectra24}(a). The momentum points in the Brillouin zone for the two system sizes are shown in Fig.\ref{figsm_Spectra24}(b) and Fig.\ref{figsm_Spectra24}(d), respectively.

While we have shown the gapped spectrum of the FQAHS phase on a $3\times 6\times 2$ torus in Fig.\ref{fig_fqahs_gs}(c), we show here the spectrum flow of the FL and the FQAHS  phase on a $3\times 4\times 2$ torus in Fig.\ref{figsm_Spectra24}(c,e).

\begin{figure}[htp!]
	\centering		
	\includegraphics[width=0.66\columnwidth]{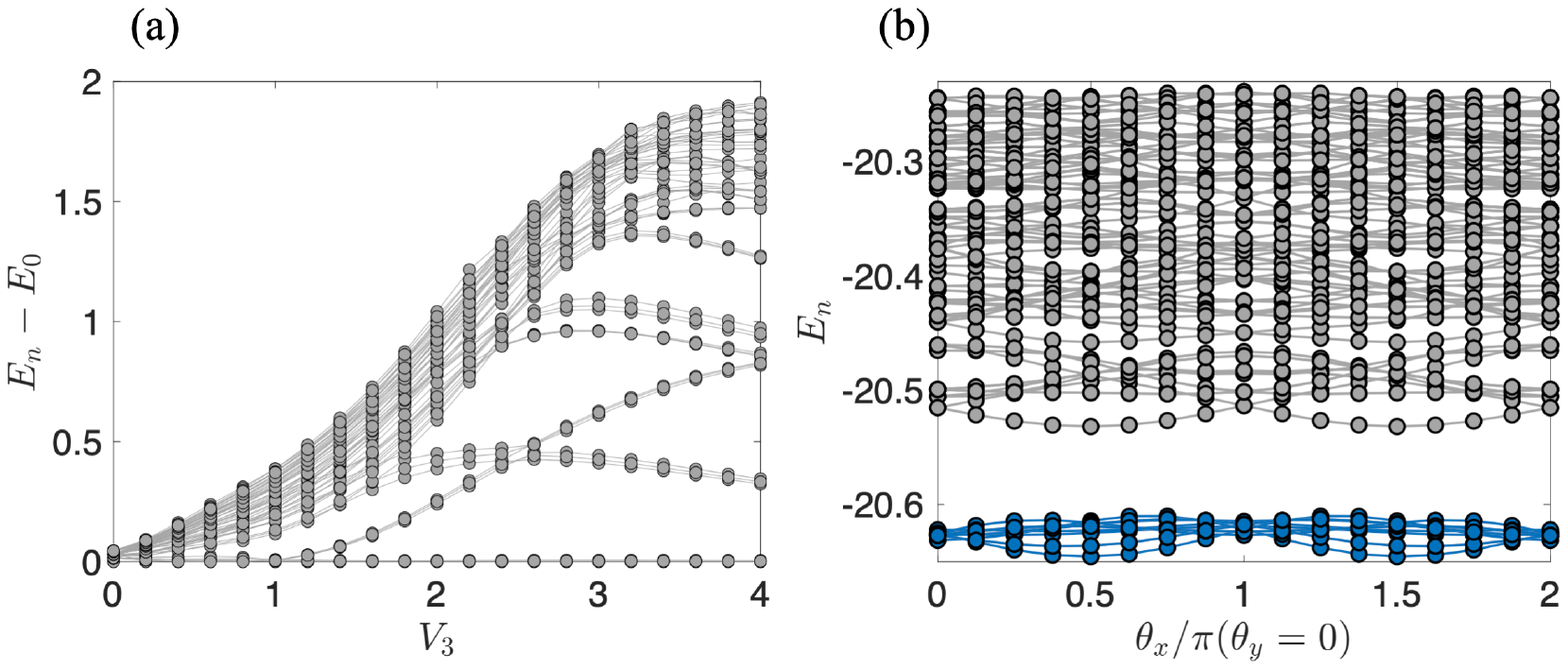}
	\caption{ (a) Energy spectrum of a $3\times5\times2$ torus with changing $V_3$. (b) Spectrum flow with $V_3=1$ and twisted boundary conditions. The blue ones represent the 15-fold degenerate ground states, consisting of the lowest energy level of each momentum sector defined in Fig.\ref{figsm_Spectra24}(f). }
	\label{figsm_ed30}
\end{figure}

Since the Bragg peaks of the density density correlation function in the FQAHS phase are either $(\pm\pi,0)$ or $(0,\pm\pi)$, the $3\times5\times2$ torus is not a suitable geometry for the charge-smectic order along any direction. However, we will show that the incompatible geometry does not change the insulating nature and value of the Hall conductivity of the FQAHS state. Here, we show the spectra of a 30-site torus in Fig.\ref{figsm_ed30}(a), and the spectrum flow at $V_3=1$ in Fig.\ref{figsm_ed30}(b). In the cases of 24 and 36 sites, the ground-state degeneracy is $6=3\times2$. While in the case of 30 sites without $(\pm\pi,0)$ momentum points, the ground-state degeneracy is $15=3\times5\ (N_y)$, and each momentum sector shown in Fig.\ref{figsm_Spectra24}(f) contributes one ground state. When calculating the Hall conductivity of each state using Eq.\ref{eq_chern} with $\phi_1(\phi_2)$ from 0 to $2\pi$, the Hall conductivity of each state is still $2/3$, in agreement with the results of FQAHS phase in the main text.

\begin{figure}[htp!]
	\centering		
	\includegraphics[width=0.36\textwidth]{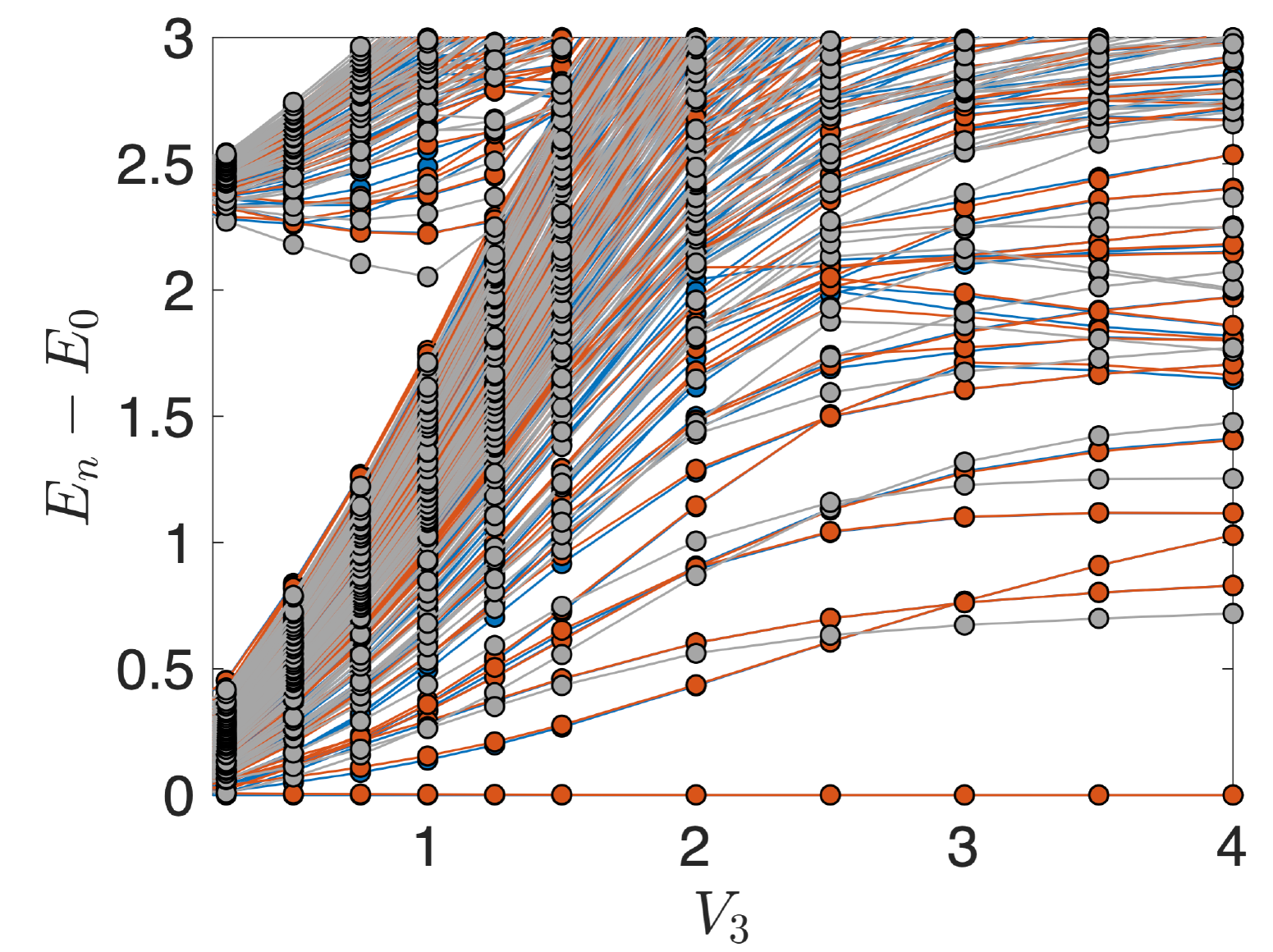}
	\caption{Spectra on a $3\times4\times2$ torus with 60 energy levels considered at each momentum sector and the definitions of colors are the same as those in Fig.\ref{figsm_Spectra24}(a).}
	\label{figsm_Eremote}
\end{figure}

\newpage

While in the previous spectra in Fig.\ref{figsm_Spectra24}(a), we consider only 10 energy levels in each momentum sector, here we show the ED spectra with around 60 energy levels in Fig.\ref{figsm_Eremote}. It is clear that, while the energy levels from the remote band are not playing a role in the FL phase, they quickly merge into the intermediate-energy levels and are playing a role in the FQAHS and PSM phases.

\begin{figure}[htp!]
	\centering		
	\includegraphics[width=0.55\textwidth]{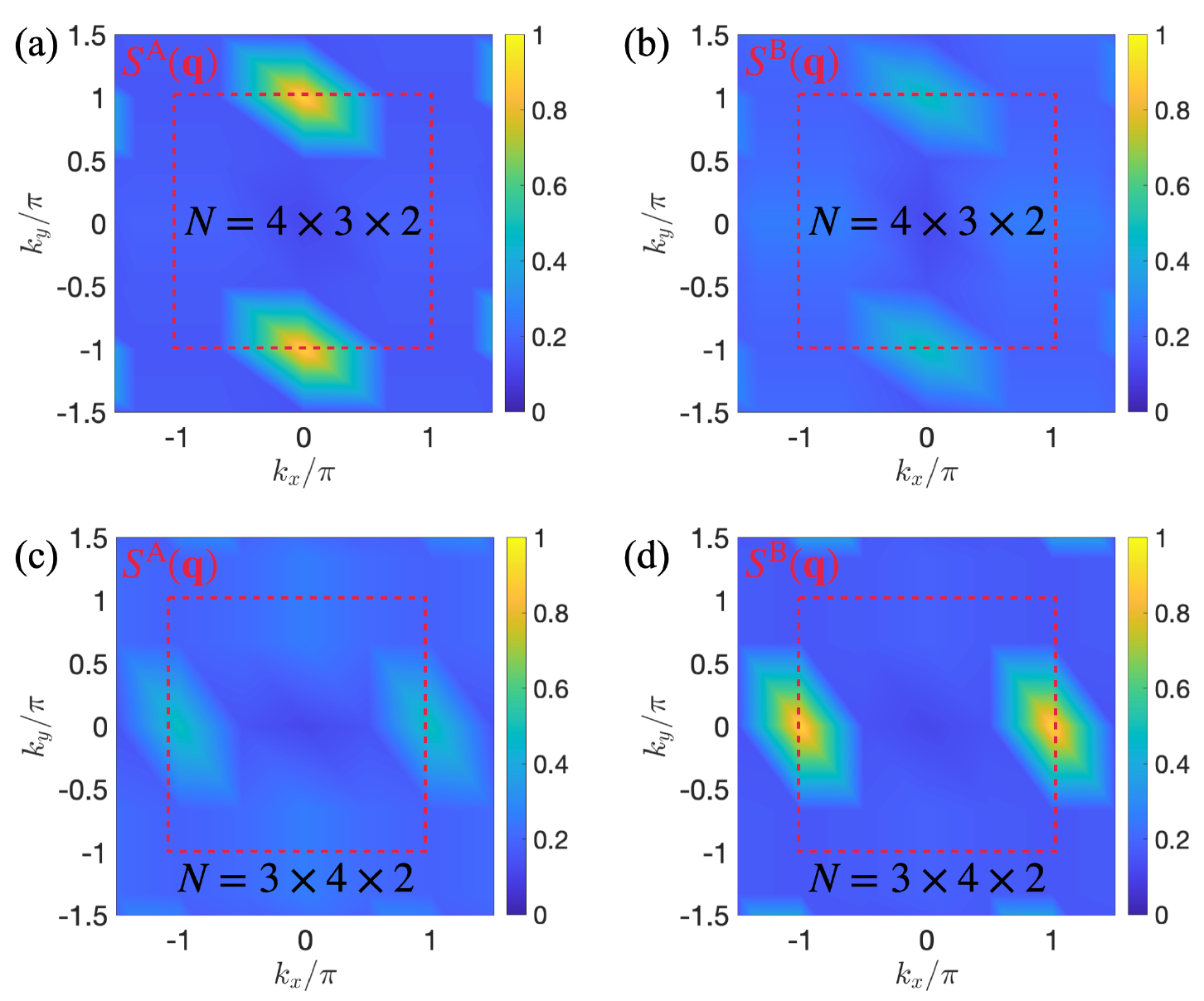}
	\caption{The structure factors of A and B sublattices in the FQAHS state with $V_3=1$ and different geometries, respectively.}
	\label{figsm_difSq}
\end{figure}

In the main text, we have shown that in the intermediate FQAHS phase, there is a large difference in the charge-smectic order between the sublattices. Here, we also plot the ED strucutre factors of $V_3=1$ with different geometries in Fig.\ref{figsm_difSq}. In the $N=3\times 4\times2$ torus, the smectic order is along the $N_x$ direction and that of the B sublattice is much stronger than that of the A sublattice. However, the results in the $N=4\times3\times2$ torus totally reverse. This supports our analysis in the main text.

\begin{figure}[htp!]
	\centering		
	\includegraphics[width=0.8\textwidth]{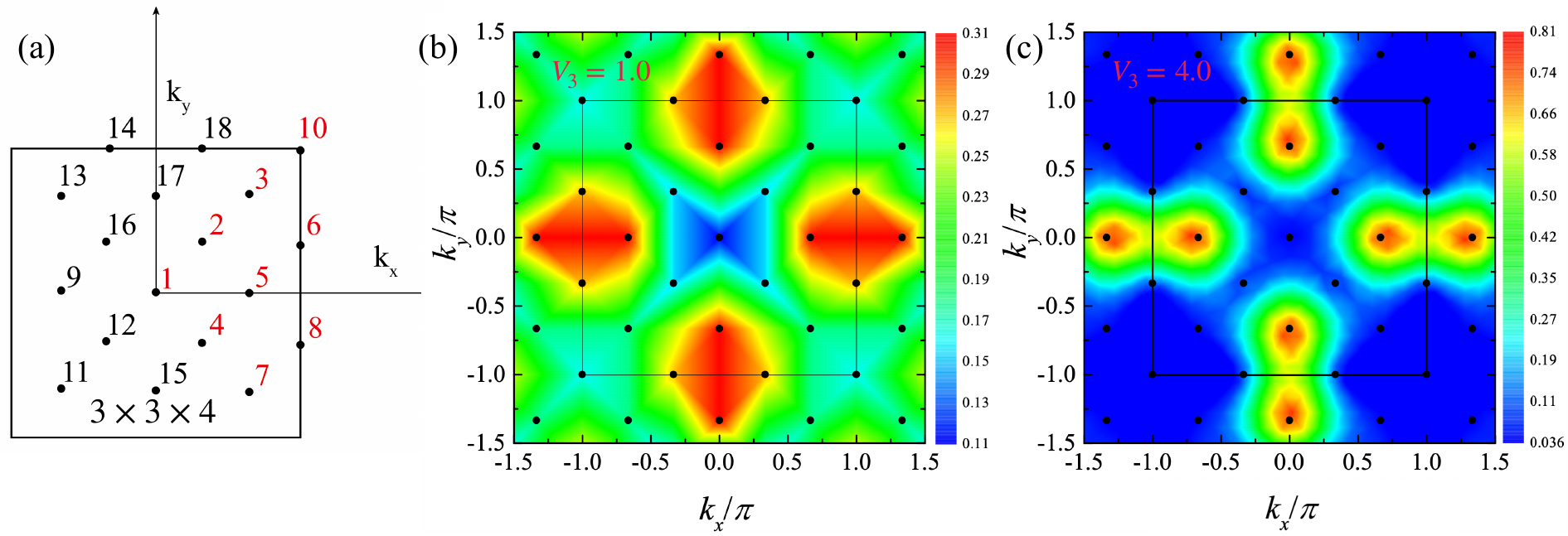}
	\caption{(a) The momentum points in the Brillouin zone of the $3\times3\times4$ lattice. The momentum points in red are simulated while those in black are obtained by symmetry.
	The structure factor $S^\mathrm{B}(\mathbf{q})$ at (b) $V_3=1$ and (c) $V_3=4$, where the black points label the momentum points in (a).}
	\label{figsm_334}
\end{figure}

In the ED study of this FQAHS state, since the compatible geometries have even and odd numbers of unit cells in the two directions, it might be a question for the ED results whether the charge order is a smectic order in the thermodynamic limit.
Since it is currently too hard for us to do ED with even by even unit cells and compatible with $\nu=2/3$ at the same time, we can choose another symmetric $N=3\times3\times4$ lattice different from all those used in the rest of our work, whose Brillouin zone is shown in Fig.\ref{figsm_334}(a). 
The Brillouin zone of this lattice does not contain the ($\pi,0$) or the ($0,\pi$) points, however, it is symmetric and contains the ($\pi,\pi$), which is absent for $3\times4\times2$ or $3\times6\times2$.
The structure factor of B lattice as an example for at $V_3=1$ (FQAHS) and $V_3=4$ (PSM) are shown in Fig.\ref{figsm_334}(b,c) respectively.
It is clear that, although there is no ($\pi,0$) or ($0,\pi$) point, the structure factors tend to suggest that there will be Bragg peaks at ($\pm\pi,0$) or ($0,\pm\pi$) points, and more importantly, the structure factor at ($\pi,\pi$) shows no peak, suggesting that the translational symmetry breaking and the doubling of unit cell will happen in only one direction, since the doubling of unit cell will require the charge order at ($\pi,\pi$).
This is in agreement with and supports our conclusion that this smectic order in the thermodynamic limit is either ($\pi,0$) or ($0,\pi$).

\subsection*{B. Detailed finite-temperature structure factors of FQAHS phase}

\begin{figure}[htp!]
	\centering		
	\includegraphics[width=0.5\textwidth]{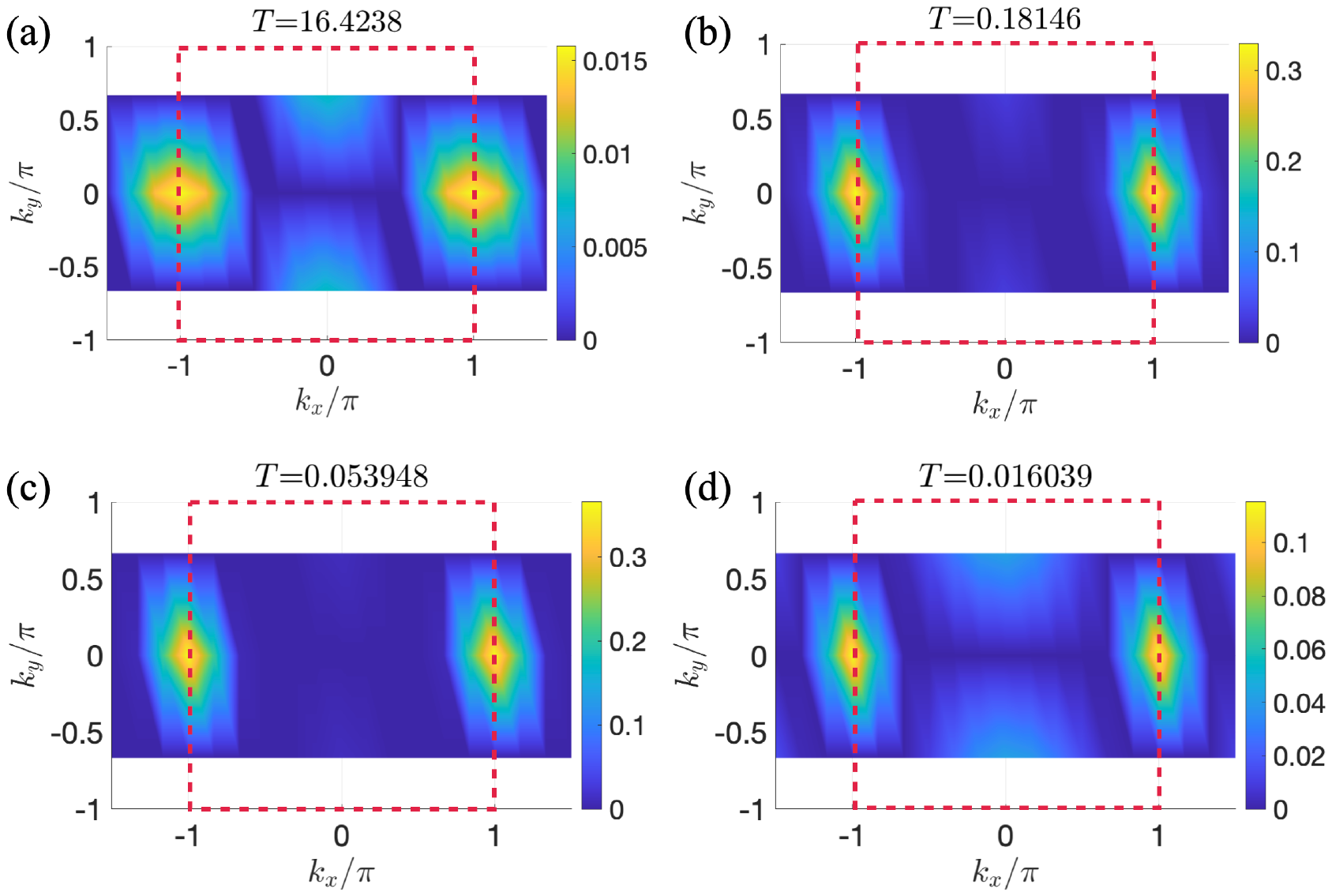}
	\includegraphics[width=0.5\textwidth]{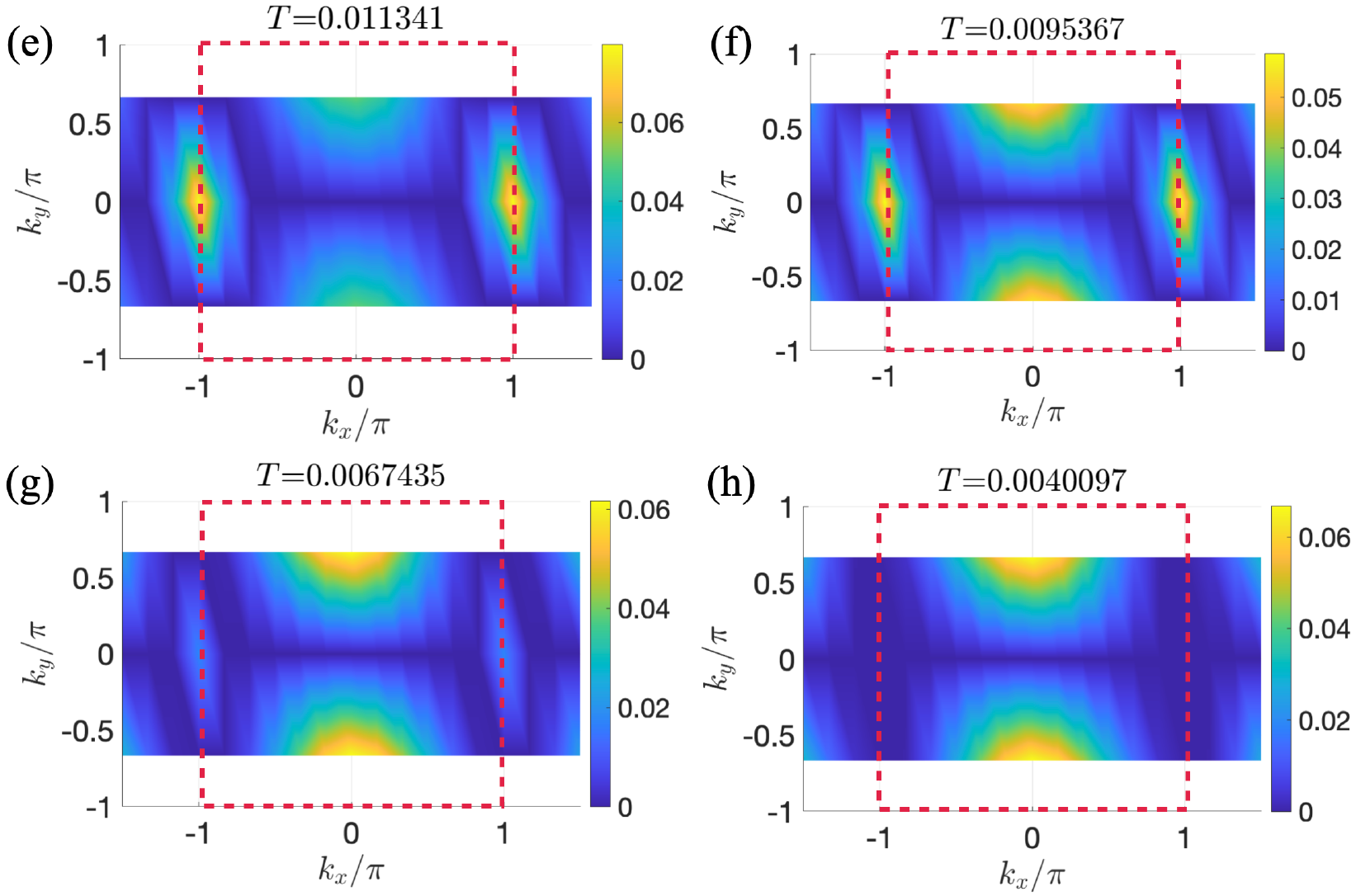}
	\caption{ Detailed structure factors in FQAHS state with $V_3=1$ from high to low temperature.} 
\label{figsm_Sq_allT}
\end{figure}

In Fig.\ref{fig_fqahs_gs} of the main text, we have shown that in the FQAHS state, when the peaks of $n(\mathbf{k})$ are at $(\pm\pi,0)$, the broad peaks of density fluctuation are at $(0,\pm\pi)$, which are the rotons. In the thermodynamic results of FQAHS, we have shown that around $T_c$ (the transition temperature of spontaneously breaking translational symmetry), the structure factor $S(\pm\pi,0)$ goes to the peak and will drop to 0 when approaching the ground state at lower temperature. Meanwhile, since the geometry of our XTRG simulations does not include $\mathbf{k}=(0,\pm\pi)$, we examine $S(0,2\pi/3)$ instead for the roton excitation, and it quickly establishes when $T<T_c$ and finally approaches the constant value around $T^\ast$. In the main text, the finite-temperature structure factors are only plotted at two distinct temperature values, so we show more detailed figures of the structure factor of the FQAHS state with $V_3=1$ in Fig.\ref{figsm_Sq_allT}


\subsection*{C. Supplementary thermodynamic data of FQAHS phase} 

Here, we show the supplementary thermodynamic data of FQAHS phase with $V_3=1$ in Fig.\ref{figsm_thermal1}.
The $\bar n=\mu$ plateau for the FQAHS state is shown in Fig.\ref{figsm_thermal1} (a), where the red dotted line refers to $\mu=2.24$, which we use for fixed-$\mu$ simulations.
The estimated charge gap is from the width of the plateau at the lowest temperature: $\Delta_{\mathrm{cg}}\approx0.15$, which is roughly the $T_{\mathrm{cg}}$ of this state.
Besides, it is in agreement with the analysis in the main text that the thermal entropy approaches 0 under the onset temperature $T^\ast$, but is still finite around $T_c$.
Moreover, the $\bar n-\mu$ plateau in Fig.\ref{figsm_thermal1}(a) shrinks when the temperature increases and it becomes compressible above $T^\ast\ll T_\mathrm{cg}$, as explained in the main text.
\begin{figure*}[htp!]
	\centering		
	\includegraphics[width=0.9\textwidth]{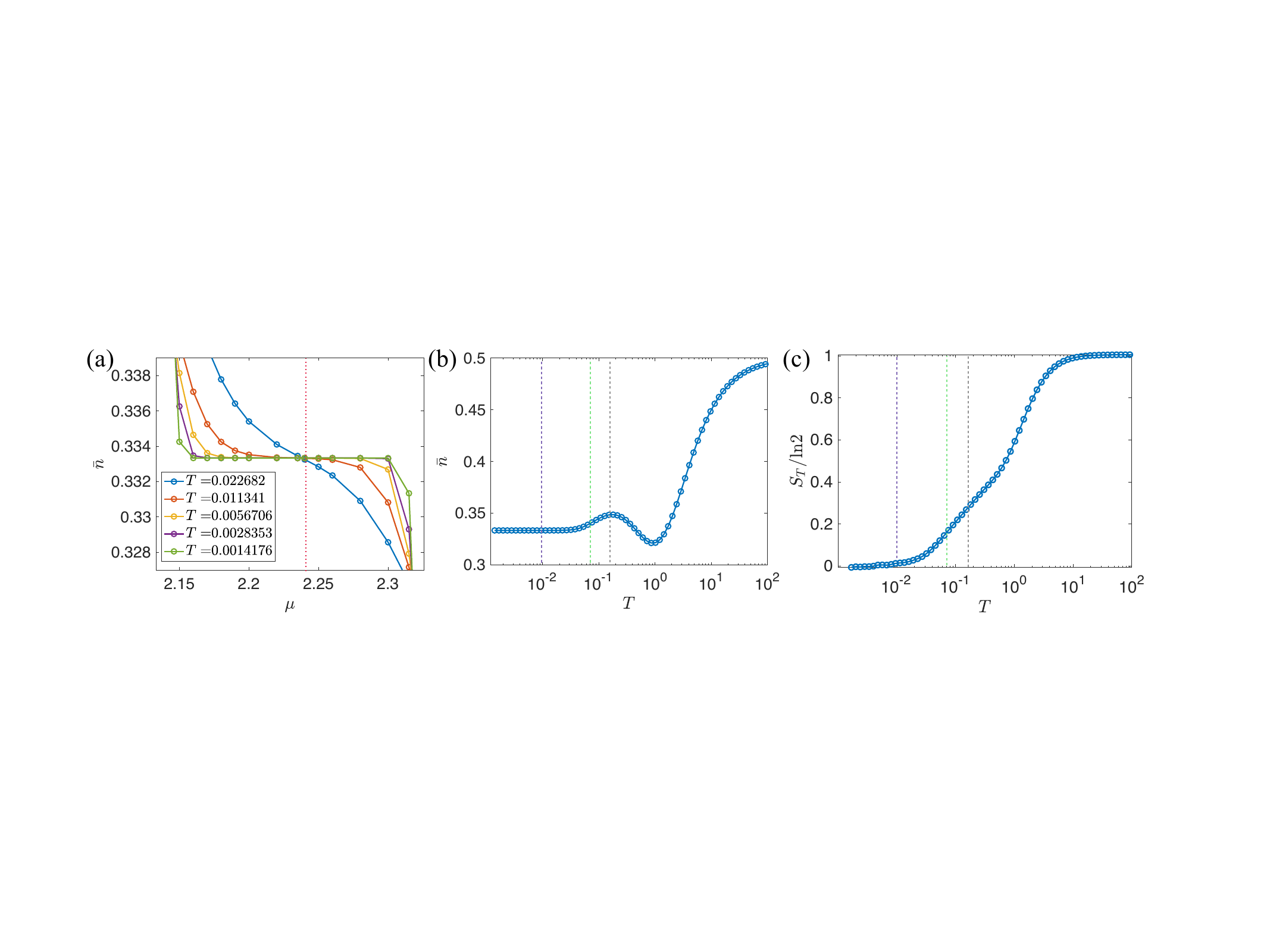}
	\caption{ Supplementary thermodynamic data with $V_3=1$ and bond dimension $D=800$. (a) $\bar n-\mu$ plateau and the estimated charge gap is $\Delta_{\mathrm{cg}}\approx0.15$ from the width of the plateau.  The red dotted line refers to $\mu=2.24$ that we use for fixed-$\mu$ simulations. Change of average density (b) and thermal entropy (c) versus T. The definitions of dashed lines are the same as Fig.\ref{fig_fqahs_thermal}.}
	\label{figsm_thermal1}
\end{figure*}

\subsection*{D. Supplementary thermodynamic data of PSM phase}
Here, we show the supplementary thermodynamic data of PSM phase with $V_3=4$ in Fig.\ref{figsm_thermal2}. Since our XTRG simulations have considered bond dimensions from $D=600$ to $D=800$ in Fig.\ref{fig_psm}, we show the $\bar n-\mu$ curves at different bond dimensions here, showing that the chemical potential is still sensitive to the bond dimension and the thermodynamic simulation of this PSM phase is challenging.

\begin{figure}[htp!]
	\centering		
	\includegraphics[width=0.8\textwidth]{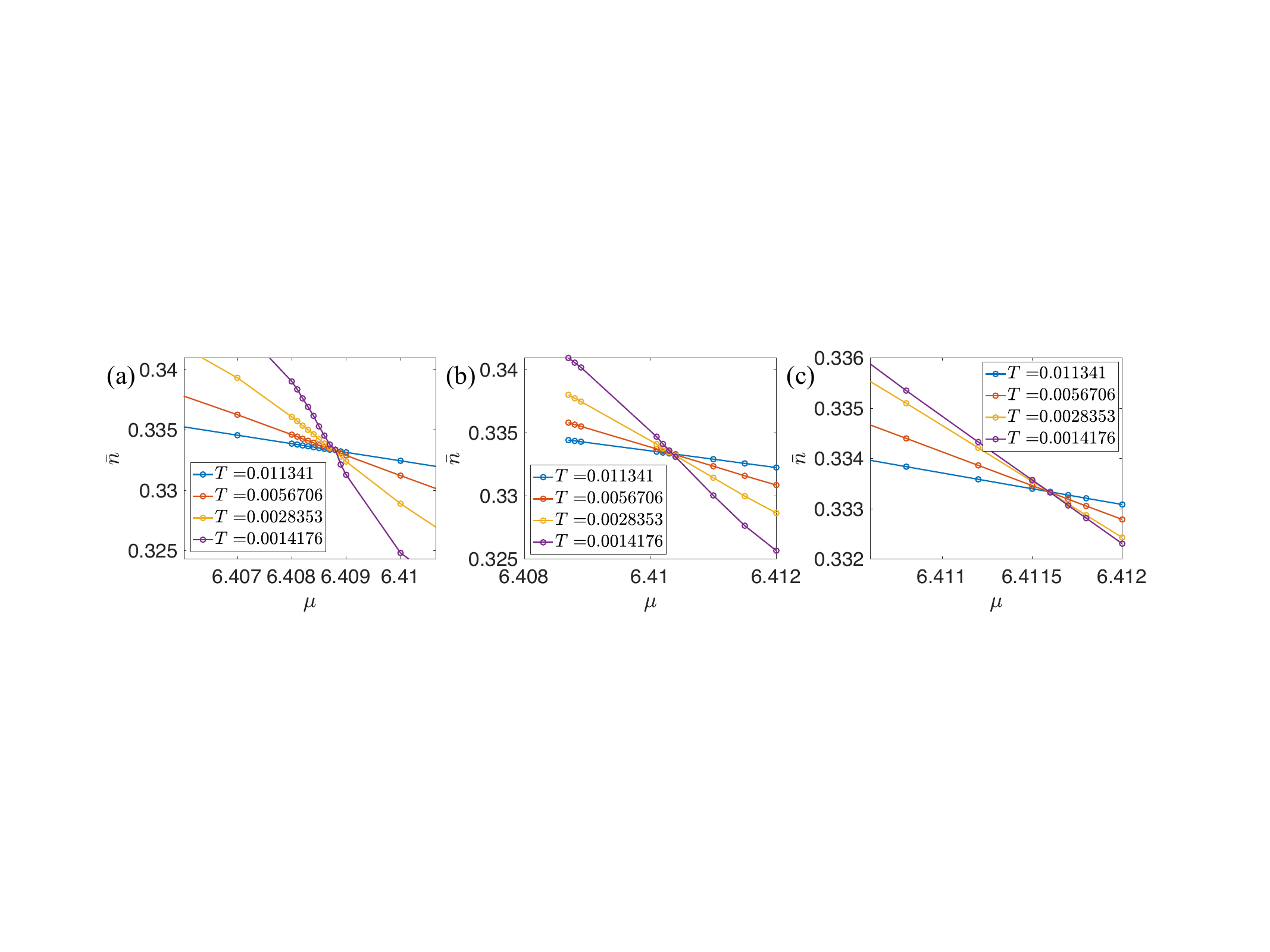}
	\caption{The $\bar n-\mu$ plots at $V_3=4$ with bond dimensions (a) $D=600$, (b) $D=700$, (c) $D=800$ respectively.}
	\label{figsm_thermal2}
\end{figure}

\newpage

\subsection*{E. Degenerate stripe patterns} 

In the main text, we showed that  the PSM state has a dipolar stripe order. Here we present the 4 degenerate charge patterns of this polar smectic order for stripes along the $\mathbf{a_1}$ direction. In the thermodynamic limit, the degeneracy is 8-fold, when the additional 4 ground states with stripes along $\mathbf{a_2}$ are taken into account.
\begin{figure}[htp!]
	\centering		
	\includegraphics[width=0.35\textwidth]{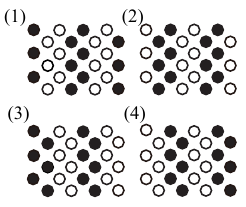}
	\caption{Four degenerate polar stripe patterns along the $\mathbf{a_1}$ direction.}
	\label{figsm_stripepattern}
\end{figure}

\subsection*{F. Order parameters, symmetry analysis and the Ginzburg-Landau theory of smectic phases} 
In this section, we analyze the symmetry breaking patterns for various stripe orders in this checkerboard lattice model, and show that this model supports two different types of smectic states, polar and non-polar. In addition, we will also present the order parameters for these two different smectic orders.

\begin{figure}[htp!]
	\centering		
	\includegraphics[width=0.33\textwidth]{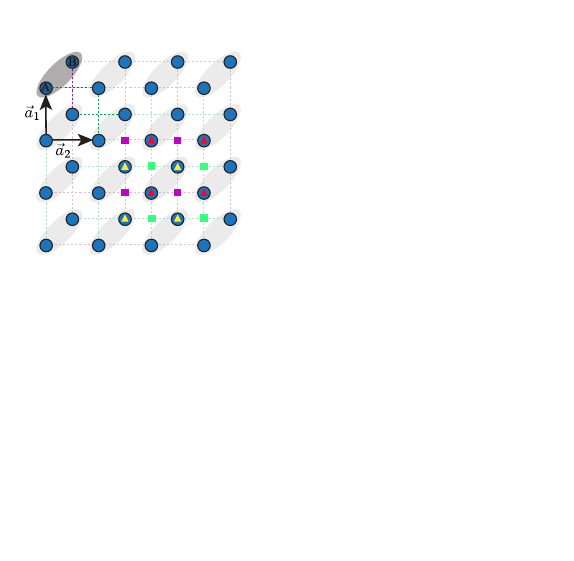}
	\caption{Space group symmetry of the checkerboard lattice model P4 (442). In the absence of any charge order, in each unit cell, there are two 4-fold rotation center (purple and green squares) and two 2-fold rotation center (yellow and red triangles).}
	\label{fig_spacegroup}
\end{figure}

For a checkerboard lattice, it turns out that to achieve this goal we need to take into account the full space-symmetry group, instead of treating translational symmetry breaking and rotational symmetry breaking separately. The 2D space group (known as the wallpaper group) of a naive checkerboard lattice is p4m (*442). However, in our model, because  the reflection and gliding-reflection symmetries are broken by the loop current pattern, the space group is reduced to P4 (442). As shown in Fig.~\ref{fig_spacegroup}, in each unit cell, this model has two 4-fold rotation centers (purple and green squares) and two 2-fold rotation centers (yellow and red triangles). 

We start from nonpolar stripe orders. For a nonpolar stripe order with an ordering wavevector $(\pi,0)$ or $(0,\pi)$, it breaks the 4-fold rotational symmetry down to two-fold. Although this naive statement on rotational symmetry breaking pattern is fully correct, as far as the point group symmetry is concerned, the full story of symmetry breaking is more complicated, once we take into account the space symmetry group. In reality, this is what happens for the two 4-fold rotation centers and two 2-fold rotation centers: (1) the stripe pattern removes one 4-fold rotational center and one 2-fold rotation center; (2) the other 2-fold rotation center remains; (3) the other 4-fold rotation center becomes a 2-fold center.

To better demonstrate this symmetry breaking pattern, here we define four charge stripe order parameters
\begin{align}
\delta^\mathrm{A}_x=\tfrac{2}{N}\sum_i (-1)^{x_i} n^\mathrm{A}_{\mathbf{r}_i}\\
\delta^\mathrm{B}_x=\tfrac{2}{N}\sum_i (-1)^{x_i} n^\mathrm{B}_{\mathbf{r}_i}
\end{align}
and
\begin{align}
\delta^\mathrm{A}_y=\tfrac{2}{N}\sum_i (-1)^{y_i} n^\mathrm{A}_{\mathbf{r}_i}\\
\delta^\mathrm{B}_y=\tfrac{2}{N}\sum_i (-1)^{y_i} n^\mathrm{B}_{\mathbf{r}_i}
\end{align}
where $i$ labels unit cells and $\mathbf{r}_i=(x_i,y_i)$ is the 2D coordinate of the unit cell. Because we set the lattice constant to be unity, $x_i$ and $y_i$ are both integers. $n^\mathrm{A}_{\mathbf{r}_i}$ and $n^\mathrm{B}_{\mathbf{r}_i}$ are average density on site A and site B respectively. The first two order parameters $\delta^\mathrm{A}_x$ and $\delta^\mathrm{B}_x$ describes stripes along $y$ (ordering wavevector along $x$), while the last two order parameters $\delta^\mathrm{A}_y$ and $\delta^\mathrm{B}_y$ describes stripes along $x$ (ordering wavevector along $y$). The superscript $A$ or $B$ indicates whether the charge density wave is from sublattice A or B.

Here, for simplicity, we will focus on stripe orders characterized by $\delta^\mathrm{A}_x$ and $\delta^\mathrm{B}_x$, setting $\delta^\mathrm{A}_y=\delta^\mathrm{B}_y=0$. The same conclusions also apply to stripe orders of $\delta^\mathrm{A}_y$ and $\delta^\mathrm{B}_y$ via a simple $90^\circ$ rotation. One crucial symmetry property of the checkerboard lattice lies in the fact that the order parameters $\delta^\mathrm{A}_x$ and $\delta^\mathrm{B}_x$ break different symmetry and thus they correspond to two totally different stripe orders:
\begin{enumerate}
\item if $\delta^\mathrm{A}_x \ne 0$ and  $\delta^\mathrm{B}_x =0$, i.e., stripe on sublattice A only, the 4-fold rotation center marked by a green square becomes a 2-fold rotation center, the 2-fold rotation center marked by a red triangle remains, and the other two rotation centers are no longer rotation centers anymore.
\item if $\delta^\mathrm{A}_x = 0$ and  $\delta^\mathrm{B}_x \ne 0$, i.e., stripe on sublattice B only, the 4-fold rotation center marked by a purple square becomes a 2-fold rotation center, the 2-fold rotation center marked by a yellow triangle remains, and the other two rotation centers are no longer rotation centers anymore.
\end{enumerate}
Note that although these two stripe ordered states share the same point group ($C_2$), their rotational centers are totally different. Thus, when space group symmetry is taken into account, these two order parameters break totally different symmetry and thus they define two different stripe orders with distinct symmetry. It is also worthwhile to emphasize that these two order parameters $\delta^\mathrm{A}_x$ and  $\delta^\mathrm{B}_x$ are {\it not} connected by any symmetry, and thus it is allowed by symmetry for the system to develop one order but not the other. 

What if both $\delta^\mathrm{A}_x$ and  $\delta^\mathrm{B}_x$ become nonzero? For $\delta^\mathrm{A}_x \ne 0$ and  $\delta^\mathrm{B}_x \ne 0$, the system breaks all rotational symmetry, and there is no rotation center anymore in this ordered state. Because all point group symmetry is broken, an electric dipole moment becomes allowed, and thus the system becomes a polar smectic state with a spontaneously generated ferroelectric order. The ferroelectric order parameter is $\delta^\mathrm{A}_x\times \delta^\mathrm{B}_x$, which is nonzero only if both $\delta^\mathrm{A}_x$and $\delta^\mathrm{B}_x$ become nonzero.

Here we summarize all possible stripe orders (for stripes along y) in this table.\ref{table1}
\begin{table*}[htp!]
	\centering
	\caption{}
\begin{tabular}{ |c| c| c| }
\hline
  & $\delta^\mathrm{A}_x=0$ & $\delta^\mathrm{A}_x\ne 0$ \\ 
\hline
 $\delta^\mathrm{B}_x=0$ & disorder & nonpolar smectic \\  
  & &  (2-fold rotation centers: green square and red triangle)\\
\hline
 $\delta^\mathrm{B}_x\ne 0$ & nonpolar smectic & polar smectic order \\  
 & (2-fold rotation centers: purple square and yellow triangle)  &  (no rotation centers)
 \\
\hline
\end{tabular}
\label{table1}
\end{table*}

With this symmetry knowledge, we can now write down the Ginzburg-Landau theory for such stripe phases. The Ginzburg-Landau free energy is
\begin{equation}
\begin{aligned}
F= &m_1[ (\delta^\mathrm{A}_x)^2+ (\delta^\mathrm{B}_y)^2]+m_2[(\delta^\mathrm{B}_x)^2 + (\delta^\mathrm{A}_y)^2] \\
&+\textrm{higher order terms}.
\end{aligned}
\end{equation}
The higher order terms include quartic terms of $\delta$'s and beyond. 
They give an energy penalty to states with both $\delta_x$ and $\delta_y$ being nonzero, and thus we only see stripe pattern with enlarged unit cells of $2\times 1$ or $1\times 2$. Because a $90^\circ$ rotation swaps A and B, as well as $x$ and $y$, the four fold rotational symmetry enforces a symmetry between $\delta^\mathrm{A}_x$ and $\delta^\mathrm{B}_y$, as well as between $\delta^\mathrm{B}_x$ and $\delta^\mathrm{A}_y$. Thus, we have only two independent quadratic coefficients $m_1$ and $m_2$. 

At small $V_3$ ($V_3<0.2$), both $m_1$ and $m_2$ are large and positive, and thus the disordered phase (all $\delta$'s being zero) is favored. As $V_3$ increases, the values of both $m_1$ and $m_2$ reduce towards zero and eventually trigger a quantum phase transition. 
Our simulation indicates that $m_2$ is likely to be smaller than $m_1$ and thus, the phase transition first leads to a nonpolar smectic order ($0.2<V_3<2.2$): either $\delta^\mathrm{B}_x\ne 0$ or $\delta^\mathrm{A}_y\ne 0$ (i.e., B-site stripes along y or A-site stripes along x). Depending on the signs of the order parameter (positive or negative), we have four degenerate charge patterns. In our systems, due to the 3-fold topological degeneracy, the total ground state degeneracy is 12-fold. 

Upon further increasing $V_3$, both $m_1$ and $m_2$ become either negative or small enough, which triggers a second phase transition $V_3>2.2$. In this phase, the ground states have both $\delta^\mathrm{A}_x\ne 0$ and  $\delta^\mathrm{B}_x\ne 0$ (or  both $\delta^\mathrm{A}_y\ne 0$ and  $\delta^\mathrm{B}_y\ne 0$. For stripes along $y$ we have two nonzero order parameters, $\delta^\mathrm{A}_x$ and $\delta^\mathrm{B}_x$, their signs can be (++), (--), (+-) and(-+), giving us four degenerate ground states. In addition, another four degenerate ground states can be found for stripes along $x$, making total degeneracy 8-fold.

To conclude this section, we introduce another sets of stripe order parameters for bond stripe order
\begin{align}
b^\mathrm{A}_x=\tfrac{2}{N}\sum_i (-1)^{x_i}|\langle c^\dagger_{\mathrm{A},\mathbf{r}_i}c^\dagger_{\mathrm{A},\mathbf{r}_i+(1,0)}\rangle|\\
b^\mathrm{B}_x=\tfrac{2}{N}\sum_i (-1)^{x_i}|\langle c^\dagger_{\mathrm{B},\mathbf{r}_i}c^\dagger_{\mathrm{B},\mathbf{r}_i+(1,0)}\rangle|
\end{align}
and
\begin{align}
b^\mathrm{A}_y=\tfrac{2}{N}\sum_i (-1)^{x_i}|\langle c^\dagger_{\mathrm{A},\mathbf{r}_i}c^\dagger_{\mathrm{A},\mathbf{r}_i+(0,1)}\rangle|\\
b^\mathrm{B}_y=\tfrac{2}{N}\sum_i (-1)^{x_i}|\langle c^\dagger_{\mathrm{B},\mathbf{r}_i}c^\dagger_{\mathrm{B},\mathbf{r}_i+(0,1)}\rangle|
\end{align}
It is easy to check that $\delta^\mathrm{A}$ and $b^\mathrm{B}$ break the same symmetry and thus they describe the same stripe order, while $\delta^\mathrm{B}$ and $b^\mathrm{A}$ break the same symmetry and describe the same stripe order.

\end{widetext}

\end{document}